\documentclass{aa}

\usepackage[dvips]{color}
\usepackage{graphicx}
\usepackage{txfonts}
\usepackage{rotating}
\newcommand{\om}{\hbox{{$\omega$}~Cen}~}

\newcommand{\masyr}{$\mbox{mas}\,\mbox{yr}^{-1}$}

\begin{document}
%
\title{Ground-based     CCD     astrometry     with     wide     field
  imagers.\thanks{Based on archive  observations with the MPI/ESO 2.2m
    telescope, located  at La Silla and  Paranal Observatory, Chile.}}
\subtitle{III.   WFI@2.2$m$  proper-motion  catalog  of  the  globular
  cluster $\omega$~Centauri.}

\author{Bellini, A.\inst{1,}\inst{2}
\and
Piotto, G.\inst{1}
\and
Bedin, L. R.\inst{2}
\and 
Anderson, J.\inst{2}
\and 
Platais, I. \inst{3}
\and
Momany, Y.\inst{4}
\and
Moretti, A.\inst{4}
\and
Milone, A. P. \inst{1}
\and
Ortolani, S. \inst{1}}

\offprints{Bellini, A.}
\institute{ Dipartimento di Astronomia, Universit\`a di Padova, Vicolo
  dell'Osservatorio        3,       I-35122        Padova,       Italy
  \\
  \email{[andrea.bellini;giampaolo.piotto;alessia.moretti;antonino.milone;sergio.ortolani]@unipd.it}
  \and  Space  Telescope Science  Institute,  3700  San Martin  Drive,
  Baltimore,                MD                21218,               USA
  \\ \email{[bellini;bedin;jayander]@stsci.edu} \and Dept.  of Physics
  and Astronomy,  The Johns  Hopkins University, Baltimore,  MD 21218,
  USA   \\    \email{imants@pha.jhu.edu}   \and   INAF:   Osservatorio
  Astronomico  di Padova,  vicolo dell'Osservatorio  5,  35122 Padova,
  Italy \\ \email{yazan.almomany@oapd.inaf.it}}

\date{Received 1 September 2008 / Accepted 3 October 2008}

\abstract
%
%
{$\omega$~Centauri is the most  well studied Galactic Globular Cluster
  because of its numerous puzzling features: significant dispersion in
  metallicity, multiple  populations, triple main-sequence, horizontal
  branch    morphology,    He-rich    population(s),   and    extended
  star-formation history.
Intensive spectroscopic follow-up  observing campaigns targeting stars
at  different  positions in  the  color-magnitude  diagram promises  to
clarify some of these peculiarities.}
%
%
{To be  able to target  cluster members reliably  during spectroscopic
  surveys  and both spatial  and radial  distributions in  the cluster
  outskirts   without   including   field   stars,  a   high   quality
  proper-motion   catalog   of    \om   and   membership   probability
  determination are required.
The only available wide field  proper-motion catalog of \om is derived
from photographic plates, and  only for stars brighter than $B\sim16$.
Using  ESO archive  data, we  create a  new,  CCD-based, proper-motion
catalog for this cluster, extending to $B\sim20$.}
%
%
{We  used high precision  astrometric software  developed specifically
  for data acquired by WFI@2.2$m$ telescope and presented in the first
  paper of this series.
We demonstrated  previously that a  7 mas astrometric  precision level
can be achieved with this  telescope and camera for well exposed stars
in  a  single  exposure,  assuming   an  empirical  PSF  and  a  local
transformation approach in measuring star displacements.}
%
%
{We achieved a good cluster-field separation with a temporal base-line
  of   only   four   years.    We   corrected   our   photometry   for
  sky-concentration  effects.  We  provide  calibrated photometry  for
  $UBVR_CI_C$ wide-band data plus  narrow-band filter data centered on
  $H_\alpha$ for almost $360\,000$ stars.
We confirm that the \om  metal-poor and metal-rich components have the
same  proper  motion,  and  demonstrate  that  the  metal-intermediate
component in addition  exhibits the same mean motion  as the other RGB
stars.  We provide membership probability determinations for published
\om variable star catalogs.}
%
%
{Our catalog  extends the proper-motion measurements to  fainter than
  the   cluster  turn-off   luminosity,   and  covers   a  wide   area
  $(\sim33\arcmin\times33\arcmin)$ around the  center of $\omega$~Cen.
  Our  catalog   is  electronically  available   to  the  astronomical
  community.}

%

\maketitle

\section{Introduction}
\label{sec_intro}

The  globular  cluster $\omega$~Centauri  ($\omega$~Cen)  is the  most
luminous and massive cluster in the Galaxy.
Observational evidence collected over the years has indicated that \om
is also the most puzzling  stellar system in terms of stellar content,
structure, and kinematics.
Probably the most well studied  of its peculiarities is one related to
its    stellar   metallicity    distribution   (Norris    \&   Bessell
\cite{norris75}; \cite{norris77}; Freeman \& Rodgers \cite{freeman75};
Bessell  \& Norris  \cite{bessell76}; Butler  et  al. \cite{butler78};
Norris \&  Da Costa \cite{norris95}; Suntzeff  \& Kraft \cite{sunt96};
Norris et al.  \cite{norris96}).
There is  a significant dispersion in the  iron abundance distribution
of      $\omega$~Cen,     with      a      primary     peak      about
$\rm{[Fe/H]}\sim-1.7$-$-1.8$   and   a   long   tail,   extending   to
$\rm{[Fe/H]}\sim-0.6$, which contains another 3-4 secondary peaks.
It  is possible  to  identify these  metallicity  peaks with  distinct
stellar  populations (Pancino  et  al.  \cite{pancino00};  Rey et  al.
\cite{rey04};  Sollima  et  al.   \cite{sollima05}; Villanova  et  al.
\cite{villanova07}).
Ground-based   (Lee    et   al.    \cite{lee99};    Pancino   et   al.
\cite{pancino00})  and \textit{Hubble Space  Telescope} (\textit{HST})
(Anderson \cite{jay97};  Bedin et al.  \cite{bedin04};  Ferraro et al.
\cite{ferraro04})  photometry show  clearly that  \om  hosts different
stellar populations.
In particular, Pancino et al. (\cite{pancino00}) demonstrated that the
\om  red  giant  branch  (RGB)  consists of  at  least  four  distinct
branches, spanning a wide range of metallicity.
On the other hand, the \om sub giant branch (SGB) has an intricate web
of 5  distinct sequences, indicating an extended  range of metallicity
and   age  (see   Bedin  et   al.   \cite{bedin04};   Hilker   et  al.
\cite{hilker04};  Sollima et  al.  \cite{sollima05};  Stanford  et al.
\cite{stanford06}; Villanova et al.  \cite{villanova07}) .

Anderson (\cite{jay97}), Bedin et al.  (\cite{bedin04}), and Villanova
et al.  (\cite{villanova07}), by  studying fainter stars with deep and
high  resolution \textit{HST} photometry,  demonstrated that  the main
sequence (MS) is divided into 3 distinct sequences.
The spectroscopic study of the MS stars of \om by Piotto et al. (2005)
showed that the  bluest of the main sequences is  more metal rich than
any of the redder sequences, which increased the ambiguity surrounding
the cluster.
An  overabundance of  He in  the blue  MS could  reproduce the  \om MS
photometric    and   spectroscopic    properties    (Bedin   et    al.
\cite{bedin04};     Norris    \cite{norris04};    Piotto     et    al.
\cite{piotto05}), although the origin of the puzzling MS morphology is
still far from being understood.

\begin{figure}[t!]
\centering
\includegraphics[width=9.0cm,height=9.0cm]{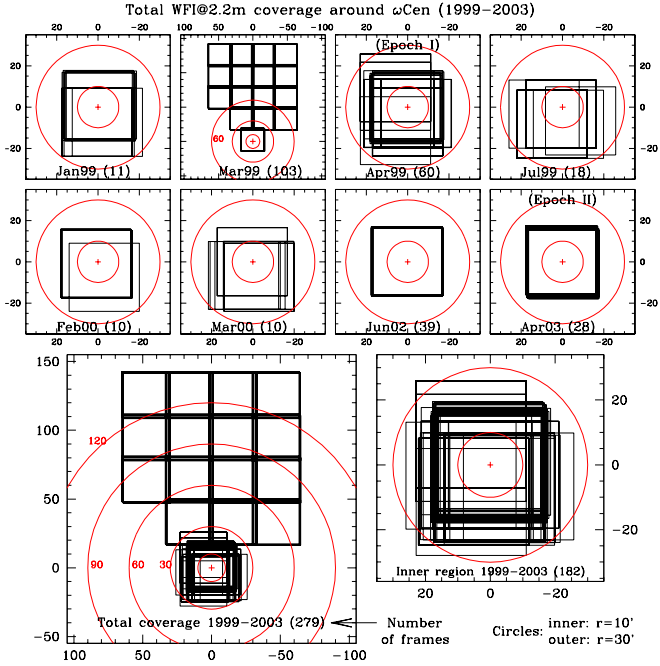}
\caption{Position footprint of the  entire sample of WFI images around
  the \om center (marked with a ``$+$'').  The first two rows show the
  covered areas sorted by month and  year.  In the lower panels of the
  figure, the total  coverage of all of the 279  images (on the left),
  and a zoom-in of the central  part of the cluster (on the right) are
  shown.   The numbers in  parenthesis after  the dates  represent the
  total number of images for that observing run.  North is up, East to
  the left.}
\label{fig_totfow}
\end{figure}

\begin{table}[th!]
\caption{Description of the data set used for the WFI@2.2$m$ catalog.}
\centering
\small{
\begin{tabular}{cccc}
\hline
\hline
filter&$t_{exp}$&seeing&airmass\\
\hline
\multicolumn{4}{c}{ }\\
\multicolumn{4}{c}{January, 1999}\\
\hline
$B_{842}$$\!\!\!\!\!\!\!$&3$\times$30s,1$\times$300s;$\!\!\!\!\!\!\!$&$1\arcsec.0$-$1\arcsec.3$$\!\!\!\!\!\!\!$&$\sim1.20$\\
$658$nm$\!\!\!\!\!\!\!$&1$\times$120s,5$\times$180s,1$\times$900s;$\!\!\!\!\!\!\!$&$1\arcsec.0$-$1\arcsec.3$$\!\!\!\!\!\!\!$&$1.15$-$1.18$\\
&&&\\
\multicolumn{4}{c}{March, 1999}\\
\hline
$V_{843}$$\!\!\!\!\!\!\!$&52$\times$200s;$\!\!\!\!\!\!\!$&$0\arcsec.8$-$1\arcsec.2$$\!\!\!\!\!\!\!$&1.1$$-$$1.2\\
$I_{845}$$\!\!\!\!\!\!\!$&51$\times$150s;$\!\!\!\!\!\!\!$&$0\arcsec.7$-$1\arcsec.5$$\!\!\!\!\!\!\!$&1.1$$-$$1.45\\
&&&\\
\multicolumn{4}{c}{April, 1999}\\
\hline
$R_{844}$$\!\!\!\!\!\!\!$&1$\times$5s,1$\times$10s,1$\times$15,1$\times$30,5$\times$60s;$\!\!\!\!\!\!\!$&$1\arcsec.0$-$1\arcsec.3$$\!\!\!\!\!\!\!$&1.3$$-$$1.6\\
$I_{845}$$\!\!\!\!\!\!\!$&1$\times$5s,1$\times$10s,1$\times$20s,1$\times$45s,4$\times$90s;$\!\!\!\!\!\!\!$&$0\arcsec.74$-$1\arcsec.7$$\!\!\!\!\!\!\!$&1.4$$-$$1.9\\
$658$nm$\!\!\!\!\!\!\!$&2$\times$30,4$\times$120,5$\times$180s;$\!\!\!\!\!\!\!$&$0\arcsec.8$-$1\arcsec.15$$\!\!\!\!\!\!\!$&$1.1$-$1.2$\\
&&&\\
\multicolumn{4}{c}{\bf(epoch~I)}\\
$B_{842}$$\!\!\!\!\!\!\!$&1$\times$15s,1$\times$30s,1$\times$60s,5$\times$120s;$\!\!\!\!\!\!\!$&$0\arcsec.75$-$1\arcsec.3$$\!\!\!\!\!\!\!$&1.2$$-$$1.5\\
$V_{843}$$\!\!\!\!\!\!\!$&3$\times$5s,3$\times$10s,1$\times$15s,2$\times$20s;$\!\!\!\!\!\!\!$&$0\arcsec.7$-$1\arcsec.0$$\!\!\!\!\!\!\!$&1.2$$-$$1.6\\
$\!\!\!\!\!\!\!$&1$\times$30s,4$\times$45s,10$\times$90s;
$\!\!\!\!\!\!\!$&$0\arcsec.76$-$1\arcsec.36$$\!\!\!\!\!\!\!$&1.1$$-$$1.5\\
&&&\\
\multicolumn{4}{c}{July, 1999}\\
\hline
$U_{877}$$\!\!\!\!\!\!\!$&2$\times$1800s;$\!\!\!\!\!\!\!$&$1\arcsec.4$-$1\arcsec.6$$\!\!\!\!\!\!\!$&$\sim1.14$\\
$B_{842}$$\!\!\!\!\!\!\!$&1$\times$10s,1$\times$30s,1$\times$40s,1$\times$300s;$\!\!\!\!\!\!\!$&$1\arcsec.4$-$1\arcsec.8$$\!\!\!\!\!\!\!$&$1.14$-$1.25$\\
$V_{843}$$\!\!\!\!\!\!\!$&1$\times$10s,1$\times$20s,1$\times$150s,1$\times$240s;$\!\!\!\!\!\!\!$&$1\arcsec.3$-$1\arcsec.6$$\!\!\!\!\!\!\!$&$1.15$-$1.25$\\
$I_{845}$$\!\!\!\!\!\!\!$&1$\times$10s,2$\times$20s,1$\times$240s;$\!\!\!\!\!\!\!$&$1\arcsec.13$-$1\arcsec.6$$\!\!\!\!\!\!\!$&$1.17$-$1.25$\\
$658$nm$\!\!\!\!\!\!\!$&3$\times$120s,1$\times$1200s;$\!\!\!\!\!\!\!$&$1\arcsec.13$-$1\arcsec.5$$\!\!\!\!\!\!\!$&$1.18$-$1.23$\\
&&&\\
\multicolumn{4}{c}{February, 2000}\\
\hline
$B_{842}$$\!\!\!\!\!\!\!$&2$\times$30s,2$\times$240s;$\!\!\!\!\!\!\!$&$1\arcsec.45$-$1\arcsec.7$$\!\!\!\!\!\!\!$&$\sim1.13$\\
$V_{843}$$\!\!\!\!\!\!\!$&1$\times$30s,2$\times$240s;$\!\!\!\!\!\!\!$&$1\arcsec.1$-$1\arcsec.2$$\!\!\!\!\!\!\!$&$\sim1.13$\\
$I_{845}$$\!\!\!\!\!\!\!$&1$\times$30s,2$\times$240s;$\!\!\!\!\!\!\!$&$\sim1\arcsec.0$$\!\!\!\!\!\!\!$&$\sim1.13$\\
&&&\\
\multicolumn{4}{c}{March, 2000}\\
\hline
$U_{841}$$\!\!\!\!\!\!\!$&4$\times$300s,1$\times$2400s;$\!\!\!\!\!\!\!$&$1\arcsec.1$-$1\arcsec.2$$\!\!\!\!\!\!\!$&1.17$$-$$1.38\\
$V_{843}$$\!\!\!\!\!\!\!$&5$\times$30s;&$0\arcsec.9$-$1\arcsec.4$$\!\!\!\!\!\!\!$&$\sim1.45$\\
&&&\\
\multicolumn{4}{c}{June, 2002}\\
\hline
$U_{877}$$\!\!\!\!\!\!\!$&4$\times$30s,7$\times$300s;$\!\!\!\!\!\!\!$&$0\arcsec.8$-$2\arcsec.0$$\!\!\!\!\!\!\!$&$1.14$-$1.18$\\
$B_{878}$$\!\!\!\!\!\!\!$&1$\times$5s,3$\times$8s,9$\times$60s;$\!\!\!\!\!\!\!$&$0\arcsec.8$-$1\arcsec.5$$\!\!\!\!\!\!\!$&$1.13$-$1.16$\\
$V_{843}$$\!\!\!\!\!\!\!$&3$\times$5s,3$\times$40s,3$\times$60s;$\!\!\!\!\!\!\!$&$0\arcsec.75$-$2\arcsec.0$$\!\!\!\!\!\!\!$&$\sim1.13$\\
$I_{845}$$\!\!\!\!\!\!\!$&3$\times$20s,3$\times$40s;$\!\!\!\!\!\!\!$&$0\arcsec.7$-$1\arcsec.4$$\!\!\!\!\!\!\!$&$\sim1.13$\\
&&&\\
\multicolumn{4}{c}{April, 2003 {\bf ~~~ (epoch~II)}}\\
\hline
$B_{878}$$\!\!\!\!\!\!\!$&7$\times$40s;$7\times120$s;$\!\!\!\!\!\!\!$&$0\arcsec.7$-$0\arcsec.9$$\!\!\!\!\!\!\!$&$1.14$-$1.16$\\
$V_{843}$$\!\!\!\!\!\!\!$&7$\times$40s;$7\times120$s;$\!\!\!\!\!\!\!$&$0\arcsec.8$-$1\arcsec.0$$\!\!\!\!\!\!\!$&$1.13$\\
&&&\\
\hline
\end{tabular}}
\label{tab:data}
\end{table}

A deeper insight into the  enigmatic stellar populations of \om should
combine a deep, high resolution analysis of the inner and most crowded
regions, with a wide field observations of the outskirts.
While  the  first  type  of  data  has  been  provided  adequately  by
\textit{HST}, wide  field coverage  of \om requires  ground-based data
which are more difficult to obtain.
Acquisition of data for a wider field-of-view will inevitably result in
higher contamination by Galactic foreground/background populations.
The only reasonable  and efficient way to decontaminate  the \om outer
stellar populations is by means of proper-motion analyses that help to
isolate the Galactic contribution.
The  only available wide field \om  proper-motion catalog (van Leeuwen
et al. \cite{vanleeuwen00}, hereafter vL00)  is based on  photographic
observations  and only provides measurements  for stars brighter than
$B\sim16$.
In this paper, we attempt to provide the first CCD-based proper motion
catalog  of $\omega$~Cen,  extending the  cleaned  stellar populations
down to $B\sim20$, i.e.  4 mag.  deeper than vL00.

In the  first paper of  this series, Anderson et  al.  (\cite{jay06}),
hereafter Paper~I, demonstrated that WFI@2.2$m$ observations,
with  a  time  base-line of  only  a  few  years, allow  a  successful
separation of cluster members from Galactic field stars in the two GCs
closest to the Sun: NGC6121 and NGC6397.
In this paper, we apply the high precision astrometric and photometric
techniques developed  by Paper~I to  all available WFI@2.2$m$
archive data of $\omega$~Cen.

In   Sect.~2,  we   describe   the  available   WFI  observations   of
$\omega$~Cen, and the data sets that we used to derive proper motions.
In Sect.~3,  we discuss our photometric data  reduction technique, the
sky-concentration    effect   minimization,   and    the   photometric
calibration.
In  Sect.~4, we  describe in  detail how  we treated  the differential
chromatic refraction (DCR) effects between the two epochs.
Membership probability  is discussed in  Sect.~5, while in  Sect.~6 we
outline possible applications of our catalog.
Finally,  in  Sect.~7,  we  summarize  our results  and  describe  the
electronic catalog.

\section{Observations}
\label{sec_obs}

\begin{table}[t!]
\caption{Characteristics of the used filters (from WFI manual)
$\lambda_c$$=$central wavelength,  $\mbox{FWHM}$$=$Full Width  at Half
Maximum,     $\lambda_p$$=$transmission         peak       wavelength,
$\mbox{T}_p$$=$transmission percentage  at peak  level.  (*) LWP means
Long Wave Pass: in this case the cutoff limit is determined by the CCD
quantum efficiency.}
\centering
\begin{tabular}{ccccc}
\multicolumn{5}{c}{\textbf{Wide-band filters}}\\
\hline
Name & $\lambda_c$ & FWHM  & $\lambda_p$  & $\mbox{T}_p$ \\
& [nm]        & [nm]  &   [nm]     & [\%]\\
\hline
$U_{877}\,(\rm{U}/50)$&340.4&73.2&350.3&82.35\\
$U_{841}\,(\rm{U}/38)$&363.7&38.3&362.5&51.6\\
$B_{878}\,(\rm{B}/123)$&451.1&133.5&502.5&88.5\\
$B_{842}\,(\rm{B}99)$&456.3&99.0&475.0&91.2\\
$V_{843}\,(\rm{V/89})$&539.6&89.4&523.0&87.0\\
$R_{844}\,(\rm{R_C})$&651.7&162.2&668.5&93.9\\
$I_{845}\,(\rm{I_C/lwp})$&783.8&LWP$^*$&1001.0&97.6\\
&&&&\\
\multicolumn{5}{c}{\textbf{Narrow-band filters}}\\
\hline
Name& $\lambda_c$ & FWHM  & $\lambda_p$  & $\mbox{T}_p$ \\
& [nm]        & [nm]  &   [nm]     & [\%]\\
\hline
$658$nm$\,(\rm{H}_\alpha)$&658.8&10.3&504.0&90.7\\
\end{tabular}
\label{tab:filters}
\end{table}

We used  a collection of  279 archive images acquired  between January
20, 1999 and April 14, 2003 at the ESO/MPI2.2$m$ telescope at La Silla
(Chile) equipped with the wide-field imager camera (WFI).
A detailed log of observations is reported in Table~\ref{tab:data}.
This camera,  which consists of  an array mosaic of  $4\times2$ chips,
$2141\times 4128$ pixels each, has a total field of view of $34\arcmin
\times 33\arcmin$, and a pixel scale of $0.238^{\prime \prime}$/pixel.
More details of the instrumental setup were given in Paper~I.
Images  were  obtained  using  $U,B,V,R_C,I_C$ wide-band  and  $658$nm
($H_\alpha$)  narrow-band  filters,  whose characteristics  are
summarized in Table~\ref{tab:filters}.

For the derivation of proper motions,  we used only $B$ and $V$ images
acquired  in April  1999 and  April 2003  (see  Sect.~\ref{sec_PM} for
further details of this choice).
The total field-of-view  covered by the entire sample  is indicated in
Fig.~\ref{fig_totfow},      where       axis      coordinates      are
($\Delta\alpha\cos\delta$,  $\Delta\delta$),  expressed  in  units  of
arcmin from the \om center (North is up, East is to the left).
Concentric circles  have diameters,  if not specified,  of $10\arcmin$
(inner circle) and $30\arcmin$ (outer circle), and are centered on the
cluster center: $\alpha=201^{\circ}.69065$, $\delta=-47^{\circ}.47855$
(Van de Ven et al. \cite{vdv06}).

The first eight plots show the covered areas sorted by month and year,
while in  the bottom part of Fig.~\ref{fig_totfow}  the total coverage
of all of the 279 images (on the left), and a zoom of the central part
of the cluster (on the right) are shown.
In the catalog presented in  this work, the proper motion measurements
are available only within the  field-of-view in common between the two
epochs used (see Fig.~\ref{fig_totfow}).

\section{Photometry, Astrometry, and Calibration}
\label{sec_photo}

\subsection{Photometric reduction}
\label{subsec_photo1}

For the reduction of the WFI@2.2$m$ photometric data, we used
the  software   {\sf  img2xym\_WFI},   a  modified  version   of  {\sf
  img2xym\_WFC.09x10}  (Anderson  \&  King  \cite{jay03}),  which  was
written  originally for \textit{HST}  images, adapted  successfully to
ground-based data, and described in detail in Paper~I.
We closely  followed the  prescription given in  Paper~I for  the data
reduction of  WFI images.  This includes standard  operations with the
pixel  data, such  as  de-biasing, flat-fielding,  and correction  for
cosmic rays hits.

At the basis of the star  position and flux measurements, there is the
fitting of  the \emph{empirical} Point Spread Function  (PSF).  In our
approach, the  PSF is represented  by a look-up  table on a  very fine
grid.
It is  well known that the shape  of the PSF changes  with position in
WFI@2.2$m$  chips.  This  variability  can be  modeled by  an
array of PSFs across the chip.
The {\sf img2xym\_WFI} software works in a fully-automated way to find
appropriate  stars to  represent  the PSF  adequately.  For  practical
purposes, the number of PSF stars  per chip can vary between 1 and 15,
depending on the richness of the star-field.
An iterative  process is  designed to work  from the brightest  to the
faintest stars and find  their precise position and instrumental flux.
A  reasonably  bright  star  can  be  measured  with  a  precision  of
$\sim0.03$ pixel ($\sim 7$ mas) on a single exposure.

Another problem of the WFI@2.2$m$ imager is a large geometric
distortion in the focal plane that effectively changes the pixel scale
across  the  field-of-view.  There  are  different  ways  to map  this
geometric distortion.
We adopted a $9\times17$ element look-up table of corrections for each
chip,  derived from multiple,  optimally-dithered observations  of the
Galactic  bulge  in Baade's  Window   (Paper~I).   
This  look-up table  provides  the most  accurate characterization  of
geometrical distortions available for the WFI@2.2$m$.  At any
given location  on the detector, a bilinear  interpolation between the
four closest grid points on the look-up table provides the corrections
for the target point.
The derived look-up table may have  a lower accuracy at the edges of a
field, because  of the way  in which the self-calibration  frames were
dithered (see Paper~I).
An  additional  source  of   uncertainty  is  related  to  a  possible
instability distortions  in the WFI@2.2$m$  reported earlier.
This  prompted us  to use  the local  transformation method  to derive
proper motions (see Sect.~\ref{sec_PM}).

\subsection{Sky-concentration correction}
\label{subsec_skyconc}

\begin{figure}[t!]
\centering
\includegraphics[width=9.0cm,height=9.0cm]{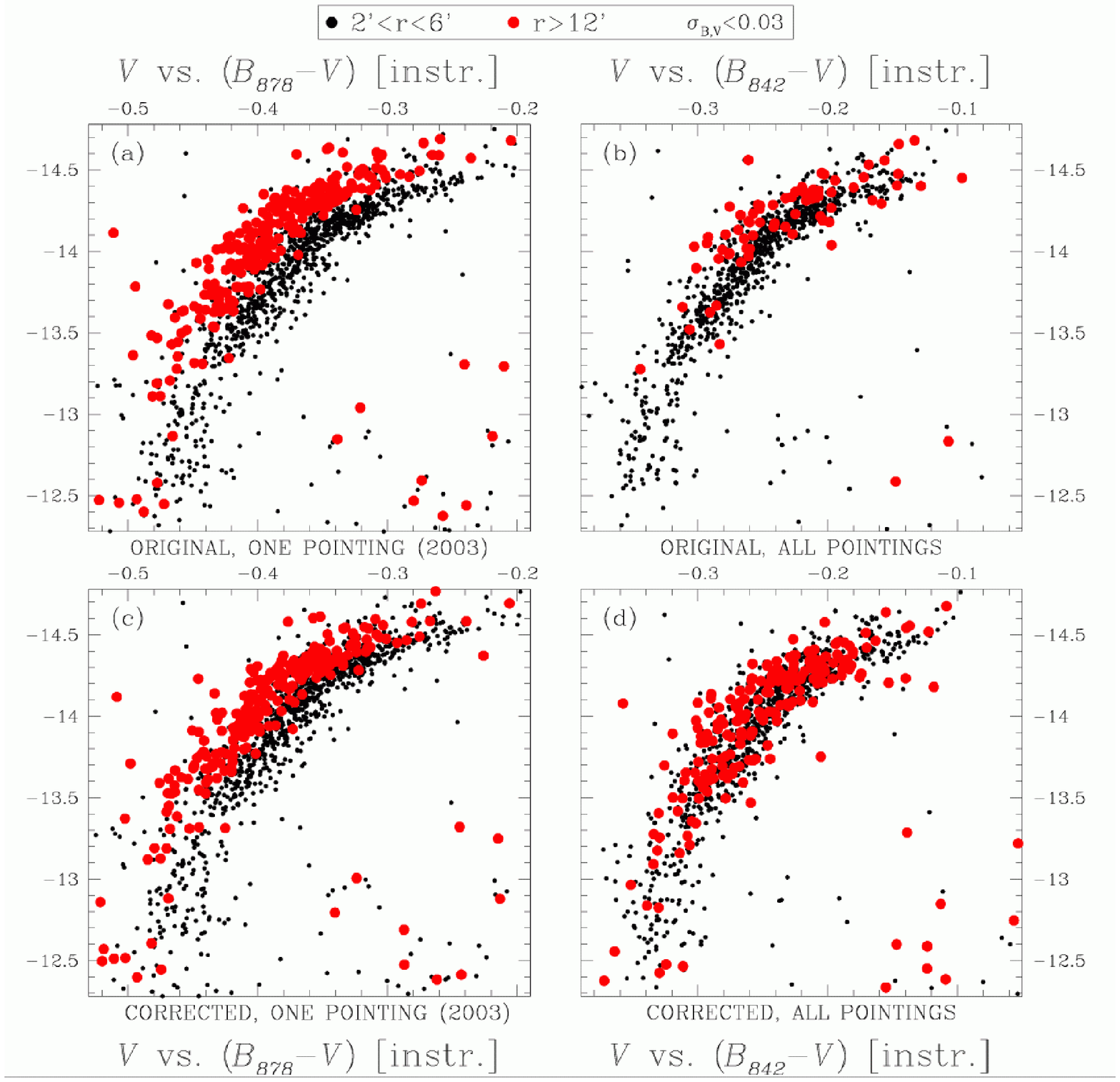}
\caption{CMDs zoomed into the  HB region of $\omega$~Cen.  All plotted
  stars have $\sigma_{B,V}<0.03$ mag. On the left, we used only images
  taken in April 2003 (i.e. only  one pointing). On the right, we show
  photometry  from  all  $V$   and  all  $B_{842}$  images  that  have
  independent  pointings.    Upper  diagrams  are   for  the  original
  photometric   catalogs.    The   lower   ones   are   derived   from
  sky-concentration-corrected catalogs.  Black  dots are stars located
  close to  the mosaic center, while  red dots are stars  close to the
  mosaic edges.}
\label{fig:sky1}
\end{figure}

\begin{figure*}[t!!]
\centering
\includegraphics[width=6cm]{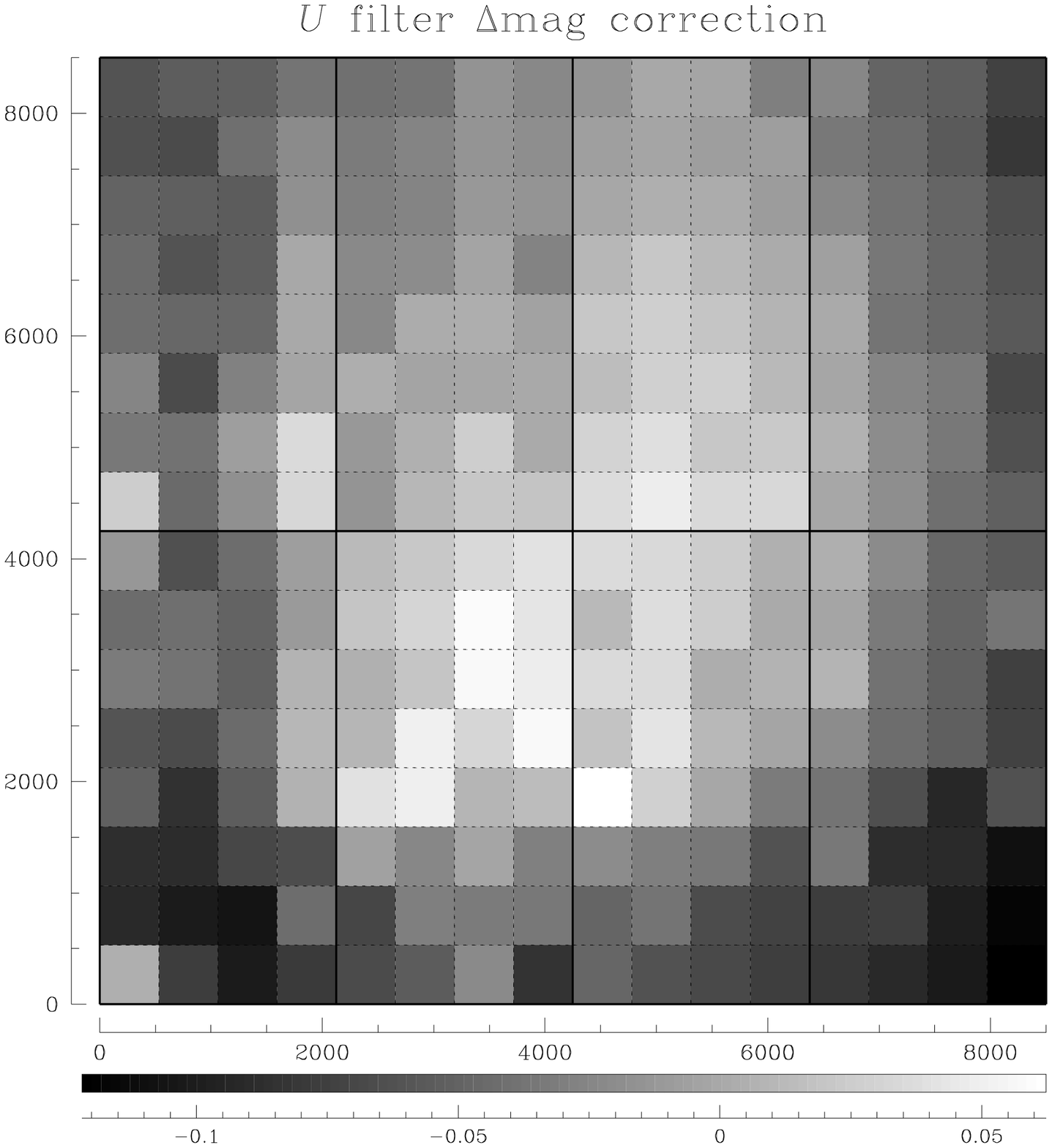}
\includegraphics[width=6cm]{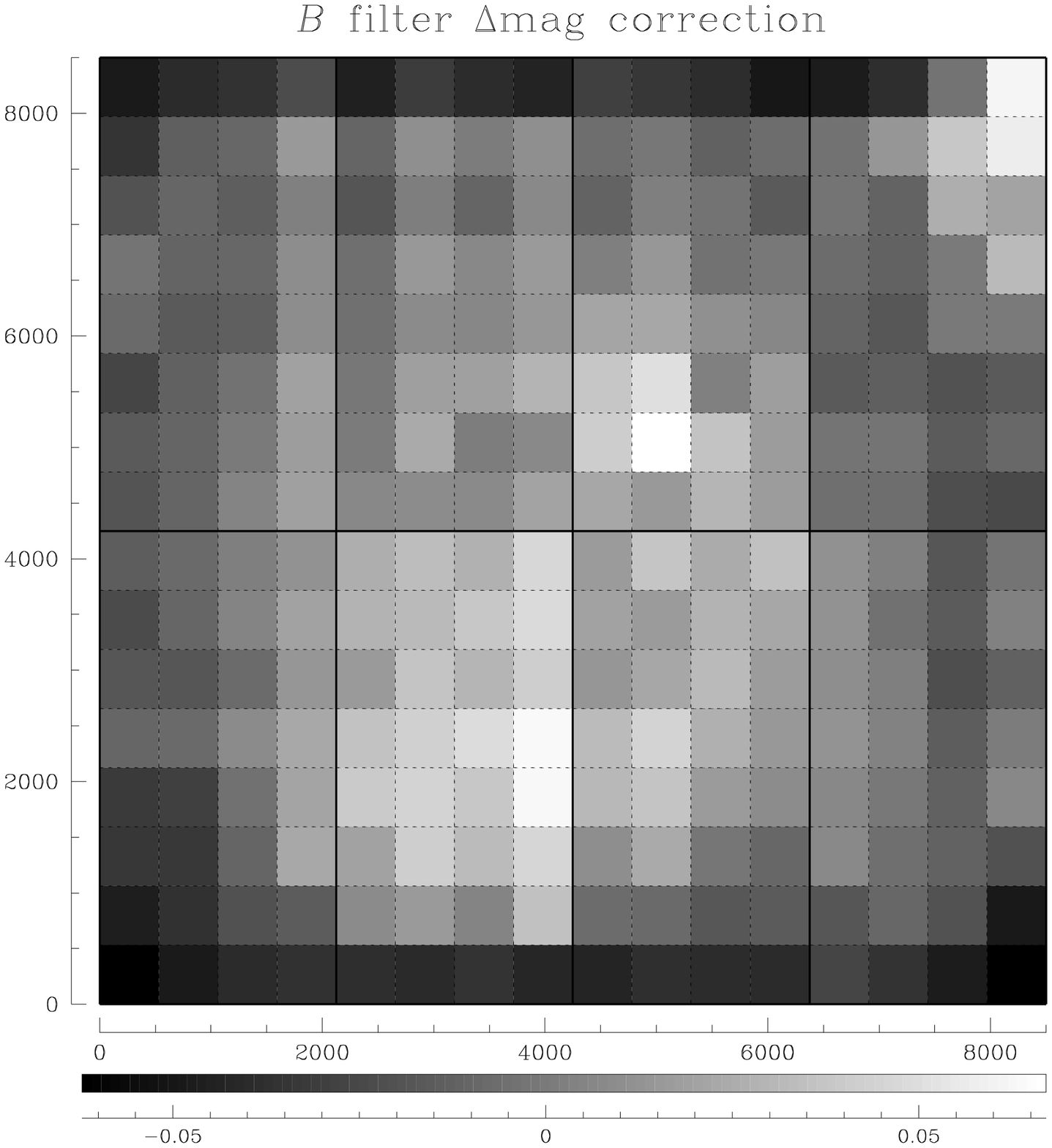}
\includegraphics[width=6cm]{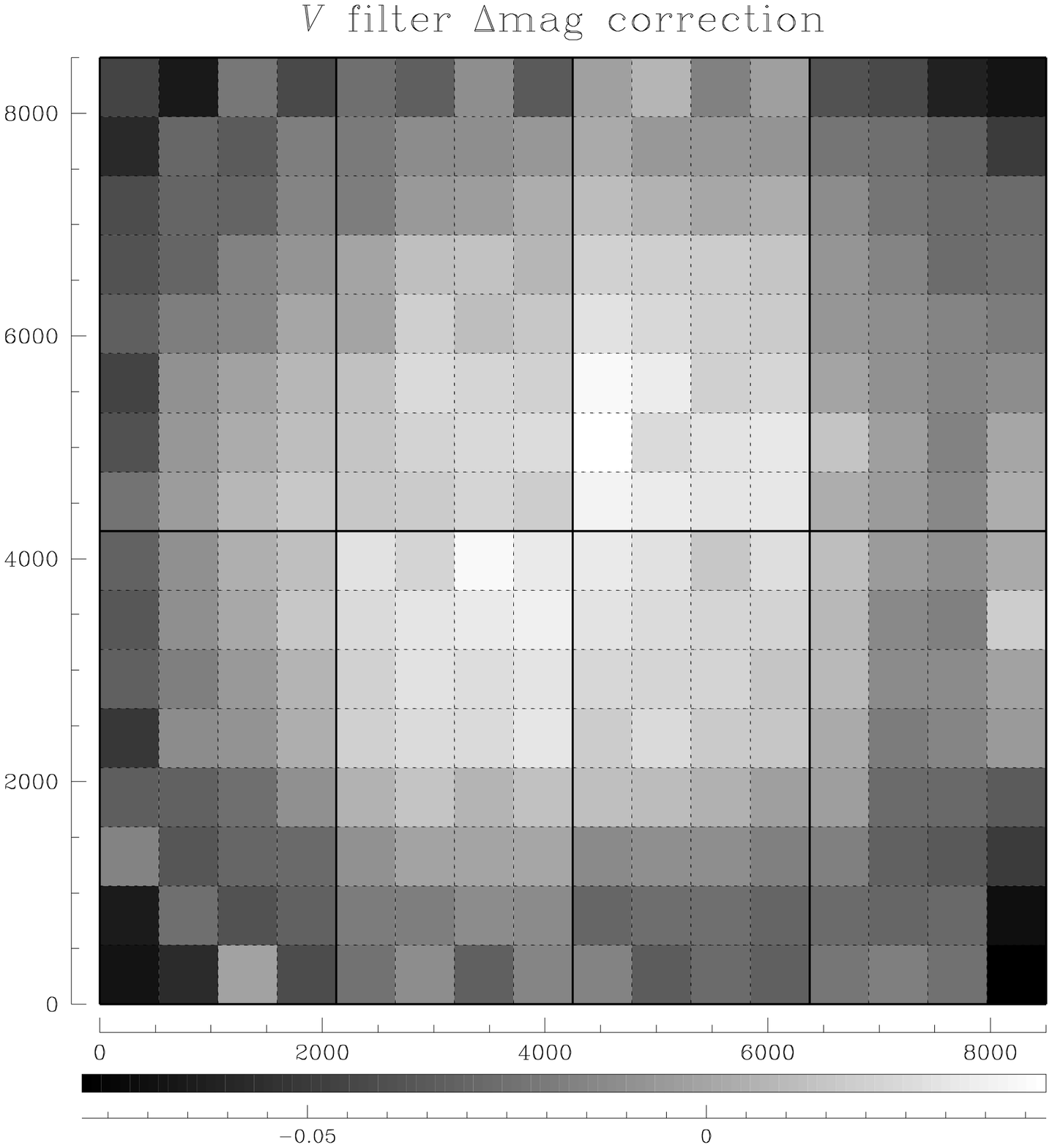}\\
\includegraphics[width=6cm]{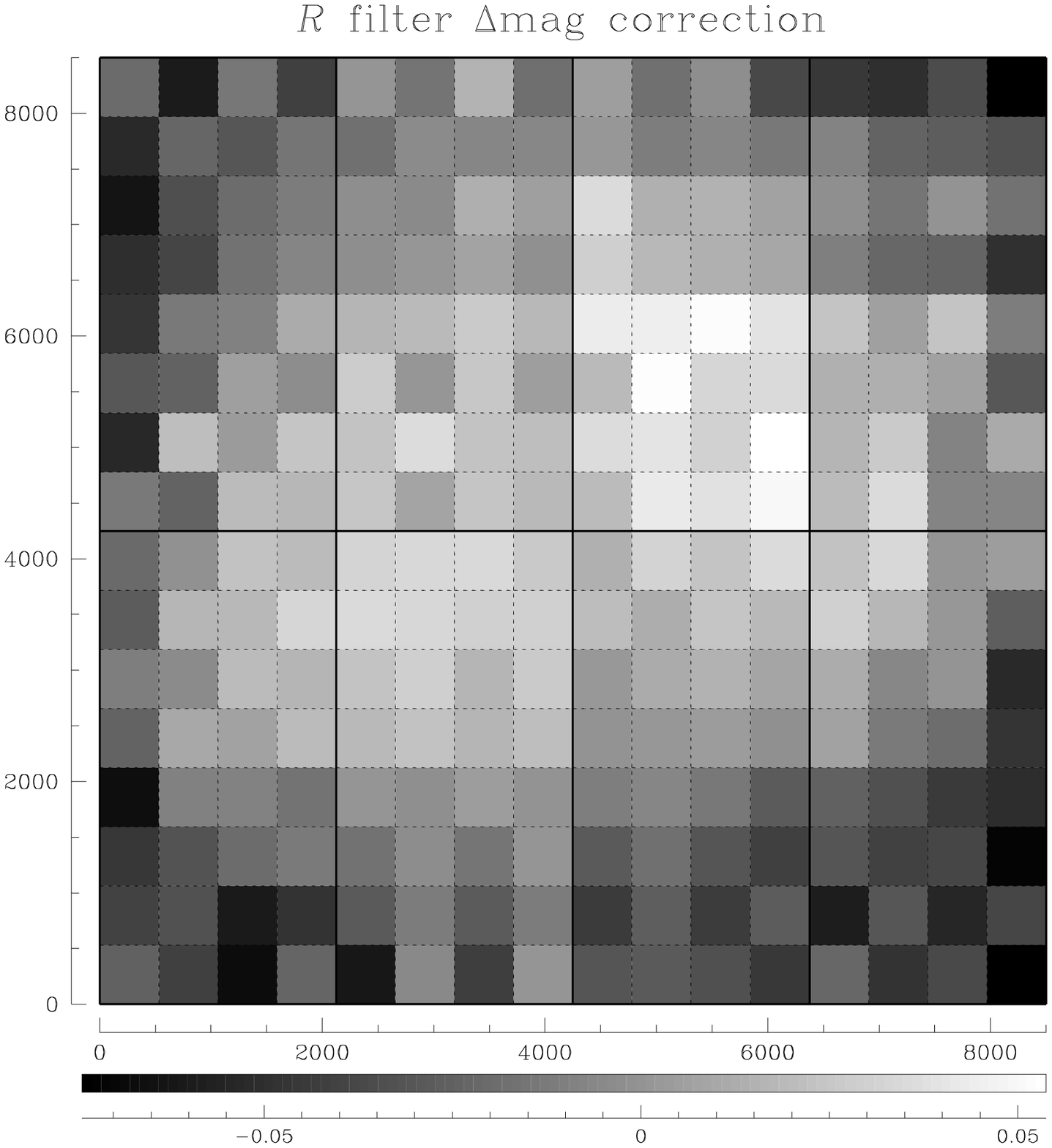}
\includegraphics[width=6cm]{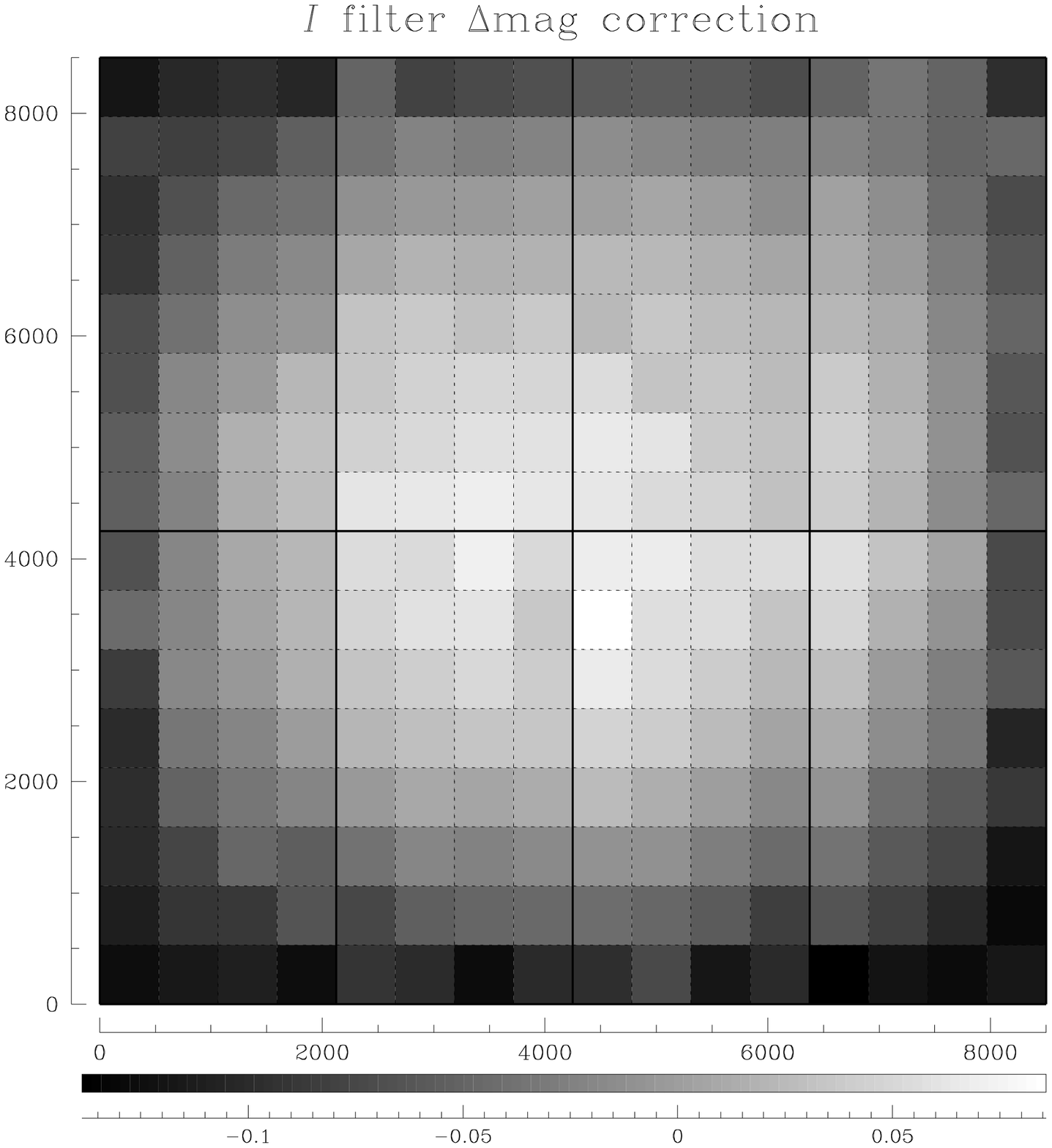}
\includegraphics[width=6cm]{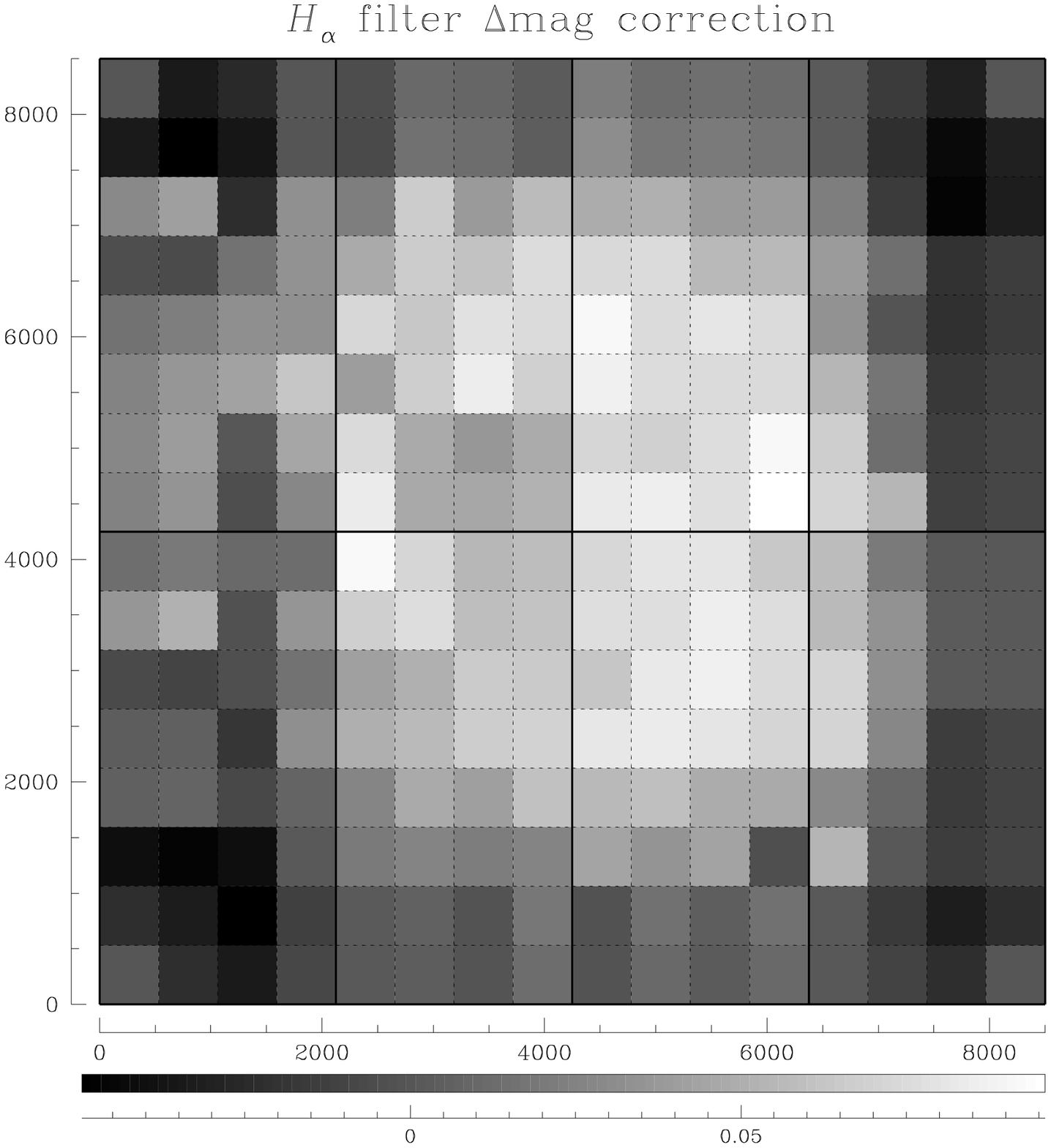}
\caption{Final $\Delta$mag correction  grids for the different filters
  ($B$ means the $B_{842}$ filter  only).  Coordinates are in units of
  WFI pixels.  Each WFI chip is highlighted by continuous black lines.
  Each element  of the  correction grids is  colored according  to the
  corresponding $\Delta$mag  correction applied.  The  grey scales are
  relative to the minimum/maximum correction for each filter.}
\label{fig:skyarray}
\end{figure*}

Once  we  obtained star  positions  and  instrumental  fluxes for  all
images, we had to minimize the so-called ``sky-concentration'' effect.
The WFI@2.2$m$ camera is  affected significantly by this kind
of light  contamination (Manfroid \&  Selman \cite{manfroid01}), which
is caused by  spurious reflections of light at  discontinuities in the
optics and the subsequent redistribution of light in the focal plane.
The  insidiousness  of  the  effect  is  due to  the  fact  that  this
redistribution of  light affects both  the science and  the flat-field
exposures.

Star  fluxes are  calculated by  considering  a local  sky value,  and
therefore may be a negligible effect.
However, since  sky contamination also  affects flat images, if  it is
not  corrected properly during  pre-reduction procedures,  the quantum
efficiency  of  the central  pixel  will  be  artificially lower  with
respect to that of the corner pixel.
Consequently, the luminosity  of a star measured in  the middle of the
mosaic  camera will be  underestimated by  $\sim0.1$-$0.2$ magnitudes
(in $V$ band) with respect to the luminosity of the same star detected
close to the mosaic edges.

In Fig.~\ref{fig:sky1}a, we plot an instrumental\footnote{Instrumental
  magnitudes are calculated to be $-2.5 \times \log({\Sigma_i DN_i})$,
  where  $DN$ are  the pixel's  Digital  Numbers above  the local  sky
  summed within  a 10-pixel circular  aperture.  For a mean  seeing of
  $0.8\arcsec$,   saturation  initiates   at   about  $\sim   -14.5$.}
color-magnitude  diagram  (CMD),  which   has  been  zoomed  into  the
horizontal-branch (HB)  region of $\omega$~Cen,  obtained by combining
all $V$ and $B_{878}$ images of April 2003.
We  chose this  particular data  set to  highlight the  effect  of sky
concentration on un dithered images.   In fact, this data set has only
one pointing, as shown in Fig.~\ref{fig_totfow}.
The positions of stars on the  CCD mosaic is almost identical from one
image to  another, implying a small  contribution to the  r.m.s of the
single star magnitude measurement due to sky-concentration effects.
In  Fig.~\ref{fig:sky1}a, we plotted  1252 stars  with $\sigma_V<0.03$
mag and  $\sigma_B<0.03$ mag, where $\sigma_V$ and  $\sigma_B$ are the
standard errors of a single measurement (r.m.s).

However,  with  only one  pointing,  sky  concentration maximizes  its
effect  on  the relative  photometry  of  stars  located at  different
positions on the image.
In  Fig.~\ref{fig:sky1}a, we  highlight the  CMD of  stars  located at
different positions  on the CCD  mosaic: with black dots  (994 objects
precisely), we  show all stars  between $2\arcmin$ to  $6\arcmin$ from
\om  center,  which  is  close  to  the  mosaic  center  ($x=4150.69$,
$y=4049.97$ on our master meta-chip).
Red  points   are  stars  (258  objects)   outside  $12\arcmin$.   The
displacement  of  the two  HBs  clearly  shows that  sky-concentration
effects affect  WFI photometry significantly  if only one  pointing is
analyzed, and therefore needs to be corrected.

In our  case, the analyzed data  sets for different  filters come from
several pointings (except the case of $B_{878}$ images).
In the process of matching all catalogs (for a given filter) to create
a   single  master-frame,   the  true   sky-concentration   effect  is
reduced. For a  given star, we considered the mean  of the star fluxes
originating  in different  positions and  for different  pointings, so
that   the  sky-concentration   effect   in  the   master  frame   was
reduced.  This process  does, however,  create systematic  errors that
affect the global photometry (see Selman \cite{selman01}).

In  Fig.~\ref{fig:sky1}b,  we  show  the  same  zoomed  HB  region  of
$\omega$~Cen derived, in this case,  by matching all the available $V$
and $B_{842}$ images obtained from the ESO archive.
This data  set contains several different pointings  for both filters,
so we  were able to  obtain photometry for  the same stars  located in
some cases close to the mosaic  center and in other pointings close to
the mosaic edges.
All plotted stars have again $\sigma_{B,V}<0.03$ mag.
In this case,  only 972 stars (with the  same previous convention, 896
black  and 76  red)  passed the  selection  criteria on  the basis  of
photometric error.
As explained before, matching catalogs for different pointing tends to
minimize  sky-concentration   effects,  but  without   an  appropriate
correction, r.m.s. of measurements for the stars are enhanced.

Andersen  et al.   (\cite{andersen95})  studied the  sky-concentration
effect,  typical of  focal  reducers, both  by  using simulations  and
analyzing data from the Danish telescope at La Silla.
Their method  for deriving the sky-concentration  correction was based
on the complex  analysis of many star-field images  taken at different
orientations and positions during the night.
Manfroid et al. (\cite{manfroid01}) applied a similar method to derive
the  sky-concentration  effect  for  WFI@2.2$m$  data,  while
Selman   (\cite{selman01})  developed   a  method   to   estimate  the
sky-concentration effect by the analysis of the zeropoint variation in
3  dithered stellar  frames,  by evaluating  this  variations using  a
Chebyshev polynomial fit.
They  were able  to  reduce the  internal  error from  0.034 to  0.009
magnitudes  in the $V$  filter, and  used the  same polynomial  fit to
correct  in addition  the photometry  in the  other filters  (see also
Selman \& Melnick \cite{selman05}).
Koch  et al.   (\cite{koch04}) provided  an analogous  prescription to
correct  for  the  sky-concentration  effect by  comparing  photometry
derived with WFI and Sloan Digital Sky Survey (SDSS) data.
Finally,  Calamida et al.   (\cite{calamida08}) used  some of  the \om
images  that we present  in this  work to  correct for  the positional
effects of the WFI camera by means of photometric comparisons with the
local standard stars of $\omega$~Cen.

The  correction  given by  the  ESO team,  based  on  the $V$  filter,
consists of  a $9^{\rm  th}$ order bidimensional  Chebyshev polynomial
that should in principle be used also for the $U$ and $B$ filters.
Selman (\cite{selman01}) found that his  solution for the $V$ band was
able to reduce the internal  photometric error from 0.029 to 0.010 mag
in $B$,  and from 0.040 to 0.014  in $U$, while for  the other filters
the $V$ correction failed to reduce the internal photometric error.
Selman (\cite{selman01}) argued that  this is probably due to problems
associated  with atmospheric variations  affecting data  for different
filters.
Unfortunately, by  using the  same polynomial coefficients  to correct
both  $V$ and $B$  magnitudes, it  is impossible  to remove  the color
degeneracy  due  to the  different  response of  the  CCD  to the  sky
concentration in the two different photometric bands.  This degeneracy
is of  the order of $\sim0.04$-$0.05$  mag in color in  our $V$ versus
$B-V$ CMD  between inner stars ($\rm{r}\sim4\arcmin$)  and outer stars
($\rm{r}>12\arcmin$).

Our adopted solution consists  of a self-consistent autocalibration of
the sky-concentration map,  and takes advantage of the  high number of
images analyzed, taken with different pointings.
Below, we provide a description of the autocalibration procedure.
We  measured the raw magnitude ${\rm{mag}}_{i,j}$  of  each $i$-star, in
each $j$-image.
We selected  an image to  be a reference  frame (at the center  of the
dither pattern), and by using common stars with frame $j$ we were able
to  compute   the  {\em  average}  magnitude  shifts   to  bring  each
image-catalog onto the magnitude reference frame ($\Delta_{j}$).

\begin{figure*}[ht!]
\centering
\includegraphics[width=9cm,height=5cm]{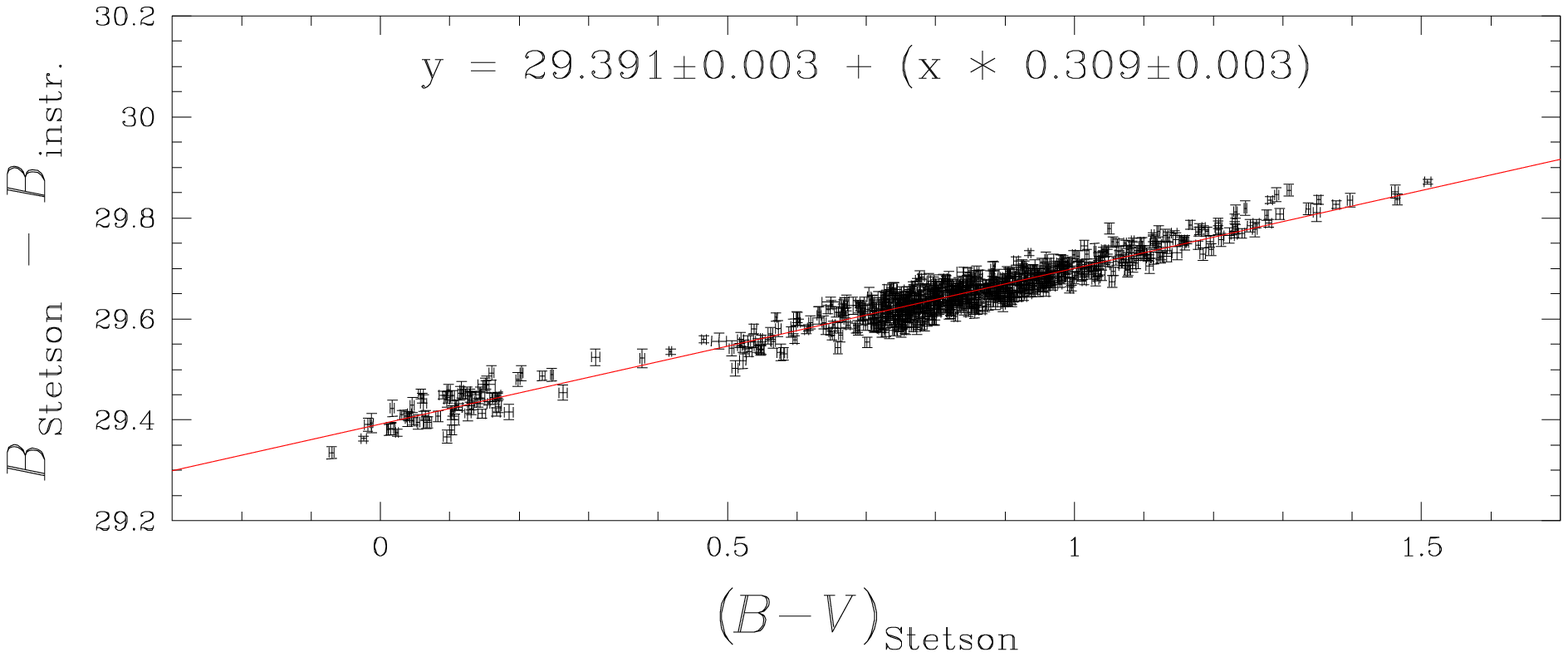}
\includegraphics[width=9cm,height=5cm]{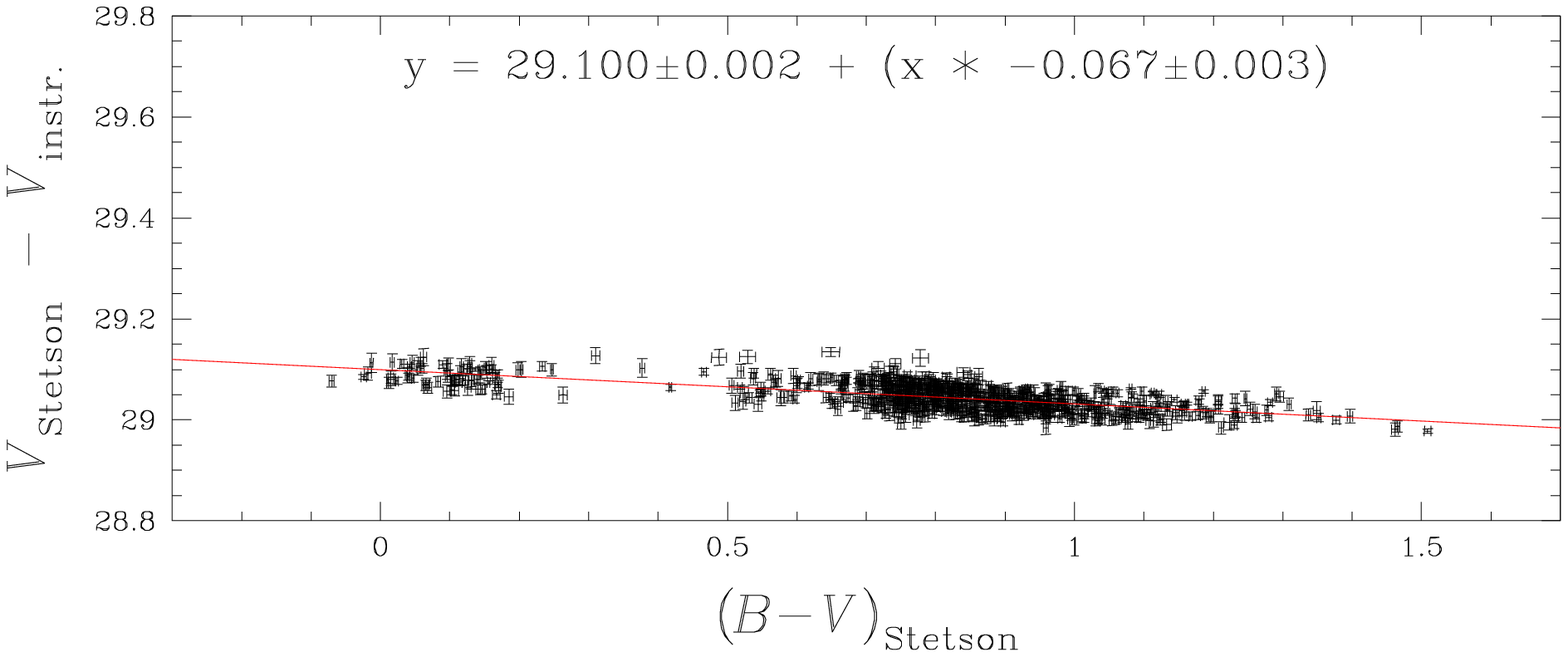}\\
\includegraphics[width=9cm,height=5cm]{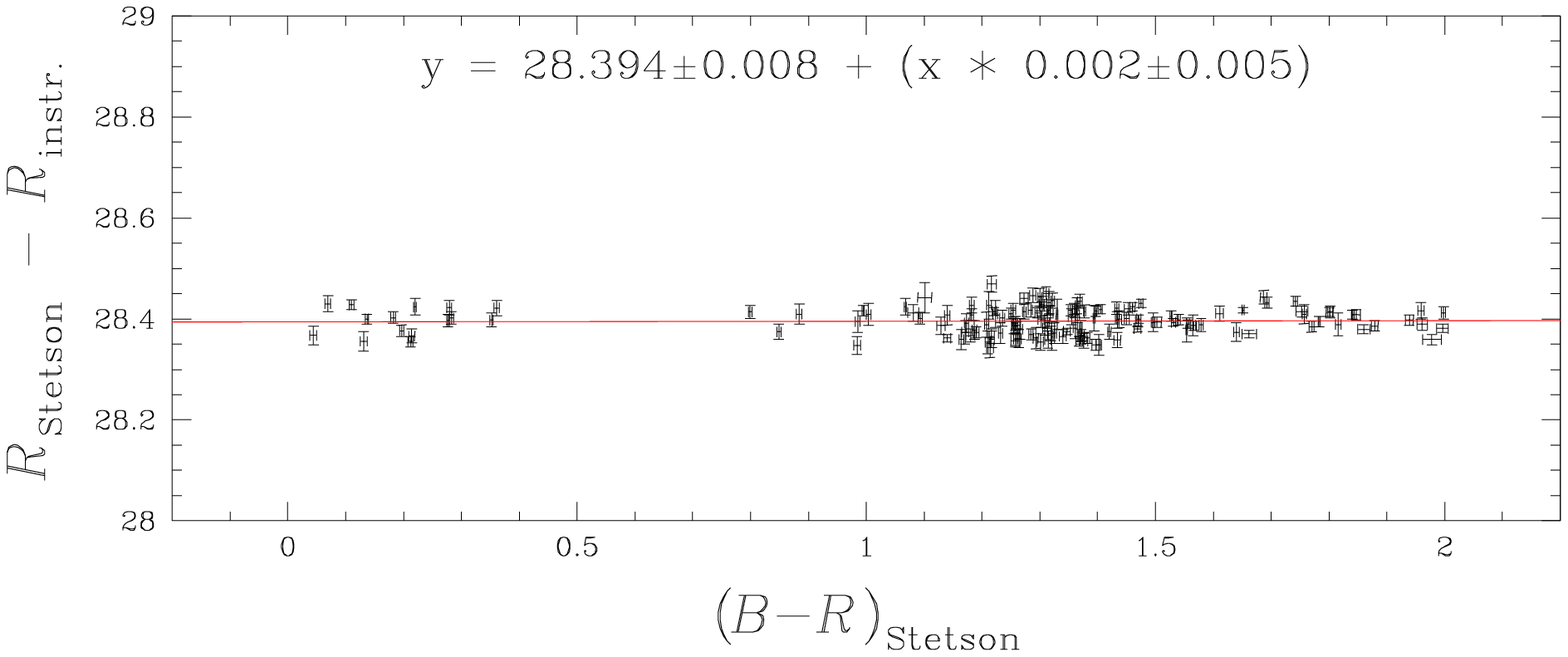}
\includegraphics[width=9cm,height=5cm]{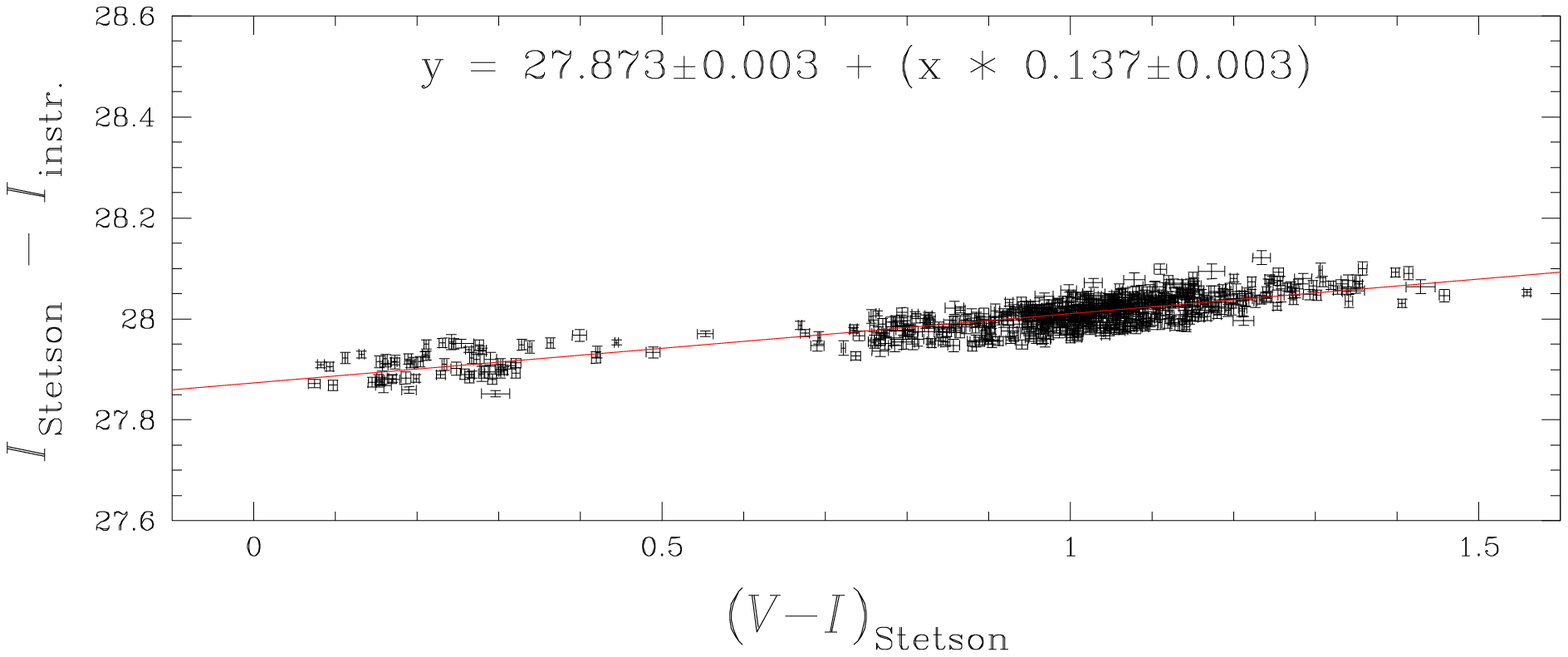}\\
\includegraphics[width=9cm,height=5cm]{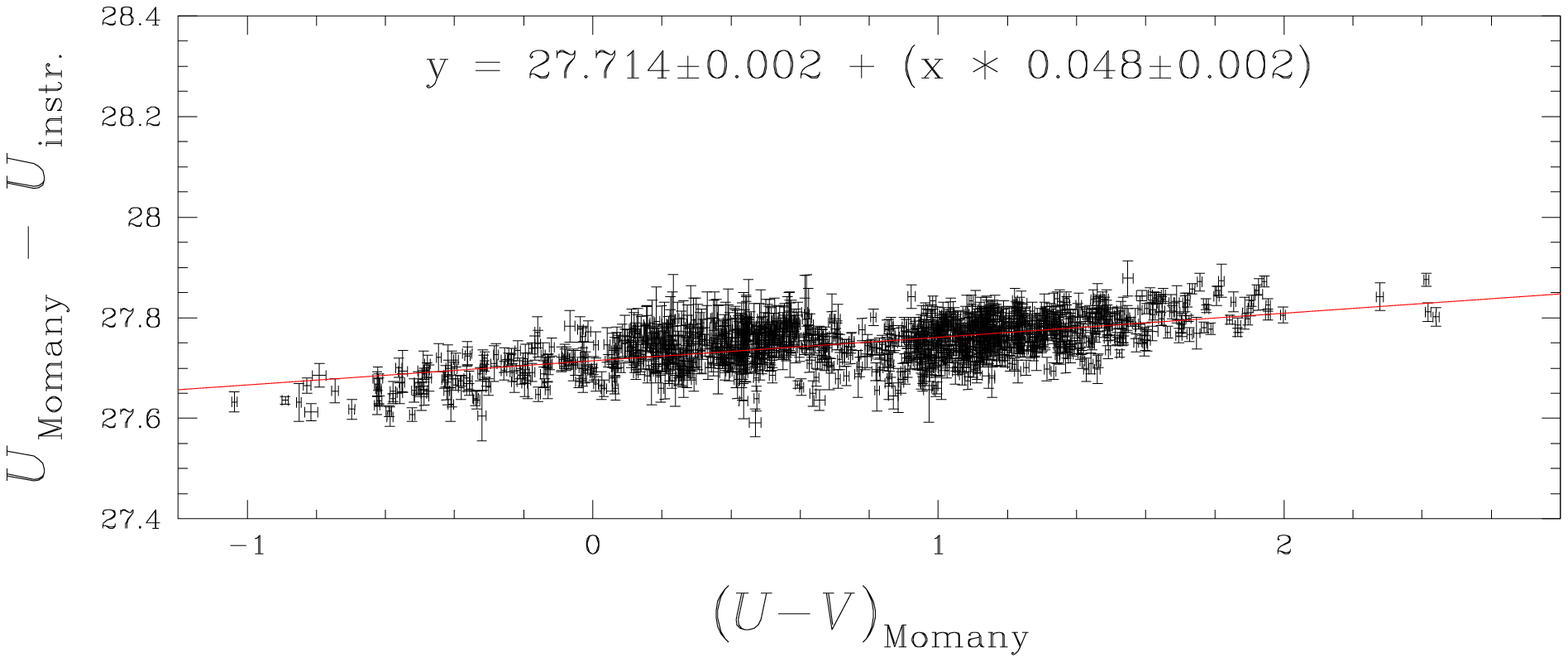}
\includegraphics[width=9cm,height=5cm]{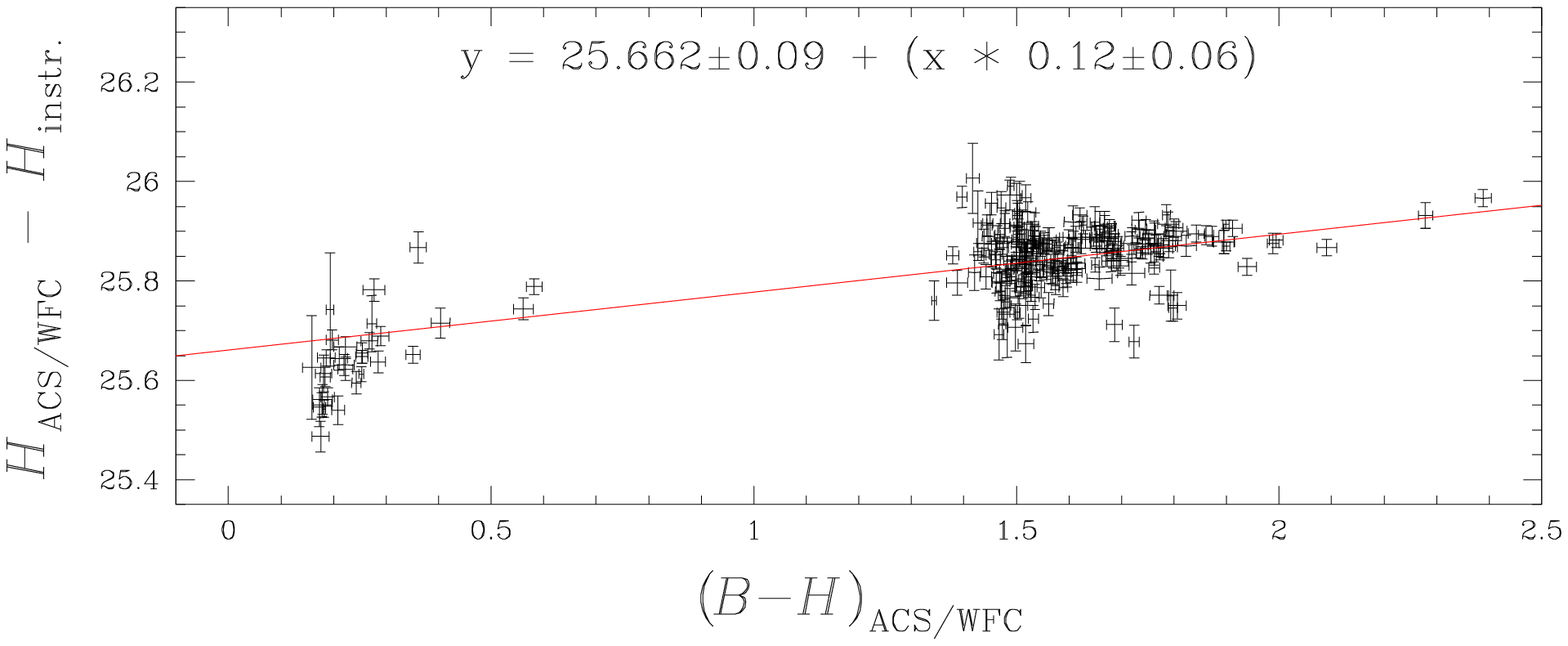}
\caption{These  figures show  our  calibration fits  with the  adopted
  color  equations.  For  the $B,  V, R_C,  I_C$ filters  we  used the
  on-line set of standard stars provided by Stetson, while for $U$ and
  $H_\alpha$  bands we  used  as  reference stars  the  Momany et  al.
  (\cite{momany03}) catalog  and the  ACS/WFC catalog of  Villanova et
  al. (\cite{villanova07}), respectively.  See text for more details.}
\label{fig_rette_cal}
\end{figure*}

\begin{figure*}[ht!]
\centering
\includegraphics[width=18.0cm]{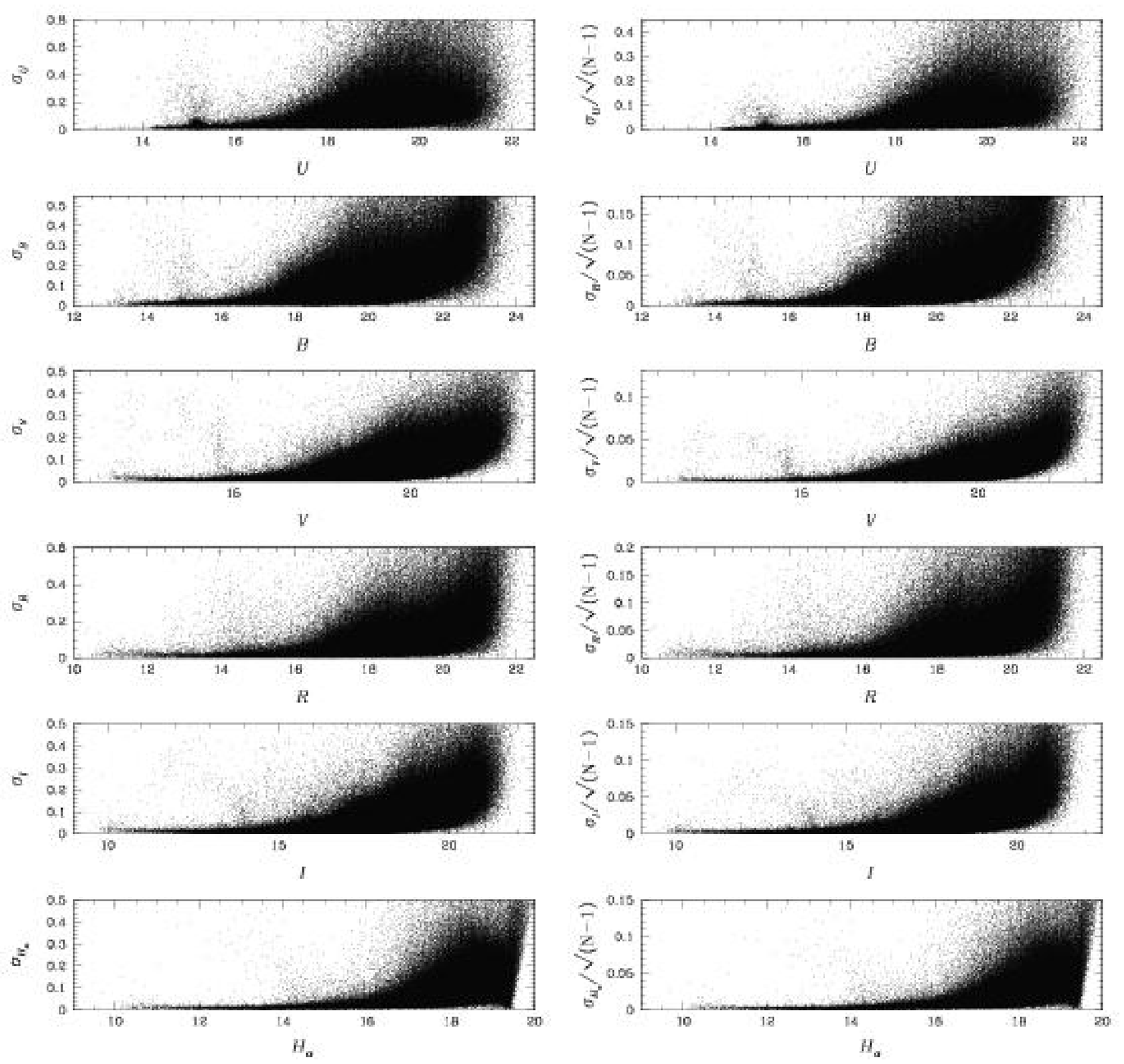}
\caption{(\textit{Left panels}): plot of magnitude r.m.s ($\sigma$) as
  a  function  of magnitude.   (\textit{Right  panels}):  plot of  the
  standard error  of the mean  ($\sigma/\sqrt{N-1}$, where $N$  is the
  number of measurements) as a function of magnitude.}
\label{fig_errori_2}
\end{figure*}

If  there  were no  systematic  errors,  the  same stars  measured  at
different  positions in  two different  frames, should  have  the same
magnitude value, within the random measurement errors:
$$ {\rm{mag}}_{i,0}-\Delta_{0} = \dots = {\rm{mag}}_{i,j}-\Delta_{j} =
\dots = {\rm{mag}}_{i,m}-\Delta_{m},$$
where  $m$ is  the total  number  of images  in that  filter, used  to
perform the autocalibration procedure.

However, the same   star  closer to the     center of the  camera   is
systematically fainter than when it  is measured closer to the  camera
edges.
For each star, measured in  several different frames, and at different
positions on the camera,  we can compute an average  of the  values of
the magnitudes in the reference system:
$$\overline{{\rm{mag}}_i}       =       \left(      \frac{1}{m}\right)
\sum_{j=1}^{m}\left( {\rm{mag}}_{i,j}-\Delta_{j} \right).$$
In  the same way, we can  compute a residual  for  the $i$-star in the
$j$-image:
$$\delta_{i,j}         =        ({\rm{mag}}_{i,j}-\Delta_{j})        -
\overline{{\rm{mag}}_i}.$$

All the residuals of stars  close to the center will be systematically
positive, and those close to the edges systematically negative.
It appears clear that  --at  any given location   on the camera--  the
average  of the  residuals from all  the stars  measured close to that
location will provide a first spatial correction to our photometry.
It also appears clear that the  determination of the sky-concentration
photometric correction will be an iterative process.

To   guarantee convergence,  we    applied  half  of  the  recommended
correction at the given location,  to all our image catalogs.  
We then  recomputed the $\Delta_j$,  and repeated the  procedure until
all the residual averages, at  any given location, became smaller than
0.001 magnitudes.
The null hypothesis of this procedure  is that the same star is imaged
several times at different locations on the detector.
To avoid systematic  error, we also select, for  each filter, the same
number of exposures  per different pointing (as much  as possible with
the existing database).

We use  only those  stars of  the image catalogs  with high  S/N ratio
(i.e.   instrumental magnitude  from $-11$  to $-14$)  with  a quality
PSF-fit   smaller    than   0.1    (as   defined   by    Anderson   et
al. \cite{anderson08}), and not too close to the cluster center (which
does not necessarily coincide with  the center of the camera) to avoid
crowding, which compromises the photometric precision.
The exact closeness to the center  of the cluster depends on the image
exposure  time; we excluded  stars within  a radius  of $1000$-$2000$
pixels from the \om center.

At this point, we define  the expression of ``given location''.  After
several tests,  we found a spatial  grid of $16\times 16$  boxes to be
the  most suitable  compromise  between a  large  number of  residuals
($\delta_{i,j}$) and  a spatial resolution  of correction sufficiently
high to be useful.
In  Fig.~\ref{fig:skyarray},  we  show  our  final  correction  grids,
respectively  for $U$,  $B_{842}$, $V$,  $R_C$, $I_C$,  and $H_\alpha$
filters.  For  each filter, we present the  final $16\times16$ element
correction grid.
Each element is colored according  to the grey scale values (black for
the minimum,  white for the  maximum).  The grey scales  vary linearly
from the minimum to the maximum grid value for each filter.
It  is  clear that  sky  concentration  affects  different filters  in
different  ways,   and  each   filter  must  therefore   be  corrected
independently.  To evaluate the correction at any point of the camera,
we completed a bilinear interpolation of the closest 4 grid points.
The adopted  correction was less  accurate close to the  mosaic edges,
where  the  peripheral  grid-points  have been  stretched  toward  the
boundaries.  The  available pointings for  the other filters  are also
lower than for $B_{842}$ and $V$, implying a less effective correction
of the sky concentration.

We emphasize  that our correction  is not completed by  a star-to-star
comparison.  After the spatial correction, a single star cannot have a
lower random error  (higher precision) than $\overline{{\rm{mag}}_i}$.
However, the  systematic errors (accuracy)  of single stars  relies on
the  quality of  individual  grid-point solutions,  which were  always
calculated to  be the  average of residuals  for several  stars within
each of the $16\times16$ cells.
Even if our random errors for individual stars are $\sim0.1$ mag, with
just 10 stars [in the worst case we still have at least 10 such stars]
we can reduce our systematic  errors to $\sim0.03$ mag.  The condition
that each star has to be observed in each of the $16\times16$ cells is
the  ideal case.  Deviations  from this  ideal case  occur frequently,
although overall, we are close to achieving the optimal solution.

The total amplitude of our correction  for the $V$ filter is 0.13 mag.
For the same filter,  Manfroid \& Selman (\cite{manfroid01}) evaluated
a 0.13 mag correction of similar spatial shape.  Nevertheless, the two
totally independent calibrations appear to be qualitatively the same.
Based   on  the   cell-to-cell   scatter  with   the  knowledge   that
sky-concentration  is relatively  flat, in  $V$ and  $I$  filters (for
which  we have  more  images) we  estimate  that the  accuracy of  our
solution is  as good as  $\sim0.03$ mag.  Although our  corrections do
not use any color information,  we note that the post-corrected CMD is
in excellent agreement with (to  within a few hundreds of a magnitude)
the HB location (Fig.~\ref{fig:sky1}).

To  verify qualitatively  the  high quality  of our  sky-concentration
correction  procedure, we  show in  Fig.~\ref{fig:sky1}d the  same CMD
region ($V$ vs.  $B_{842}-V$), derived using all the available images,
after applying our correction.
Stars located at different positions on the meta-chip are not affected
by  the sky-concentration  effect, and  are  located in  the same  CMD
region.
The total number of  plotted stars (1227, all with $\sigma_{B,V}<0.03$
mag)  is comparable  with that  of Fig.~\ref{fig:sky1}a;  this implies
that  we were  able to  remove  the systematic  contribution from  our
photometric r.m.s values.

Our solution works well for  the available \om archive data sets (used
to derive  the corrections), but archive observations,  in general, do
not  map  every  chip  in  a  way  that  enables  a  sky-concentration
correction that is universally applicable to be derived.
We cannot  guarantee that our solution  can be applied  to achieve the
same positive results with other data sets.
As  proof  of  this  issue,  we applied  our  $B_{842}$  solution,  to
$B_{878}$ images with only one pointing.

Images collected using the $B_{842}$  and $B_{878}$ filters are not so
different, in term of central wavelength (see Table~\ref{tab:filters}):
sky-concentration effects  appear to  be similar for  almost identical
filters (since they are  related to atmospheric variations that affect
the  data  for  a  range  of  different  filters),  so  a  photometric
improvement  is expected  after correcting  $B_{878}$ images  with our
$B_{842}$-derived solution.
If our  solution fails  to correct the  $B_{878}$ photometry,  this is
probably due to the different pointings of the two $B$ filters instead
of the filters themselves.
As  shown  in  Fig.~\ref{fig:sky1}c,   we  found  that  a  photometric
improvement is present, with respect to Fig.~\ref{fig:sky1}a, but that
the correction is not satisfactory.

\subsection{Instrumental $UBVR_CI_C$-$H_\alpha$ photometric catalog}
\label{subsec_instr_cat}

We  derived instrumental  single-filter catalogs  using  all available
images, by  matching each chip individually to  minimize the zeropoint
differences  between WFI chips.   Included stars  were measured  in at
least three distinct images.
Photometric single-filter catalogs were then linked to the astrometric
one.  Linked  star positions  agreed with those  in the  proper-motion
catalog within  1 pixel for $BVR_CI_CH_\alpha$ filters,  while for the
$U$ filter we had to adopt a larger matching radius (3.5 pixels).
This is mainly due to the poorer distortion solution in the $U$ band.

Due to the aforementioned sky-concentration minimization problem with
$B_{878}$ images, our $B$ photometry refers to the $B_{842}$ filter.

As for  the $U$  photometry, we used  only $U_{877}$ images:  in fact,
$U_{841}$ is  a ``medium''  rather than a  wide-band filter,  of quite
different central  wavelength and  a low transmission  efficiency with
respect  to  $U_{877}$  (Table~\ref{tab:filters}).  If  not  specified
otherwise, we  refer to $B_{842}$  and $U_{877}$ simply using  $B$ and
$U$, respectively.

\begin{figure}[t!]
\centering
\includegraphics[width=9.0cm]{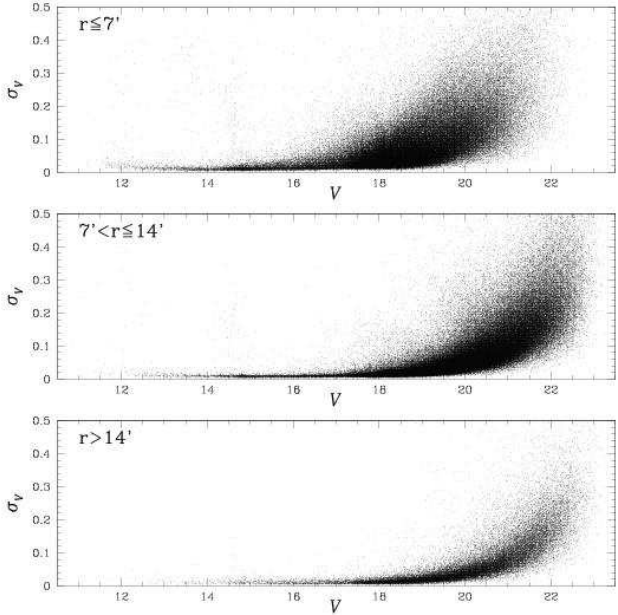}
\caption{Photometric r.m.s for stars within $7\arcmin$ (\textit{top
    panel}), from $7\arcmin$ to $14\arcmin$ (\textit{middle panel}),
  and outside $14\arcmin$ (\textit{bottom panel}).}
\label{fig_errv}
\end{figure}

\subsection{Photometric calibration}
\label{subsec_photocal}

The  photometric calibration  of  the WFI@2.2$m$  data for  $BVR_CI_C$
bands  was performed  using  a  set of  $\sim 3000$  online wide  field
photometric \om  Secondary Standards stars  (Stetson \cite{stetson00},
\cite{stetson05}).
The Secondary  Standards   star  catalog  covers  an  area   of   about
$30\arcmin\times30\arcmin$ around the cluster center.
We calibrated our $U$ instrumental photometry by cross-correlating our
photometry  with  Momany  et  al.   (\cite{momany03})  $U$  calibrated
catalog (Stetson does not  provide $U$ photometry for the $\omega$~Cen
Secondary Standards).
For $H_\alpha$ calibration, we used as reference-standard stars
the  $3\times3$ central  ACS/WFC mosaic  photometric catalog  in F658N
band   (GO  9442),   which   was  presented   by   Villanova  et   al.
(\cite{villanova07}).
This \textit{HST} catalog  was obtained using {\sf img2xym\_WFC.09x10}
software;  instrumental  magnitudes  were  transformed  onto  the  ACS
Vega-mag flight  system following Bedin et  al.  (\cite{bedin05}), and
by using the zero points of Sirianni et al. (\cite{sirianni05}).

For  $BVR_CI_C$  bands, we  matched  our  instrumental magnitudes  and
colors to the Stetson standard ones, and derived calibration equations
by means of an iterative least squares fitting of a straight line (see
Fig.~\ref{fig_rette_cal}).
For these filters, we found that only a  first-order dependency of the
color term affects our instrumental magnitudes.
The linearity of  our calibration equations, which cover  a wide range
of colors (being  derived from both HB and RGB  stars) is evident from
the plots of Fig.~\ref{fig_rette_cal}.

As in the  calibration of our $U$ instrumental  photometry, the Momany
et   al.    (\cite{momany03})    catalog   was   not   corrected   for
sky-concentration effects.  We found a magnitude dependence related to
the star positions.
We adopted  a straight  line fit to  derive the  calibration equation,
because  we  were unable  to  consider  the different  color/magnitude
dependencies individually.
Therefore, our  calibrated $U$ magnitudes were not  more reliable than
the 0.15 magnitude level (maximal error).

For the  $H_\alpha$ filter, we again performed  a straight line
fit to derive  the calibration equations, even though  the data appear
to suggest a second order color effect (see the corresponding panel of
Fig.~\ref{fig_rette_cal}).
ACS/WFC  data cover  only the  inner $\sim  10\arcmin\times 10\arcmin$
region of  our catalog, and  are therefore taken in  extremely crowded
conditions.
This  effect might  strongly  influence our  photometry mimicking  the
aforementioned second-order effect.
In Fig.~\ref{fig_errori_2}, we show in the left panels our photometric
errors for each filter, as a function of the corresponding magnitude.
The photometric  errors (standard  deviation) have been  computed from
multiple observations, all reduced to the common photometric reference
frame  in  the chosen  bandpass.   
In  the   right  panels   of  Fig.~\ref{fig_errori_2},  we   plot  the
photometric   standard  error  of   the  mean   --to  be   defined  as
$\sigma/\sqrt{N-1}$, where  $N$ is the total  number of observations--
versus the magnitude, for each filter.

To illustrate more clearly the  dependence of our photometric r.m.s on
crowding, we show (for the  $V$ filter only) in Fig.\ref{fig_errv} the
photometric  r.m.s $\sigma_V$  with respect  to $V$  for  stars within
$7\arcmin$ (top panel), from $7\arcmin$ to $14\arcmin$ (middle panel),
and  outside $14\arcmin$ (bottom  panel).  Stars  located in  the most
crowded  region  of  the  field  suffer higher  uncertainty  in  their
photometry.

Our final  catalog consists of about $360\,000$  stars, in $UBVR_CI_C$
wide-band  and  $658$nm  narrow-band  filters, covering  a  wide  area
($\sim33\arcmin\times33\arcmin$) centered on $\omega$~Cen.
We   reach 3 magnitudes in  $V$   band below  the  TO  point with  a
photometric r.m.s  of 0.03 mag.

\subsection{Zeropoint residuals}
\label{subsec_zp_res}

Even  if our  sky-concentration correction  works well,  residuals are
still present, especially  close to the corners of  our final catalog.
Due to  the wide field area analyzed  in this work, there  is also the
possibility of a contribution from differential reddening.
Using   $ubvy$   Str\"omgren   and $V$ $I$  photometry,  Calamida   et
al. (\cite{calamida05}) developed an  empirical method to estimate the
differential  reddening of  $\omega$~Cen. The  authors found  that the
reddening can  vary in the  range $0.03\lesssim E(B-V)  \lesssim 0.15$
from Str\"omgren filters, and $0.06\lesssim E(B-V) \lesssim 0.13$ from
$V$   $I$   filters,    within   their   analyzed   field-of-view   of
$14\arcmin\times  14\arcmin$,  which was  centered  on  the center  of
$\omega$~Cen.    However,    the   results   by    Calamida   et   al.
(\cite{calamida05})    were   questioned    by   Villanova    et   al.
(\cite{villanova07}), and  the quantitative value  of the differential
reddening still needs to be confirmed.

In the case of the $B-V$  color, the maximum zeropoint residual in our
final  catalog  is less  than  0.1  mag.   To minimize  any  zeropoint
variations, we used  a method similar to that  described by Sarajedini
et al. (\cite{sara07}).
Briefly,  we defined the  fiducial ridge-line  of the  most metal-poor
component of the \om RGB and  tabulate, at a grid of points across the
field, how the  observed stars in the vicinity of  each grid point may
lie systematically  to the red or  the blue of  the fiducial sequence;
this systematic  color offset is indicative of  the local differential
reddening.

Our  online  catalog magnitudes  are  not  corrected for  differential
reddening  to enable  the  user to  adopt  their preferred  correction
method in removing differential reddening and zeropoint residuals.

\section{Proper-motion measurements}
\label{sec_PM}

To complete the  proper-motion analysis, we used only  the $B$ and $V$
images taken in April 1999 (epoch~I) and April 2003 (epoch~II).
This  choice was  due  to the  fact  that: (i)  we  have a  fine-tuned
geometric distortion  correction map  for $V$ filter  (Paper~I), which
has been proven  to work well for the two  $B$ filters (Paper~I); (ii)
it  offers the widest  possible time  base-line of  $\sim 4$  yrs; and
(iii) we have  a relatively high number of images  in both epochs, and
with relatively deep exposures.

We first photometrically selected  probable cluster members in the $V$
versus $B-V$ color-magnitude diagram.   These stars are located on the
RGB (see the RGB selections in the top-panel of Fig.~\ref{fig_rgbamp},
within the magnitude interval $14.6<V<17.2$.
We used these  stars only as a local reference  frame to transform the
coordinates  from one  image  to the  system  of the  other images  at
different epochs and therefore derive relative proper motions.
By  using predominantly cluster stars, we   ensure that proper motions
will be measured relative to the bulk motion of cluster stars.
The expected  intrinsic velocity dispersion  of \om stars for which we
can   measure   reliable proper  motions,    is   between 10  and   15
km$\,$s$^{-1}$ (Merritt, Meylan \& Mayor \cite{merritt97}).
If we  assume a distance of 5.5  kpc for $\omega$~Cen,  as reported by
Del   Principe   et   al.    (\cite{delprincipe06}),   and   isotropic
distribution of stars (good to first order), then these translate into
an internal dispersion of $0.4$-$ 0.6$ mas yr$^{-1}$.

Over  the   four-year  epoch,  the   difference  would  result   in  a
displacement of only $1.5$-$2.3$ mas,  which is a factor of 3 smaller
than the random measurement errors ($\sim7$ mas).
Conversely,  the tangential velocity  dispersion of  field stars  is a
factor of  $\sim10$ larger than  the intrinsic velocity  dispersion of
the cluster.
For  field  stars, proper  motions  are  clearly  not negligible  with
respect to measurement  errors, and this has an  adverse effect on the
coordinate transformations.
We removed  iteratively stars from the  preliminary photometric member
list  that  had  proper  motions  clearly  inconsistent  with  cluster
membership, even though their colors placed them close to the fiducial
cluster sequence.

To minimize the  effects of geometric-distortion-solution residuals on
proper motions, we used local  transformations based on the closest 20
reference stars, typically extending over $\sim30$ arcsec.
These were well-measured cluster stars of any magnitude selected to be
on the  same CCD chip, as  long as their preliminary  proper motion is
consistent with cluster membership.
No systematic errors  larger than our random errors  are visible close
to the corners or edges of chips.

To  avoid  possible filter-dependent  systematic  errors, we  measured
proper  motions in  the $V$  and $B$  bandpasses only,  for  which the
geometrical distortion corrections were derived originally (Paper~I).
Individual errors of proper motions for single stars were estimated as
described  in Sect.~7.3 of  Paper~I.  For  both epochs  separately, we
estimated  the  intra-epoch r.m.s  error  from  all same-epoch  plates
transformed locally to the same reference frame.
The proper-motion  errors were computed to be the  r.m.s of the proper
motion, obtained  by solving locally each first-epoch  frame into each
second-epoch frame.
These errors,  however, were not  entirely  independent because the same
frames were used more than once.
Therefore, to obtain  our most reliable estimate of  the proper-motion
standard error, we  added in quadrature the intra-epoch  r.m.s of each
epoch.

\begin{figure}[t!]
\centering
\includegraphics[width=9.0cm,height=9.0cm]{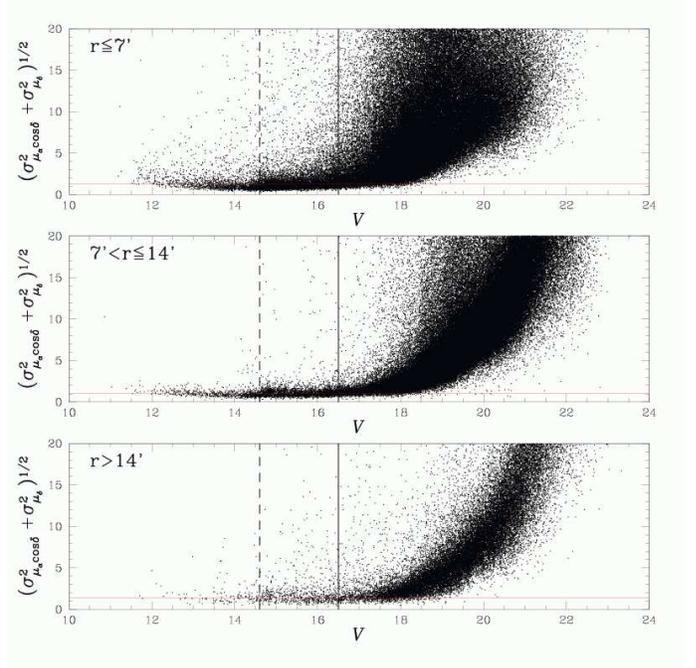}
\caption{Proper-motion errors for stars within $7\arcmin$ (\textit{top
    panel}), from  $7\arcmin$ to $14\arcmin$  (\textit{middle panel}),
  and   outside  $14\arcmin$  (\textit{bottom   panel}).   The
    proper-motion   errors    are   expressed   in    the   units   of
    mas$\,$yr$^{-1}$.}
\label{fig_errori_1}
\end{figure}

In Fig.~\ref{fig_errori_1}, we show our  proper-motion r.m.s,  in units
of     \masyr,    versus      $V$     magnitudes,   calculated      as
$\sigma_{\mu}=\sqrt{\sigma^2_{\mu_{\alpha}\cos\delta}+\sigma^2_{\mu_\delta}}$.
The  top panel  presents  stars  within $7\arcmin$  of  the center  of
$\omega$~Cen,  the middle  panel is for  stars between  $7\arcmin$ and
$14\arcmin$, while the lower panel  shows the errors for stars outside
$14\arcmin$.
The vertical dashed line indicates the saturation limit of the deepest
exposures ($V=14.6$),  while the continuous  line is at  $V=16.5$, the
vL00 faintness limit.
The precision of our proper-motion measurement is $\lesssim 0.03$ WFI
pixels in  4 yrs down  to $V\sim 18$  mag (i.e. $\sigma  \lesssim 1.9$
mas~yr$^{-1}$).  At fainter magnitudes, the errors gradually increase,
reaching $\sim 5$ \masyr at $V=20$.
The stars brighter than $V\sim  13$ magnitude show a higher dispersion
because of the image saturation even in the shortest exposures.
Horizontal  lines  in   Fig.~\ref{fig_errori_1}  indicate  the  median
proper-motion r.m.s of unsaturated stars brighter than $V=16.5$.
We have  1.3  \masyr, 1.1 \masyr, and 1.3 \masyr  for the top, middle,
and bottom panel,  respectively. The higher value for  the inner stars
is due to crowding while, for  the outer stars, there is a combination
of  three factors:  (1) our  geometrical distortion  solution  is less
accurate close  to the WFI mosaic  edges; (2) there  are fewer cluster
members,  per unit  area, usable  as a  reference for  deriving proper
motions; (3)  we have a lower  number of images that  overlap with the
external areas of the field-of-view.

\begin{figure*}[ht!]
\centering
\includegraphics[width=18cm]{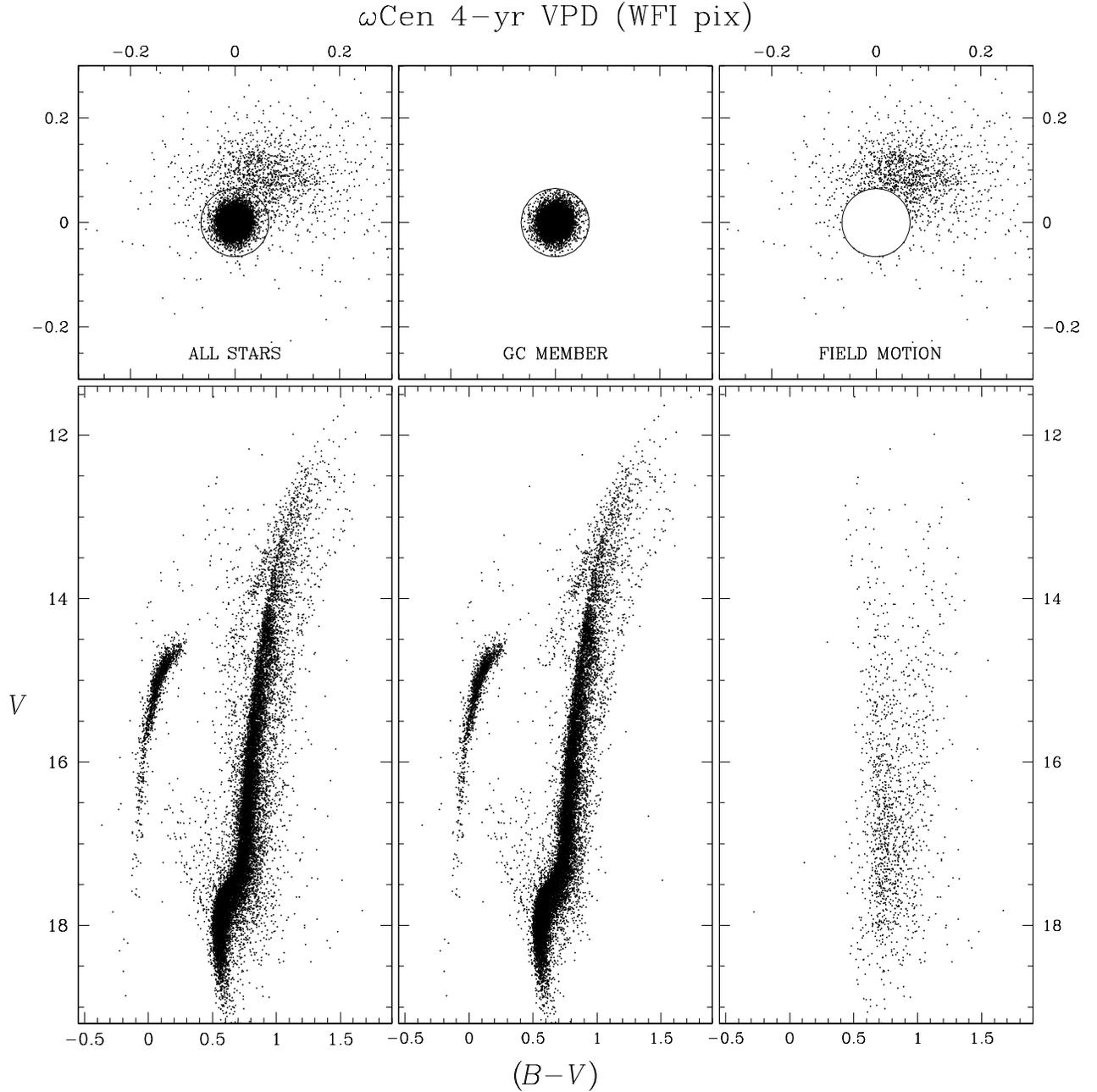}
\caption{(\emph{Top-panels}):  proper-motion  vector-point  diagram.
  Zero point  in VPD is the  mean motion of  cluster stars candidates.
  (\emph{Bottom  panels}):  calibrated  $V$, ($B-V$),  color-magnitude
  diagram.  (\emph{Left}):  the entire sample.  (\emph{Center}): stars
  in   VPD   with   proper    motion   within   0.065   pixels   (i.e.
  $\sim3.9\rm{\,mas\,yr}^{-1}$)     around    the     cluster    mean.
  (\emph{Right}):  probable background/foreground  field stars  in the
  area of \om  studied in this paper.  All plots  show only stars with
  proper-motion   $\sigma$   smaller   than   0.032   pixels   (i.e.
  $\sim1.9\rm{\,mas\,yr}^{-1}$) and $V$ magnitude r.m.s.  smaller than
  0.02.}
\label{fig_PMsel}
\end{figure*}

\subsection{Cluster CMD decontamination}
\label{subsec:cmd_decon}

To probe the effectiveness of our proper motions in separating cluster
stars  from  the field  stars,  we  show  in Fig.~\ref{fig_PMsel}  the
vector-point diagrams  (VPDs, top  panels), and the  CMDs in  the $V$
vs. $B-V$ plane (bottom panels).
In the  left panels, we  show the entire  sample of stars;  the middle
panels display what we considered  to be probable cluster members; the
right panels show predominantly the field stars.  Plotted stars have a
$V$ r.m.s. lower than 0.03 mag.

In the  VPDs, we draw a  circle around the cluster  centroid of radius
3.9 \masyr.  Provisionally, we define as cluster members all points in
the VPD within this circle.
The chosen  radius is the  optimal compromise between  missing cluster
members with uncertain proper  motions, and including field stars that
have velocities equal to the cluster mean proper motion.
Even  this  approximate separation  between  cluster  and field  stars
demonstrates the power of proper motions derived in this study.
A   description    of  membership     probability    is   given     in
Sect.~\ref{sec_memb}.

\subsection{Differential chromatic refraction (DCR)}
\label{subsec_DCR}

\begin{figure*}[ht!]
\centering
\includegraphics[width=17cm]{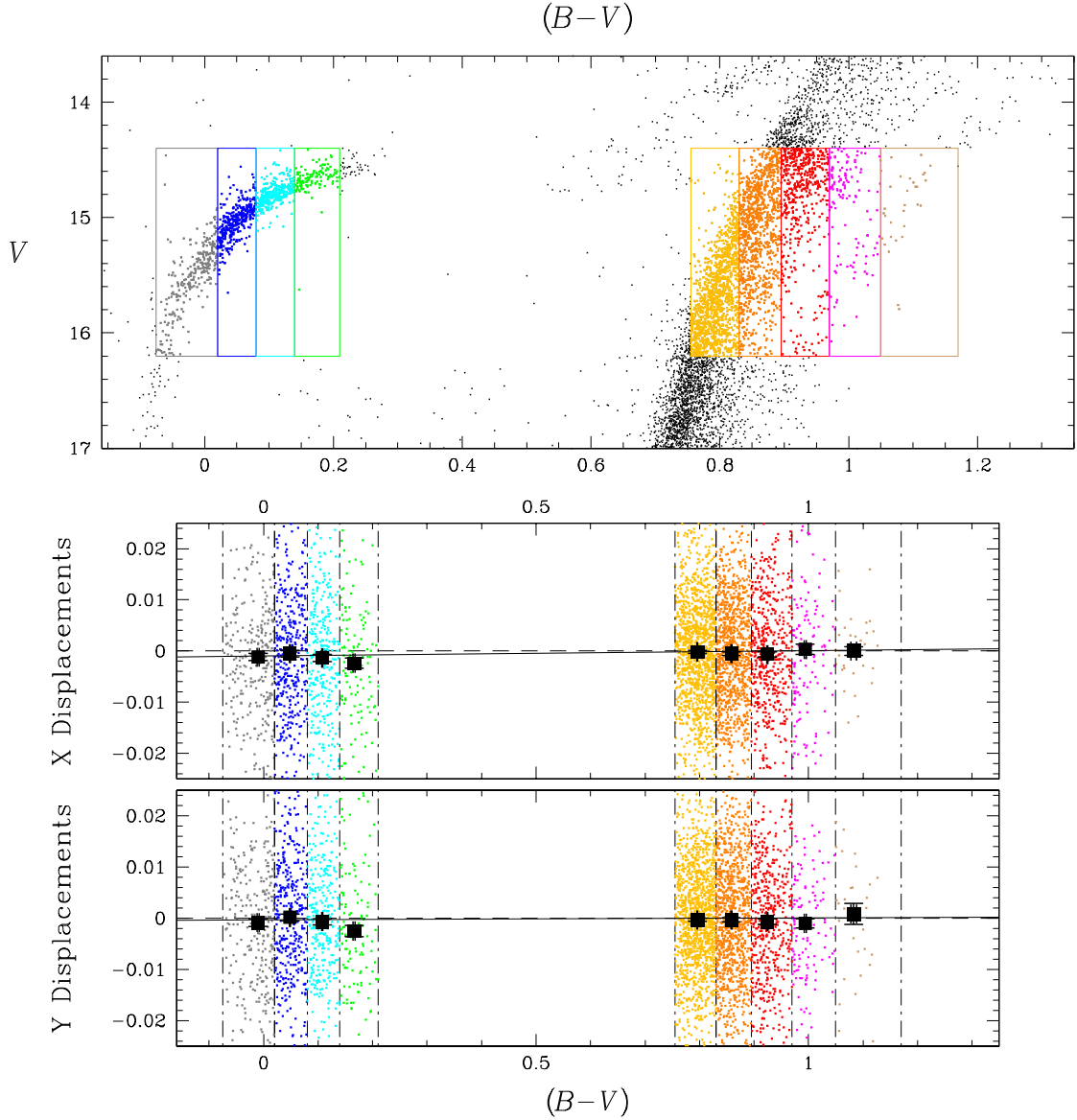}
\caption{(\textit{Top panel}):  selected stars in  the CMD of  \om for
  DCR  effect correction.  (\textit{Middle  and bottom  panels}): star
  displacements along $X$ and $Y$ axes, as a function of stars ($B-V$)
  color.  The median  shift of the nine samples  are also showed, with
  errors.  The continuous lines show the adopted fits used to quantify
  the DCR effects.}
\label{fig_dcr1}
\end{figure*}

The DCR  effect causes  a shift  in the photon  positions in  the CCD,
which  is proportional  to their  wavelength,  and a  function of  the
zenithal distance:  blue photons will  occupy a position  that differs
from  that of  red photons.  The DCR  effect is  easier to  detect and
remove from CCDs, due to their linearity.

Unfortunately,  within each  epoch, the  available data  sets  are not
optimized   to  perform   the  DCR   correction  directly   (Monet  et
al.  \cite{monet92}),  because  the  images  have not  been  taken  at
independent zenithal distances.
We can, however,  check  if  possible differences  in  the DCR  effect
between the  two epochs could  generate an apparent proper  motion for
blue stars relative to red stars.

We selected four  samples of stars located on the HB,  and five on the
RGB,  as  shown in  Fig.~\ref{fig_dcr1}  (top  panel), with  different
colors to estimate the DCR effect,  in a magnitude interval of 1.8 mag
in $V$  (14.4$\le$$V$$\le$16.2), with proper motions  $\le 3.8$ \masyr
and r.m.s.  $\le 1.9$ \masyr.
We  chose  this magnitude  range  to:  (i) avoid  luminosity-dependent
displacements  (if any);  (ii) include  stars  with a  low r.m.s.   in
positions      and      fluxes      (see      Fig.~\ref{fig_errori_2},
\ref{fig_errori_1}),  excluding saturated stars;  and (iii)  cover the
widest possible color  baseline according to the above  points (i) and
(ii).
For each  of the  nine samples,  we derived the  median color  and the
median proper motion  along $\mu_{\alpha}\cos\delta$ and $\mu_\delta$,
and  their respective  errors. Proper  motions were  expressed  in the
terms of  a displacement  over 4 years,  that is  in the units  of WFI
pixels along  the $X$ and $Y$ axes  of a detector (parallel  to the RA
and Dec directions).

In Fig.~\ref{fig_dcr1}, we show, in  the top panel, the selected stars
on the CMD used to examine the DCR effect; the linear fits adopted for
the  $X$ and  $Y$ displacements  are shown  in the  lower part  of the
figure.
We found  a negligible DCR  effect along both  $X$ and $Y$  axes.  For
this reason, we have not corrected our measurements for this effect.

\begin{figure*}[ht!]
\centering
\includegraphics[width=18cm]{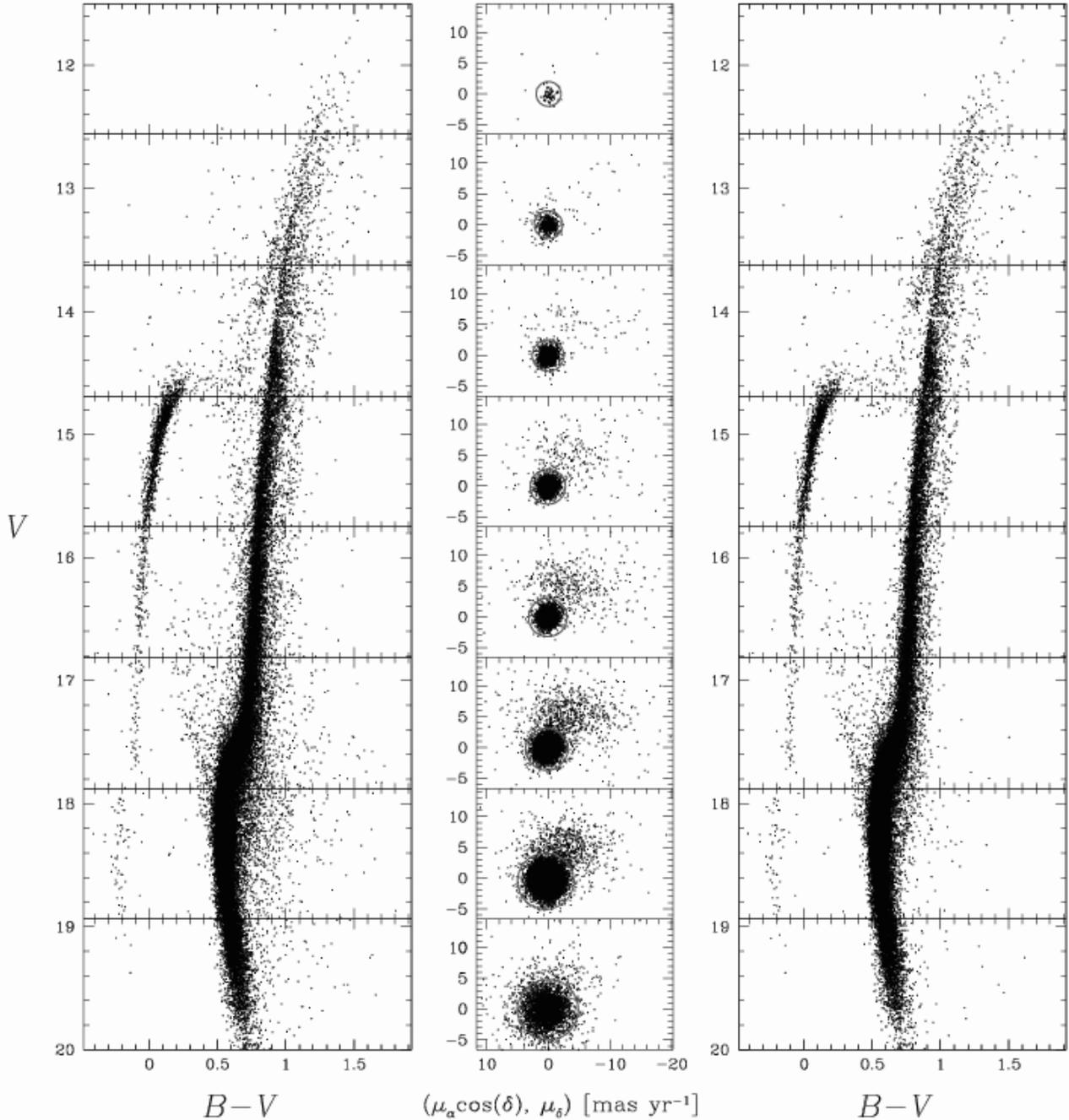}
\caption{(\emph{Left}):  color-magnitude diagram, splitted  into eight
  magnitude  bins, for  all the  stars having  a proper-motion r.m.s.
  increasing from 1.8 \masyr for the brightest bin to 5 \masyr for the
  faintest one,  and photometric  r.m.s.  from 0.02  mag to  0.05 mag,
  generous  enough to  include  main sequence  stars  down to  $V=20$.
  (\emph{Middle}): proper  motions for the stars  in the corresponding
  magnitude  intervals.  A circle  in each  diagram shows  the adopted
  membership  criterion.  (\emph{Right}): color-magnitude  diagram for
  assumed cluster members.}
\label{fig_PMgo6}
\end{figure*}

In  Fig.~\ref{fig_PMgo6}, we show,  on the  left, our  calibrated $V$,
$B-V$ CMD, divided into eight magnitude bins.
In  each bin,  we  adopted different  selection  criteria to  identify
cluster  members,  which  were  more  stringent for  stars  with  more
reliable  measurements from  data of  high signal-to-noise  ratio, and
less restrictive for star with less precise measurements.
Plotted stars have a proper-motion r.m.s.  of $<1.8$, \masyr\ for the
brightest bin, to 5 \masyr\ for the faintest one.
The Photometric r.m.s.  for both  the bands range between 0.02 mag for
the brightest bin to 0.05 for the faintest one, which is sufficient to
include main-sequence stars down to $V=20$.
For each  magnitude bin, we considered as   cluster members those stars
with a proper motion within the circle shown in the middle column 
of Fig.~\ref{fig_PMgo6}.

On the right side of Fig.~\ref{fig_PMgo6}, we show the color-magnitude
diagram  for  stars assumed  to  be  cluster  members.  The  available
archive  images are  again not  sufficiently deep  to  derive reliable
proper motions below the TO.
The  proper  motions presented  in  this  paper  are not  sufficiently
accurate to study the internal motion of $\omega$~Cen.
The main  purpose of the  proper motion presented  in this work  is to
provide a reliable  membership probability for spectroscopic follow-up
projects, star counts, and the study of the radial distribution of the
different branches (Bellini et al. in preparation).

\subsection{Astrometric calibration}
\label{subsec_astrocal}

To   translate the pixel   coordinates  into the equatorial coordinate
system,    we     adopted  the    UCAC2     catalog   (Zacharias   et
al. \cite{zacharias04}) as a reference frame. 
Due  to  the  severe  crowding   in  images  of  the  inner  parts  of
$\omega$~Cen,  this  catalog was  however  inadequate for  calibration
purposes    close    to   the    center    of    \om   (the    central
$10\arcmin\times10\arcmin$ area corresponds  almost entirely to a void
in UCAC2).   Another possible reference frame, especially  for the
cluster center, is the vL00 catalog.
However, the precision of published coordinates is lower than $\sim20$
mas and no analysis was provided by vL00 for the presence of potential
systematic  errors  in  the  positions.  Examination  of  vL00  proper
motions  by Platais et  al. (\cite{platais03})  indicated that  {\it a
  priori} these systematic errors could not be discounted.
These deficiencies  in the vL00 positional catalog  were eliminated by
re-reducing  the original Cartesian  coordinates (of  formal precision
equal to 2~mas), kindly provided by F. van Leeuwen.

First, we selected only Class  0-1 stars from vL00 (i.e. their images
were isolated or only slightly disturbed by an adjacent image). Second,
a  trial equatorial  solution was  obtained for  vL00 stars  using the
UCAC2 catalog as a reference frame.
Third, the  new set of vL00  coordinates was tested  against the UCAC2
positions  as  a function  of  coordinates,  magnitude,  and color  of
stars. There are $\sim3000$ stars  in common between these two sets of
coordinates. Assuming that the  UCAC2 positions are free of magnitude-
color-related  systematic  errors, we  found  that  the original  vL00
Cartesian coordinates were biased by up to 16~mas~mag$^{-1}$.
There is  also a detectable  quadratic color-dependent bias  along the
declination.   Both magnitude- and  color-related biases  were removed
from  the  vL00  Cartesian  coordinates before  the  final  equatorial
solution was obtained.

The new  reference catalog, covering a  region of $1\fdg5\times1\fdg5$
and magnitudes to $V$$\sim$16.5, contains $10\,291$ stars and consists
of approximately equal parts of  the UCAC2 (trimmed down to stars with
positional accuracies  of higher quality than 75~mas)  and the updated
vL00 coordinates on the system of ICRS and epoch J2000.0.

This  new  reference  catalog   was  used  to  obtain  the  equatorial
coordinates  of our  \om stars.   The WFI  pixel coordinates  of these
stars  were translated  into global  Cartesian system  coordinates and
corrected for geometric distortions.
A simple  low-term-dominated plate  model was sufficient  to calculate
equatorial coordinates. The standard error of this solution, employing
$\sim5500$  reference  stars, was  $45$-$50$~mas  in each  coordinate.
These  errors  were   higher  than  those  listed  in   Yadav  et  al.
(\cite{yadav08},    Paper~II),    which    is   based    on    similar
WFI@2.2$m$ data for the open cluster M67.
We  understand  that  image  crowding  remains a  dominant  source  of
increased  scatter  in our  solution  for  $\omega$~Cen.  Although  we
removed all stars with obviously poor astrometry, even a close but not
overlapping  image might  slightly  distort the  position  of a  star,
especially in  photographic plates.  The J2000 positions  of all stars
for the epoch 2003.29 are given in Table~6.

The  proper motions in  this work  have not  been translated  into the
absolute  values,  because  there  are  too  few  background  galaxies
suitable for defining an absolute reference frame.

\subsection{Comparison with other $\omega$~Cen proper-motion catalogs}
\label{subsec_comparison}

\begin{figure}[ht!]
\centering
\includegraphics[width=9cm]{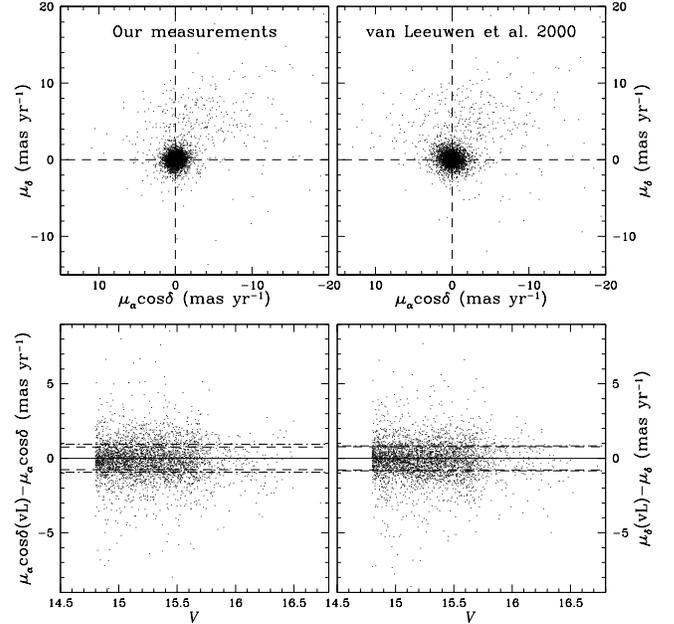}
\caption{(\textit{Top panels}): vector-point  diagrams of common stars
  in our catalog (left) and those of vL00 (right), with respect to the
  mean  cluster  motion.   (\textit{Bottom panels}):  right  ascension
  (\textit{left})  and   declination  (\textit{right})  proper-motion
  residuals as  a function of  the $V$ magnitudes.   Horizontal dashed
  and  point-dashed lines  show the  $3\sigma$-clipped  median  of the
  proper-motion  dispersion for the selected cluster  members, for our
  measurements and for vL00, respectively.}
\label{fig_vl_check}
\end{figure}

We compare our  results with the proper-motion  catalog by van Leeuwen
(\cite{vanleeuwen00}).  First,  we considered the  common, unsaturated
stars in our catalog ($V>14.6$) to the vL00 faint limit ($V\sim16.5$).
The selected samples contained $\sim3400$ stars.
Since vL00 proper  motions are given in an  absolute reference system,
we  subtracted from the  individual vL00  proper motions  the absolute
mean   motion  of   the   cluster  provided   by   the  same   authors
[($\mu_{\alpha}\cos\delta,\mu_\delta$)=($-3.97,-4.38$) \masyr].

In the top panels of  Fig.~\ref{fig_vl_check}, we show on the left the
vector-point diagram from our measurements, while on the right we show
the vL00 values.
In both diagrams, a concentration of stars at (0,0) \masyr\ corresponds
to the cluster  members, while a diffuse distribution  of stars around
($-3$,~5) \masyr, consists of the  field objects in the foreground and
possibly the background of $\omega$~Cen.

The  size of  the  proper-motion dispersions  of  the cluster  members
reflect  both  internal motions  in  the  cluster  and random  errors.
However, the internal motions are expected to be negligible.
Assuming   a  distance   of   5.5   kpc,  and   the   Meylan  et   al.
(\cite{meylan95})  measurement  of the  dispersion  in the  transverse
velocity of  $\sim$10 $\rm{km}\,\rm{s}^{-1}$ (in the  outskirts of the
cluster that  we are  probing) the expected  dispersion in  the proper
motions would be $\sim0.4$ \masyr.
Our  estimated errors for the  selected  sample was  0.74 \masyr\  for
$\mu_{\alpha} \cos\delta$, and 0.77 \masyr\ for $\mu_\delta$.
Therefore,  the internal proper  motions should  not affect  more than
$10$-$15$\% of the observed dispersions.

To estimate the observed proper-motion dispersion in the two samples
(this paper and vL00), we adopted the 68.27-th percentile of the
distribution ($\sigma$) about the median (estimated iteratively with
a 3$\sigma$-clipping), and for each coordinate independently.
Due to  the significance differences between cluster  and field object
motion, this procedure allowed us to isolate a sub sample of members.

%
Our results were:
$$ \rm{this~work:~}\left\{
\begin{array}{ll}
\sigma (\mu_{\alpha} \cos\delta) = 0.76 
\rm{~mas}\,\rm{yr}^{-1}&\\
\sigma (\mu_\delta) = 0.78  \rm{~mas}\,\rm{yr}^{-1}&\\
\end{array}
\right.
$$
$$
\rm{vL00:~}\left\{
\begin{array}{ll}
\sigma (\mu_{\alpha} \cos\delta) = 0.94 
\rm{~mas}\,\rm{yr}^{-1}&\\
\sigma (\mu_\delta) = 0.83  \rm{~mas}\,\rm{yr}^{-1}&\\
\end{array}
\right.
$$

For the selected sample, it is clear that our distribution is tighter,
rounder, and in good agreement with our estimate of the errors.
Even if our  proper motions originate in images  representing half the
total number of plates used by vL00, we note that we study more than 3
mag  fainter  in  $V$,  and  use  a time  baseline  that  equals  only
$\sim1/12$ of that used by vL00.

The  above  performed test  could  be a  bit  unfair  versus the  vL00
catalog, because we  used our non-saturated stars only,  which are the
faintest in the vL00 catalog.  vL00 demonstrated emphatically that not
all  stars are  suitable  for astrometric  measurements. We  therefore
performed a second  test in which we chose the stars  in vL00 with the
most reliable measurements (belonging to class 0 and 1 only) that were
brighter than  $V=16$, and had a  good probability of  being a cluster
member ($P_\mu (\mbox{vL00})>75\%$);  we compared the measurements for
these stars with those in catalog.

The results were as follows:
$$ \rm{this~work:~}\left\{
\begin{array}{ll}
\sigma (\mu_{\alpha} \cos\delta) = 0.68 
\rm{~mas}\,\rm{yr}^{-1}&\\
\sigma (\mu_\delta) = 0.69  \rm{~mas}\,\rm{yr}^{-1}&\\
\end{array}
\right.
$$
$$
\rm{vL00:~}\left\{
\begin{array}{ll}
\sigma (\mu_{\alpha} \cos\delta) = 0.71
\rm{~mas}\,\rm{yr}^{-1}&\\
\sigma (\mu_\delta) = 0.62  \rm{~mas}\,\rm{yr}^{-1}&\\
\end{array}
\right.
$$

These  results illustrate the  slightly higher  precision of  the vL00
catalog, but  the dispersions  n the measurements  are, in  any cases,
comparable.  The dispersion obtained  with our catalog, which now also
includes saturated star measurements, is more reliable than found with
the first test.
The explanation  of this  apparent paradox is  that, by  selecting the
best proper-motion measured stars in  the vL00 catalog (class 0 and 1,
among the  most isolated  stars), we also  selected the  most isolated
stars in our catalog, making the PSF-wings fitting more effective.
Once again, it  appears clear the huge potential  that wide  field CCD
imager will have in astrometry in the future.

\section{Membership probability}
\label{sec_memb}

In   the   vector-point    diagrams   of   Figs.~\ref{fig_PMsel}   and
\ref{fig_PMgo6}, two distinct groups  of stars are clearly detectable:
the  bulk  of  stars  belong  to $\omega$~Cen,  with  no  mean  motion
$(\mu_\alpha  \cos(\delta)=\mu_\delta=0.0$  $\rm{mas}\,yr^{-1})$,  and
there is a secondary broad group, which corresponds to field stars.

Vasilevskis  et al. (\cite{vasi58})  were the  first to  formulate the
proper-motion membership probability.
This   method   was  later   developed   by   many  authors   (Sanders
\cite{sanders71},   Zhao   \&    He   \cite{zhao90},   Tian   et   al.
\cite{tian98}, Balaguer-N{\'u}nez et al. \cite{bala98}, and references
therein) for several open and globular clusters.
To derive  our membership probability,  we followed a method  based on
proper motions described by Balaguer-N{\'u}nez et al. (\cite{bala98}).

\begin{figure}[ht!]
\centering
\includegraphics[width=9.0cm]{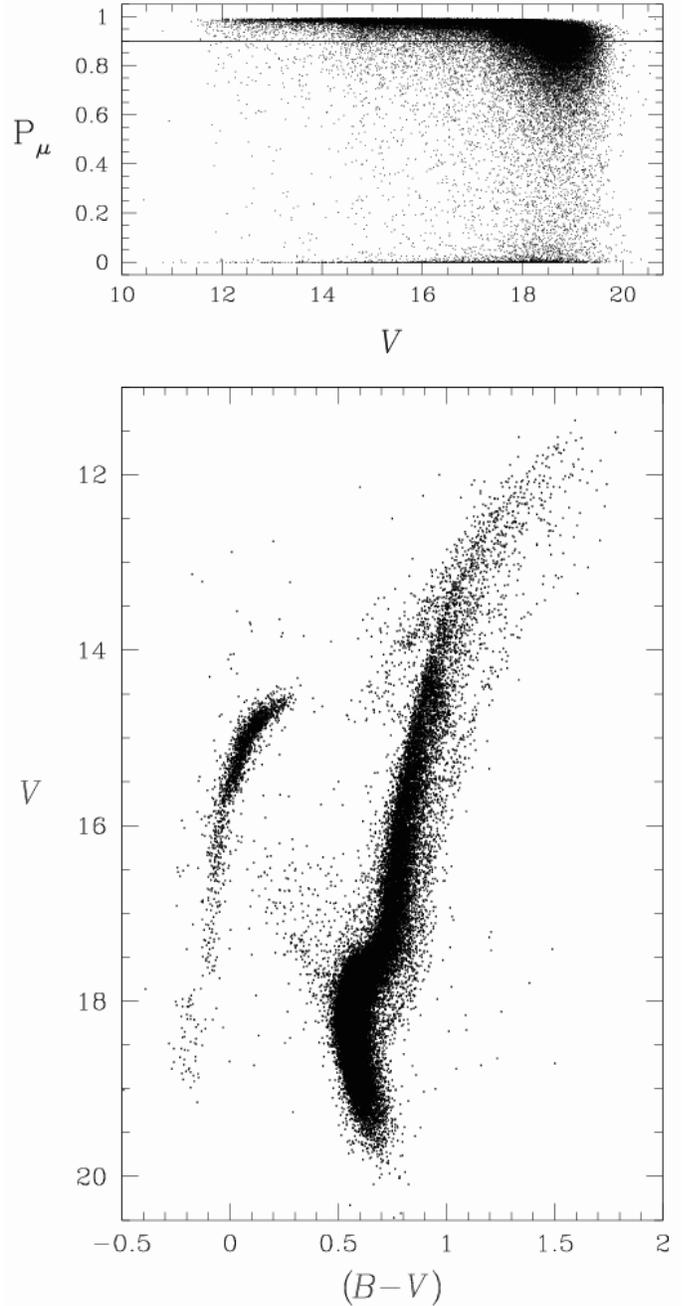}
\caption{Membership  probability $P_{\mu}$  versus  $V$ magnitude  for
  stars with $\sigma_{V,B}<0.05$ mag  and proper-motion r.m.s $<5$ mas
  yr$^{-1}$.   Horizontal  line show  the  $P_{\mu}=90$\% level.   The
  resulting CMD for  stars with $P_{\mu}>90$\% is shown  on the bottom
  panel.}
\label{fig_mp90}
\end{figure}

First of all, we constructed  frequency functions for both cluster and
field stars,  $\Phi^{\nu}_c$ and  $\Phi^{\nu}_f$ respectively, derived
from the  cluster and  field star distribution  in the VPD.   
For the $i^{\rm th}$ star, these functions were as follows: 
$$
\Phi^{\nu}_c=\frac{\exp\left\{-\frac{1}{2}\left[
\frac{(\mu_{x_i}-\mu_{x_c})^2}{\sigma^2_{x_c}+\epsilon^2_{x_i}}+
\frac{(\mu_{y_i}-\mu_{y_c})^2}{\sigma^2_{y_c}+\epsilon^2_{y_i}}\right]\right\}
}
{2\pi(\sigma^2_{x_c}+\epsilon^2_{x_i})^{1/2}
(\sigma^2_{y_c}+\epsilon^2_{y_i})^{1/2}},
$$
and
$$
\Phi_f^{\nu}=\frac{
\exp \left\{-\frac{1}{2(1-\gamma^2)}\cdot
\left[
\frac{(\mu_{x_i}-\mu_{x_f})^2}{\sigma^2_{x_f}+\epsilon^2_{x_i}}
-\frac{2\gamma      (\mu_{x_i}-\mu_{x_f})       (\mu_{y_i}-\mu_{y_f})}
{(\sigma^2_{x_f}+\epsilon^2_{x_i})^{1/2}
(\sigma^2_{y_f}+\epsilon^2_{y_i})^{1/2}} +
\frac{(\mu_{y_i}-\mu_{y_f})^2}{\sigma^2_{y_f}+\epsilon^2_{y_i}}\right]\right\}
}     {2\pi (1-\gamma^2)^{1/2} (\sigma^2_{x_f}+\epsilon^2_{x_i})^{1/2}
(\sigma^2_{y_f}+\epsilon^2_{y_i})^{1/2}}; $$
where $(\mu_{x_i},\mu_{y_i})$ are the $i^{\rm th}$ star proper motions
along  the  $X$   and  $Y$  axes,  $(\epsilon_{x_i},  \epsilon_{y_i})$
represent the  respective displacement r.m.s., $(\mu_{x_f},\mu_{y_f})$
and $(\mu_{x_c},\mu_{y_c})$  are the central  points of the  field and
cluster   star   proper   motion,  $(\sigma_{x_f},\sigma_{y_f})$   and
$(\sigma_{x_c},\sigma_{y_c})$   are  the   field   and  cluster   star
proper-motion  intrinsic dispersion, and  $\gamma$ is  the correlation
coefficient, calculated to be
$$
\gamma=\frac{(\mu_{x_i}-\mu_{x_f})(\mu_{y_i}-\mu_{y_f})}{\sigma_{x_f}\sigma_{y_f}}.
$$

The spatial distribution was ignored  due to the relatively small size
of our field ($\sim33\arcmin\times33\arcmin$)  with respect to the \om
radial    extent    of    $r_t\simeq57\arcmin$    (Harris    et    al.
\cite{harris96}).
For  our calculations,  we considered  only  stars with  an r.m.s.  in
proper   motion   $<1.9$   \masyr   to   define   $\Phi^{\nu}_c$   and
$\Phi^{\nu}_f$.
The center  of the proper-motion  distribution in the VPD  for cluster
stars was  found to  be at $x_c=0.00$  \masyr, and  $y_c=0.00$ \masyr,
with    r.m.s.    values    of    $\sigma_{x_c}=0.83$   \masyr,    and
$\sigma_{y_c}=0.83$ \masyr.
For  field  stars, we  have:  $x_f=-3.57$  \masyr, $y_f=5.12$  \masyr,
$\sigma_{x_f}=5.06$    \masyr,    and   $\sigma_{y_f}=2.86$    \masyr,
respectively.

%
The distribution  function  for all  the  stars  can  be  computed  as
follows:
$$
\Phi=(n_c\cdot\Phi^{\nu}_c)+(n_f\cdot\Phi^{\nu}_f)=\Phi_c+\Phi_f;
$$
where $n_c$ and  $n_f$ are the normalized number  of stars for cluster
and field $(n_c+n_f=1)$.
Therefore, for the $i^{\rm th}$ star the resulting membership probability
is
$$ P_\mu(i)=\frac{\Phi_c(i)}{\Phi(i)}.  $$

In  Fig.~\ref{fig_mp90}, we  show  in the  upper  panel the  $P_{\mu}$
distribution  versus $V$ magnitude.  To include  also faint  stars and
reach $V=20$, plotted stars have proper-motion r.m.s. $<5$ \masyr\ and
$\sigma_{V,B}<0.05$ mag.
The horizontal line marks the 90\% probability level.  The lower panel
contains the CMD for stars with $P_{\mu}>90$\%.

When calculating formal  membership probabilities for $\omega$~Cen, we
have a paradoxical situation in which  the main concern is to assign a
reasonable probability to  field stars. This is because  in our sample
the number  of cluster  stars is significantly  higher than  the small
number of field stars.
In  addition,  proper-motion  errors   have  a  strong  dependence  on
magnitude  (Figs.~\ref{fig_errori_1},~\ref{fig_PMgo6}),  which is  not
accounted for in our membership probability $P_{\mu}$ calculation.  We
therefore,  also  used the  so-called  local  sample  method (e.   g.,
Paper~II) for membership probability calculation.  In this method, for
each target star a sub-sample of stars was selected to reflect closely
the properties of a target.   This assures a smooth transition in the
calculated $P_{\mu}$ as a function of the chosen parameter.
For $\omega$~Cen, the  obvious choice of parameter was  the mean error
$\sigma_{\mu}$  of  the  proper  motions.   Given the  wide  range  of
$\sigma_{\mu}$,    we   chose    to   consider    only    stars   with
$\sigma_{\mu}$$<$7  mas~yr$^{-1}$. Less  accurate  proper motions  are
marginally useful for membership studies.
We  note that  $\sigma_{\mu}$  is calculated  in  the same  way as  in
Sect.~4 and Fig.~\ref{fig_dcr1}.

For each  target star,  we then  selected a star  a subsample  of 3000
stars almost with  identical proper-motion errors to that  of a target
star.   The trial  calculations  indicated that  we  cannot model  the
distribution  of field  stars with  a Gaussian  because the  number of
potential field stars in the  vector-point diagram is extremely low in
the vicinity  of the cluster-star  centroid (Fig.~\ref{fig_vl_check}).
A reasonable alternative to a Gaussian is a flat distribution of field
stars.

In  essence, the  membership probability  $P_{\mu}$ is  driven  by the
distribution of  cluster stars.  If  the modulus of  the proper-motion
vector  of a  star exceeds  $2.5$-$3\sigma_{\mu}$,  the corresponding
$P_{\mu}$ is less than 1\%.
We   provide  these   alternative  estimates   of   $P_{\mu}$  (called
$P_{\mu}(2)$,  to distinguish  this  from the  first mentioned  method
$P_{\mu}(1)$) for $120\,259$ stars.
For  the majority  of stars,  both  values of  $P_{\mu}$ are  similar.
However, there are a number  of cases in which the $P_{\mu}$ estimates
for the two methods differ radically.
In the case  of high proper-motion errors for cluster stars, the local
sample   method   clearly  provides   a   more  realistic   membership
probability.
A closer  inspection of these  cases indicates indirectly  a potential
problem in calculating the proper motion.
If the error  of proper motion along one axis  is several times larger
than the error along the other axis,  or if this error is too high for
a particular  magnitude, the chances are that  our proper-motion value
is  corrupted  and,  hence,   its  formal  membership  probability  is
meaningless.

Unless  specified otherwise, we  mean $P_{\mu}(1)$  determination when
referring to $P_{\mu}$.

\section{Applications}
\label{sec_apps}

Our \om proper-motion catalog can be used for different purposes.  The
first application was  the selection of the most  probable \om members
for spectroscopic follow-up studies.
In Villanova  et al.   (\cite{villanova07}), we used  the proper-motion
catalog of the present paper to pre-select sub-giant branch stars.
This helped us   to avoid the  Galactic  field stars  close  to the TO
level.  
All resultant radial velocities were close to the \om mean value, which
confirmed their membership.
On the other  hand, the high photometric quality  and the availability
of several filters covering a wide area around the cluster, imply that
this catalog is  an excellent photometric reference frame  on which to
register photometry from different  telescopes and cameras (Bellini et
al. in preparation).

Besides these obvious applications, our \om proper-motion catalog also
provides an  observational constraint of  the origin of  the composite
stellar populations in $\omega$~Cen.
Our catalog provides the necessary wide-field coverage and photometric
accuracy to  investigate the radial distribution of  the different \om
sub-population from the center of  the cluster to $\sim 22\arcmin$ (in
the corners).  We  will report on this analysis  in a subsequent paper
(Bellini et al. in preparation).

\subsection{The proper motion of the RGB sub-populations}
\label{subsec_rgb_a_mp}

Ferraro    et   al.  (\cite{ferraro02})     cross-correlated   the WFI
photometric  catalog of Pancino   et al.  (\cite{pancino00})  with the
vL00 photographic proper-motion catalog.
Their goal  was to investigate the nature  of the anomalous metal-rich
RGB of $\omega$~Cen, the so-called RGB-a.
In particular,  they investigated  the presence of  proper-motion mean
differences between the \om bulk population (metal-poor RGB, so-called
RGB-MP) and the minor,  but yet important, metal-rich RGB-a population
(Pancino et al. \cite{pancino00}).
Their  Fig.~2  showed significant  variation  in  the relative  proper
motion  of  RGB-a  stars  with  respect  to  \om  RGB-MP  motion.   In
particular, they found  that the RGB-a stars had  a mean proper motion
of $|\delta\mu_{{\rm{tot}}}|=0.8$ \masyr\ that  is offset from that of
the RGB-MP population.
Therefore, they concluded that the RGB-a subpopulation must have had an
independent origin with respect to the RGB-MP one.

Unsurprisingly,  the    Ferraro  et  al.  (\cite{ferraro02})  study
triggered new interest  in the \om proper  motion, and  Platais et al.
(\cite{platais03})  presented   a detailed   reanalysis  of  the vL00
catalog.
Platais  et   al.  (\cite{platais03})  concluded   that  the  reported
proper-motion  offset   between  the  \om   sub-populations  could  be
attributed to instrumental bias.
However, Hughes  et al. (\cite{hughes04}) commented that  there was no
residual color  term in the omega  Cen proper motions,  and that these
authors misinterpreted the  observed offsets.  Specifically, Hughes et
al.  (\cite{hughes04}) asserted that the summary effect of color terms
(before   the    corrections)   amounted    to   no   more    than   1
\masyr$\,$mag$^{-1}$ in $B-V$, while  the offset in the proper motions
for the anomalous  omega Cen stars reached 2 \masyr\  and did not have
the same direction as the color term.
Regardless of the reason for  the reported offset in the vL00 catalog,
the  presence   of  this   offset  was  not   confirmed  by   the  new
\textit{HST}-based     preliminary     proper    motions     (Anderson
\cite{jay03a}).

Two    spectroscopic   studies   completed    by   Pancino    et   al.
(\cite{pancino07})  and Johnson  et al.   (\cite{Johnson08}) indicated
that there was  no evidence for any of the  RGB stellar populations to
have  an offset in  the mean  radial velocity,  or a  different radial
velocity dispersion.
This result applied also to the RGB-a sub-population.

\begin{figure}[t!]
\centering
\includegraphics[width=8.50cm, height=7cm]{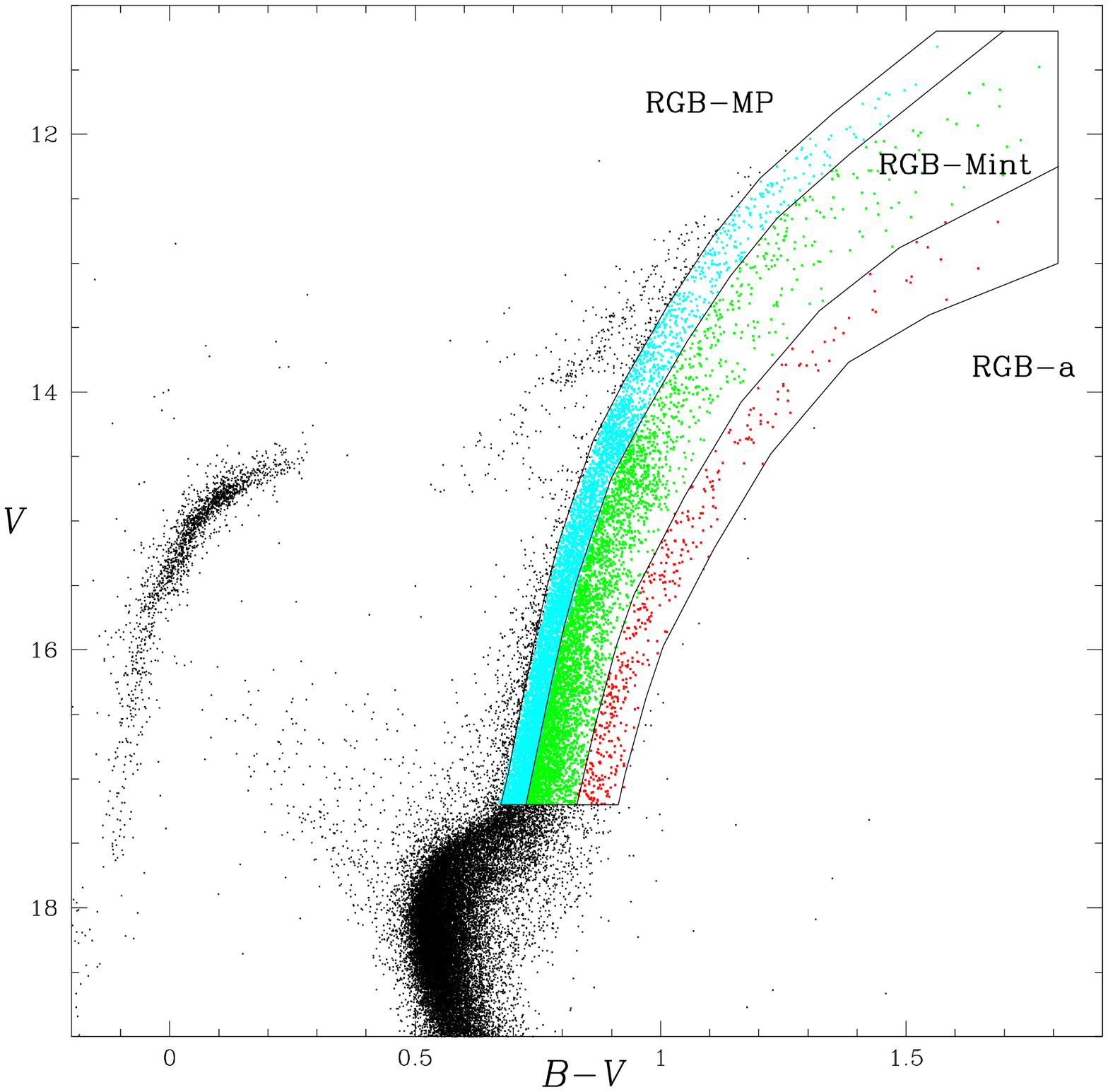}\\
\includegraphics[width=9.0cm]{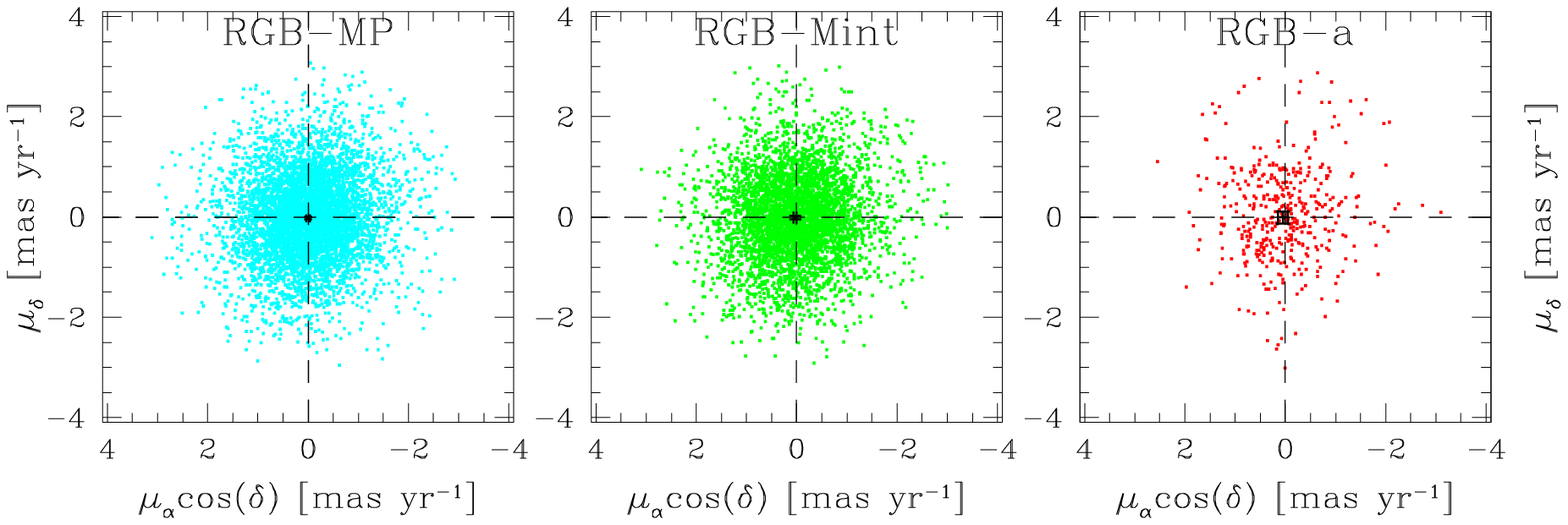}
\caption{(\textit{Top    panel}):    $V$    vs.     $B-V$    CMD    of
  proper-motion-selected member  stars on  which we defined  three RGB
  subsamples: RGB-MP  (cyan), RGB-Mint  (green), and RGB-a  (red). See
  text  for the  definition of  the  subpopulations.  (\textit{Bottom
    panels}):  Proper  motions  of  RGB-MP  (\textit{left}),  RGB-Mint
  (\textit{middle}), and RGB-a (\textit{right}) subpopulations.}
\label{fig_rgbamp}
\end{figure}

Our astrometric catalog provided an independent data set with which we
could test the Ferraro et al.  (\cite{ferraro02}) conclusions.
We   repeated  the  same   analysis  performed   in  Ferraro   et  al.
(\cite{ferraro02}).   First, we  divided the  \om RGB  population into
three subpopulations (see  top panel of Fig.~\ref{fig_rgbamp}): RGB-MP
(cyan), RGB-a (red), and RGB-Mint (RGB stars between RGB-MP and RGB-a,
green).
All plotted stars had  high membership probability ($P_\mu>90\%$), and
photometric errors ranging from 0.02  mag for bright stars to 0.05 mag
for faintest ones, in both filters.

The RGB samples include  all stars with $V\le17.2$.   
MP and Mint  RGB stars merge with  each other at  fainter magnitudes.
Sollima et   al.   (\cite{sollima05})  and Villanova et   al.
(\cite{villanova07}) clearly showed that the  SGB region of \om includes
five subpopulations (see again Fig.~\ref{fig:cmd_all}).
Three of these merge into each other and  form the RGB-MP and RGB-Mint
samples.
We  emphasize that  our  population  sub-division extends to   fainter
magnitudes  than in Ferraro et   al. (\cite{ferraro02}), and therefore
provides a higher RGB sampling  and  more robust statistics.  
Our samples contain 5182 RGB-MP stars, 3127 RGB-Mint stars,
 and 313 RGB-a stars.

A  closer look  at  the  three RGB  sub-population  proper motions  is
presented in the bottom panels of Fig.~\ref{fig_rgbamp}: RGB-MP on the
left, RGB-Mint in the middle, and RGB-a on the right.
The red crosses report the median value (estimated iteratively) of the
proper motion of the three sub-samples.   
For the RGB-MP sample, we have:
$$
\left\{
\begin{array}{ll}
<\mu_{\alpha} \cos\delta> = (0.00 \pm 0.01)
\rm{\,\,mas\,}\rm{yr}^{-1}&\\ <\mu_\delta> = (-0.02 \pm 0.01)
\rm{\,\,mas\,}\rm{yr}^{-1},&
\end{array}
\right.
$$ 
for RGB-Mint stars:
$$
\left\{
\begin{array}{ll}
<\mu_{\alpha} \cos\delta> = (0.03 \pm 0.01)
\rm{\,\,mas\,}\rm{yr}^{-1}&\\ <\mu_\delta> = (-0.01 \pm 0.01)
\rm{\,\,mas\,}\rm{yr}^{-1},&
\end{array}
\right.
$$ 
and for the RGB-a sample:
$$
\left\{
\begin{array}{ll}
<\mu_{\alpha} \cos\delta> = (0.05 \pm 0.03) \rm{\,\,mas\,}\rm{yr}^{-1}&\\
<\mu_\delta> = (-0.01 \pm 0.04) \rm{\,\,mas\,}\rm{yr}^{-1}.&
\end{array}
\right.
$$

We found no evidence for the presence of differences among the relative
proper motions  of the three RGB  sub-populations at the  level of 0.05
\masyr\ in $\mu_\alpha \cos \delta$  and of 0.04 \masyr\ in $\mu_\delta$
(i.e.   relative  tangential   velocities  of   $\lesssim   1.3$,  and
$\lesssim1.1$  km  s$^{-1}$,  assuming  a  distance  of  5.5  kpc  for
$\omega$~Cen).  All  three RGB  sub-samples exhibit the  same mean
proper motion, within the errors.

We  therefore,  agree  with  the   results  of  both  Platais  et  al.
(\cite{platais03}) and  Johnson et al.   (\cite{Johnson08}) for RGB-a,
and show that  the RGB-Mint proper motion is  also consistent with the
other \om sub-populations.

A  final word on  this issue requires   an internal stellar proper-motion
analysis, but suitable catalogs are not yet available.

\begin{figure*}[ht!]
\centering
\includegraphics[width=9.0cm,height=13cm]{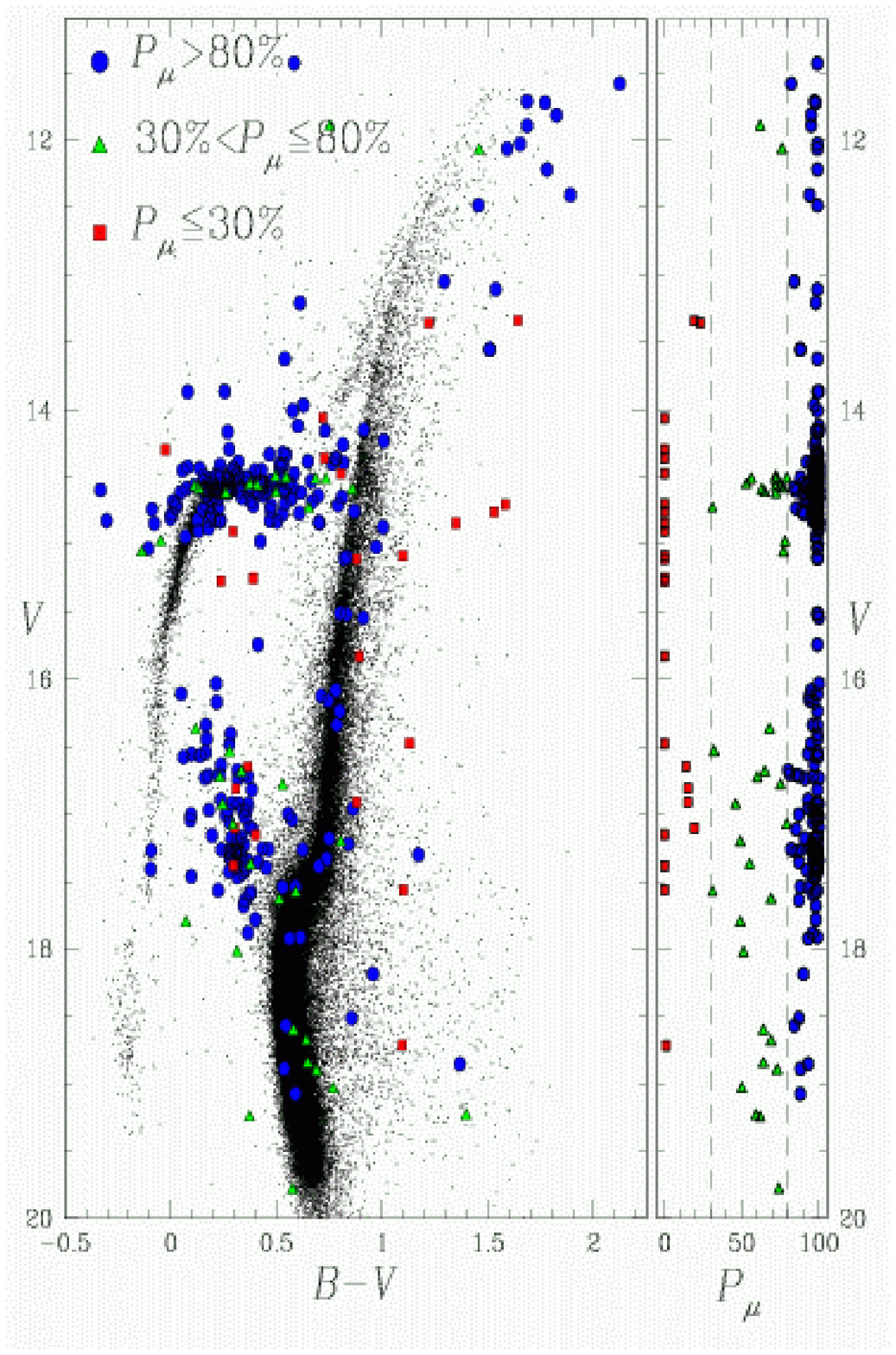}
\includegraphics[width=9.0cm,height=13cm]{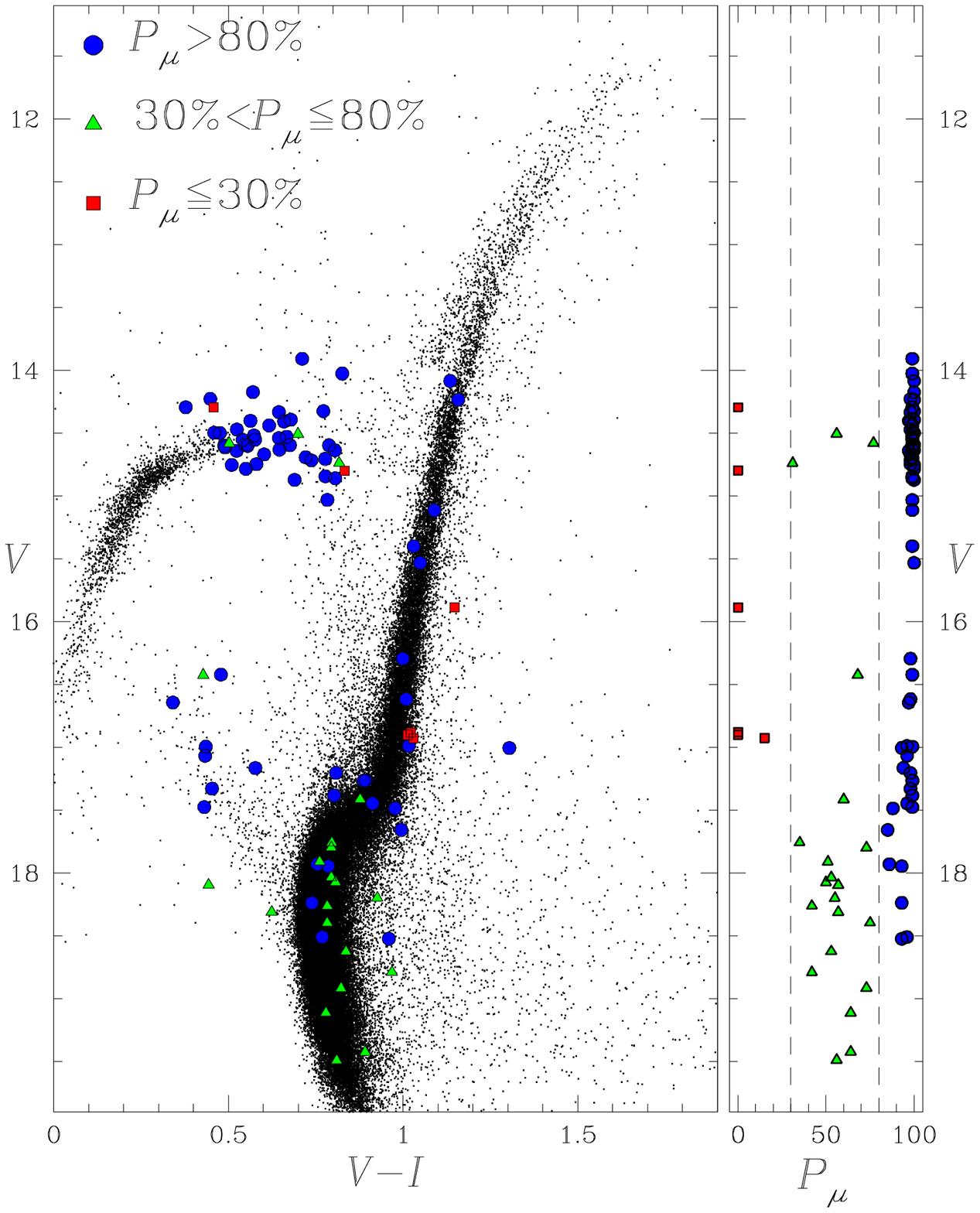}\\
\caption{(\textit{Left}): \om CMD with the cross-checked variable stars
 from the Kaluzny et al.  (\cite{kaluzny04}) catalog; red squares mark
 stars with a  membership probability $P_{\mu}<30$; green triangles are
 stars with  $30\le  P_{\mu}<80$, while  blue circles  are those with a
 probability to be  cluster  members  $P_{\mu}\ge80$.  The   membership
 probability    $P_{\mu}$ versus the    $V$ magnitude  is also  shown.
 (\textit{Right}): comparison with Weldrake et al.  (\cite{weldrake07})
 catalog.}
\label{fig_variable}
\end{figure*}

\subsection{Membership probability of published \om variable stars}
\label{subsec_var}

The study  of variable  stars in  \om can certainly  benefit from  our
proper-motion catalog and cluster membership derivation.

Using the  1.0$m$ Swope Telescope, Kaluzny  et al.  (\cite{kaluzny04})
measured light  curves of $\sim  400$ variable stars  in $\omega$~Cen,
117 of which were new identifications.
We cross-checked our proper-motion catalog with their own, and found a
total of  338 variable  stars in  common, which had  both $V$  and $B$
measurements in our catalog.
In particular,  there  were 90  variable  stars  for which  Kaluzny  et
al. (\cite{kaluzny04}) did  not  provide $B$ and/or   $V$ magnitudes.  
Our proper-motion catalog was also useful to locate these stars in the
color-magnitude diagram.
The position of the Kaluzny et al. (\cite{kaluzny04}) variables in our
CMD, as  well as their  membership probability,  is shown  in the left
panel of Fig.~\ref{fig_variable}.
Depending on our measured   membership   probability, we divide    the
Kaluzny  et al. (\cite{kaluzny04})  sample into  three categories: (1)
$P_{\mu}<30\%$    (red  squares);  (2)  $30\%\le   P_{\mu}<80\%$  (green
triangles); and (3) $P_{\mu}\ge80\%$ (blue circles).
Of the  117 new variable identifications, 112  are  in common with our
catalog.   Of  these 112 stars,  15  have $P_{\mu}<30\%$, and therefore
these are probably not cluster members.
On  the   other  hand,  19  of  these  new   variables have  $30\%  \le
P_{\mu}<80\%$ and their membership remains uncertain.  The remaining 78
stars ($P_{\mu}\ge80\%$) are almost certainly \om members.
In Table~\ref{tab_kalu},  we report the  membership probability values
for all  Kaluzny et al.  (\cite{kaluzny04})  variable stars identified
in our catalog.
New Kaluzny et al. (\cite{kaluzny04})  identifications have their  IDs
starting with ``N''.

%
The  Kaluzny  et al.   (\cite{kaluzny04})  variable  star catalog  was
cross-checked  with  the  X-ray  sources detected  by  the  XMM-Newton
analysis presented by Gendre et al.  (\cite{gendre03}).
For  the  9 stars  in  common  (see also  Table~2  of  Kaluzny et  al.
\cite{kaluzny04}),  we  can now  provide  (see Table~\ref{tab_xmm})  a
membership probability  based on our proper-motion  analysis.  A quick
glance at  Table~4 allows us  to infer that  only NV383 and  NV369 are
very likely  cluster members, whereas  the remaining 7 stars  are most
probably field population.

\begin{table*}[ht!]
\centering
\caption{Membership    probability   for    the    Kaluzny   et    al.
  (\cite{kaluzny04})  variable star  catalog. ID$_K$  are  the Kaluzny
  identification labels, while ID$_{tw}$ refer to this~work.} \tiny{
\begin{tabular}{rclrclrclrclrclrcl}
\hline                                                           \hline
\multicolumn{18}{c}{\phantom{z}}\\
\multicolumn{1}{l}{ID$_K$}&\multicolumn{1}{c}{$P_{\mu}$}&\multicolumn{1}{l}{ID$_{tw}$\phantom{zzz}}&
\multicolumn{1}{l}{ID$_K$}&\multicolumn{1}{c}{$P_{\mu}$}&\multicolumn{1}{l}{ID$_{tw}$\phantom{zzz}}&
\multicolumn{1}{l}{ID$_K$}&\multicolumn{1}{c}{$P_{\mu}$}&\multicolumn{1}{l}{ID$_{tw}$\phantom{zzz}}&
\multicolumn{1}{l}{ID$_K$}&\multicolumn{1}{c}{$P_{\mu}$}&\multicolumn{1}{l}{ID$_{tw}$\phantom{zzz}}&
\multicolumn{1}{l}{ID$_K$}&\multicolumn{1}{c}{$P_{\mu}$}&\multicolumn{1}{l}{ID$_{tw}$\phantom{zzz}}&
\multicolumn{1}{l}{ID$_K$}&\multicolumn{1}{c}{$P_{\mu}$}&\multicolumn{1}{l}{ID$_{tw}$}\\
\multicolumn{18}{c}{\phantom{z}}\\                               \hline
\multicolumn{18}{c}{\phantom{z}}\\    V1    $\!\!\!\!$&$\!\!\!\!$   99
$\!\!\!\!$&$\!\!\!\!$   290466    &   V66   $\!\!\!\!$&$\!\!\!\!$   99
$\!\!\!\!$&$\!\!\!\!$   311494   &   V132   $\!\!\!\!$&$\!\!\!\!$   96
$\!\!\!\!$&$\!\!\!\!$   184072   &   V214   $\!\!\!\!$&$\!\!\!\!$   97
$\!\!\!\!$&$\!\!\!\!$   42989   &   NV296   $\!\!\!\!$&$\!\!\!\!$   99
$\!\!\!\!$&$\!\!\!\!$   356415   &   NV354  $\!\!\!\!$&$\!\!\!\!$   97
$\!\!\!\!$&$\!\!\!\!$   263543    \\   V2   $\!\!\!\!$&$\!\!\!\!$   74
$\!\!\!\!$&$\!\!\!\!$   271987    &   V67   $\!\!\!\!$&$\!\!\!\!$   99
$\!\!\!\!$&$\!\!\!\!$   352882   &   V135   $\!\!\!\!$&$\!\!\!\!$   97
$\!\!\!\!$&$\!\!\!\!$   181199   &   V216   $\!\!\!\!$&$\!\!\!\!$   99
$\!\!\!\!$&$\!\!\!\!$   155556   &   NV297  $\!\!\!\!$&$\!\!\!\!$   32
$\!\!\!\!$&$\!\!\!\!$   250290   &   NV355  $\!\!\!\!$&$\!\!\!\!$   93
$\!\!\!\!$&$\!\!\!\!$   196052    \\   V3   $\!\!\!\!$&$\!\!\!\!$   99
$\!\!\!\!$&$\!\!\!\!$   248511   &   V68   $\!\!\!\!$&$\!\!\!\!$   100
$\!\!\!\!$&$\!\!\!\!$   345639   &   V136   $\!\!\!\!$&$\!\!\!\!$   99
$\!\!\!\!$&$\!\!\!\!$   213549   &   V217   $\!\!\!\!$&$\!\!\!\!$   98
$\!\!\!\!$&$\!\!\!\!$   218342  &   NV298   $\!\!\!\!$&$\!\!\!\!$  100
$\!\!\!\!$&$\!\!\!\!$   87382   &   NV356   $\!\!\!\!$&$\!\!\!\!$   49
$\!\!\!\!$&$\!\!\!\!$   206069    \\   V4   $\!\!\!\!$&$\!\!\!\!$   97
$\!\!\!\!$&$\!\!\!\!$   279540    &   V70   $\!\!\!\!$&$\!\!\!\!$   98
$\!\!\!\!$&$\!\!\!\!$   93459    &   V137   $\!\!\!\!$&$\!\!\!\!$   99
$\!\!\!\!$&$\!\!\!\!$   225149   &   V218   $\!\!\!\!$&$\!\!\!\!$   96
$\!\!\!\!$&$\!\!\!\!$   358729   &   NV299  $\!\!\!\!$&$\!\!\!\!$   98
$\!\!\!\!$&$\!\!\!\!$   310461   &   NV357  $\!\!\!\!$&$\!\!\!\!$   92
$\!\!\!\!$&$\!\!\!\!$   158687   \\   V5   $\!\!\!\!$&$\!\!\!\!$   100
$\!\!\!\!$&$\!\!\!\!$   298983    &   V71   $\!\!\!\!$&$\!\!\!\!$   99
$\!\!\!\!$&$\!\!\!\!$   209884   &   V139   $\!\!\!\!$&$\!\!\!\!$   90
$\!\!\!\!$&$\!\!\!\!$   215311   &   V219   $\!\!\!\!$&$\!\!\!\!$   88
$\!\!\!\!$&$\!\!\!\!$   347441   &   NV300  $\!\!\!\!$&$\!\!\!\!$   98
$\!\!\!\!$&$\!\!\!\!$   284928   &   NV358  $\!\!\!\!$&$\!\!\!\!$   99
$\!\!\!\!$&$\!\!\!\!$   129801    \\   V8   $\!\!\!\!$&$\!\!\!\!$   99
$\!\!\!\!$&$\!\!\!\!$   200194   &   V74   $\!\!\!\!$&$\!\!\!\!$   100
$\!\!\!\!$&$\!\!\!\!$   360819   &   V141  $\!\!\!\!$&$\!\!\!\!$   100
$\!\!\!\!$&$\!\!\!\!$   177819   &   V220  $\!\!\!\!$&$\!\!\!\!$   100
$\!\!\!\!$&$\!\!\!\!$   321724   &   NV301  $\!\!\!\!$&$\!\!\!\!$   96
$\!\!\!\!$&$\!\!\!\!$   170554   &   NV359  $\!\!\!\!$&$\!\!\!\!$   46
$\!\!\!\!$&$\!\!\!\!$   222238   \\   V9   $\!\!\!\!$&$\!\!\!\!$   100
$\!\!\!\!$&$\!\!\!\!$   238167    &   V75   $\!\!\!\!$&$\!\!\!\!$   99
$\!\!\!\!$&$\!\!\!\!$   352935   &   V142   $\!\!\!\!$&$\!\!\!\!$   65
$\!\!\!\!$&$\!\!\!\!$   193092   &   V221   $\!\!\!\!$&$\!\!\!\!$   96
$\!\!\!\!$&$\!\!\!\!$   320315   &   NV302  $\!\!\!\!$&$\!\!\!\!$   19
$\!\!\!\!$&$\!\!\!\!$   252220  &   NV360   $\!\!\!\!$&$\!\!\!\!$  100
$\!\!\!\!$&$\!\!\!\!$   118132   \\   V10   $\!\!\!\!$&$\!\!\!\!$   98
$\!\!\!\!$&$\!\!\!\!$   273997   &   V76   $\!\!\!\!$&$\!\!\!\!$   100
$\!\!\!\!$&$\!\!\!\!$   340408   &   V143   $\!\!\!\!$&$\!\!\!\!$   90
$\!\!\!\!$&$\!\!\!\!$   217472    &   V223   $\!\!\!\!$&$\!\!\!\!$   0
$\!\!\!\!$&$\!\!\!\!$   355759   &   NV303  $\!\!\!\!$&$\!\!\!\!$   98
$\!\!\!\!$&$\!\!\!\!$   145690   &   NV361  $\!\!\!\!$&$\!\!\!\!$   64
$\!\!\!\!$&$\!\!\!\!$   234998   \\   V11   $\!\!\!\!$&$\!\!\!\!$   99
$\!\!\!\!$&$\!\!\!\!$   302079    &   V77   $\!\!\!\!$&$\!\!\!\!$   99
$\!\!\!\!$&$\!\!\!\!$   316247   &   V144   $\!\!\!\!$&$\!\!\!\!$   95
$\!\!\!\!$&$\!\!\!\!$   201059   &   V224   $\!\!\!\!$&$\!\!\!\!$   99
$\!\!\!\!$&$\!\!\!\!$   349215   &   NV304  $\!\!\!\!$&$\!\!\!\!$   98
$\!\!\!\!$&$\!\!\!\!$   341374   &   NV363  $\!\!\!\!$&$\!\!\!\!$   93
$\!\!\!\!$&$\!\!\!\!$   321474   \\   V12   $\!\!\!\!$&$\!\!\!\!$   99
$\!\!\!\!$&$\!\!\!\!$    283465   &   V78    $\!\!\!\!$&$\!\!\!\!$   0
$\!\!\!\!$&$\!\!\!\!$   243506   &   V145   $\!\!\!\!$&$\!\!\!\!$   91
$\!\!\!\!$&$\!\!\!\!$   143665   &   V225   $\!\!\!\!$&$\!\!\!\!$   96
$\!\!\!\!$&$\!\!\!\!$   274338   &   NV305   $\!\!\!\!$&$\!\!\!\!$   0
$\!\!\!\!$&$\!\!\!\!$   343091   &   NV364  $\!\!\!\!$&$\!\!\!\!$   93
$\!\!\!\!$&$\!\!\!\!$   243121   \\   V13   $\!\!\!\!$&$\!\!\!\!$   99
$\!\!\!\!$&$\!\!\!\!$   259326    &   V81   $\!\!\!\!$&$\!\!\!\!$   99
$\!\!\!\!$&$\!\!\!\!$   270509   &   V146   $\!\!\!\!$&$\!\!\!\!$   72
$\!\!\!\!$&$\!\!\!\!$   177889   &   V226   $\!\!\!\!$&$\!\!\!\!$   99
$\!\!\!\!$&$\!\!\!\!$   167657   &   NV306  $\!\!\!\!$&$\!\!\!\!$   96
$\!\!\!\!$&$\!\!\!\!$   198996   &   NV365   $\!\!\!\!$&$\!\!\!\!$   0
$\!\!\!\!$&$\!\!\!\!$   71679    \\   V14   $\!\!\!\!$&$\!\!\!\!$   97
$\!\!\!\!$&$\!\!\!\!$    26311   &   V82    $\!\!\!\!$&$\!\!\!\!$   99
$\!\!\!\!$&$\!\!\!\!$   236315   &   V147   $\!\!\!\!$&$\!\!\!\!$   97
$\!\!\!\!$&$\!\!\!\!$   143269   &   V227   $\!\!\!\!$&$\!\!\!\!$   98
$\!\!\!\!$&$\!\!\!\!$   287148   &   NV307  $\!\!\!\!$&$\!\!\!\!$   82
$\!\!\!\!$&$\!\!\!\!$   211960   &   NV366  $\!\!\!\!$&$\!\!\!\!$   97
$\!\!\!\!$&$\!\!\!\!$   132355   \\   V15  $\!\!\!\!$&$\!\!\!\!$   100
$\!\!\!\!$&$\!\!\!\!$   273754    &   V83   $\!\!\!\!$&$\!\!\!\!$   99
$\!\!\!\!$&$\!\!\!\!$   323500   &   V148   $\!\!\!\!$&$\!\!\!\!$   95
$\!\!\!\!$&$\!\!\!\!$   209005   &   V228   $\!\!\!\!$&$\!\!\!\!$   96
$\!\!\!\!$&$\!\!\!\!$   92375   &   NV308   $\!\!\!\!$&$\!\!\!\!$   98
$\!\!\!\!$&$\!\!\!\!$   327308   &   NV367   $\!\!\!\!$&$\!\!\!\!$   0
$\!\!\!\!$&$\!\!\!\!$   350053   \\   V16  $\!\!\!\!$&$\!\!\!\!$   100
$\!\!\!\!$&$\!\!\!\!$    40312   &   V85    $\!\!\!\!$&$\!\!\!\!$   98
$\!\!\!\!$&$\!\!\!\!$   292312   &   V150  $\!\!\!\!$&$\!\!\!\!$   100
$\!\!\!\!$&$\!\!\!\!$   58317    &   V229   $\!\!\!\!$&$\!\!\!\!$   97
$\!\!\!\!$&$\!\!\!\!$   285982   &   NV309  $\!\!\!\!$&$\!\!\!\!$   85
$\!\!\!\!$&$\!\!\!\!$   193994   &   NV368   $\!\!\!\!$&$\!\!\!\!$   0
$\!\!\!\!$&$\!\!\!\!$   349013   \\   V17   $\!\!\!\!$&$\!\!\!\!$   99
$\!\!\!\!$&$\!\!\!\!$   261135    &   V86   $\!\!\!\!$&$\!\!\!\!$   99
$\!\!\!\!$&$\!\!\!\!$   242948   &   V152   $\!\!\!\!$&$\!\!\!\!$   82
$\!\!\!\!$&$\!\!\!\!$   177559   &   V231   $\!\!\!\!$&$\!\!\!\!$   99
$\!\!\!\!$&$\!\!\!\!$   59688   &   NV310   $\!\!\!\!$&$\!\!\!\!$   15
$\!\!\!\!$&$\!\!\!\!$   127228   &   NV369  $\!\!\!\!$&$\!\!\!\!$   94
$\!\!\!\!$&$\!\!\!\!$   124330   \\   V18   $\!\!\!\!$&$\!\!\!\!$   98
$\!\!\!\!$&$\!\!\!\!$   267824    &   V87   $\!\!\!\!$&$\!\!\!\!$   97
$\!\!\!\!$&$\!\!\!\!$   255154   &   V153  $\!\!\!\!$&$\!\!\!\!$   100
$\!\!\!\!$&$\!\!\!\!$   238719   &   V233   $\!\!\!\!$&$\!\!\!\!$   99
$\!\!\!\!$&$\!\!\!\!$   336053   &   NV311  $\!\!\!\!$&$\!\!\!\!$   91
$\!\!\!\!$&$\!\!\!\!$   170132   &   NV370   $\!\!\!\!$&$\!\!\!\!$   0
$\!\!\!\!$&$\!\!\!\!$   358442   \\   V19  $\!\!\!\!$&$\!\!\!\!$   100
$\!\!\!\!$&$\!\!\!\!$   205260    &   V88   $\!\!\!\!$&$\!\!\!\!$   94
$\!\!\!\!$&$\!\!\!\!$   261450   &   V154   $\!\!\!\!$&$\!\!\!\!$   96
$\!\!\!\!$&$\!\!\!\!$   155744    &   V234   $\!\!\!\!$&$\!\!\!\!$   0
$\!\!\!\!$&$\!\!\!\!$   157490   &   NV312  $\!\!\!\!$&$\!\!\!\!$   95
$\!\!\!\!$&$\!\!\!\!$   183516   &   NV372  $\!\!\!\!$&$\!\!\!\!$   93
$\!\!\!\!$&$\!\!\!\!$   290428   \\   V20   $\!\!\!\!$&$\!\!\!\!$   99
$\!\!\!\!$&$\!\!\!\!$   204861    &   V90   $\!\!\!\!$&$\!\!\!\!$   89
$\!\!\!\!$&$\!\!\!\!$   239287   &   V155   $\!\!\!\!$&$\!\!\!\!$   98
$\!\!\!\!$&$\!\!\!\!$   272492   &   V235   $\!\!\!\!$&$\!\!\!\!$   87
$\!\!\!\!$&$\!\!\!\!$   309505   &   NV313  $\!\!\!\!$&$\!\!\!\!$   98
$\!\!\!\!$&$\!\!\!\!$   134090   &   NV373  $\!\!\!\!$&$\!\!\!\!$   59
$\!\!\!\!$&$\!\!\!\!$   300775   \\   V21   $\!\!\!\!$&$\!\!\!\!$   99
$\!\!\!\!$&$\!\!\!\!$   246994    &   V91   $\!\!\!\!$&$\!\!\!\!$   92
$\!\!\!\!$&$\!\!\!\!$   241579   &   V156   $\!\!\!\!$&$\!\!\!\!$   99
$\!\!\!\!$&$\!\!\!\!$   128871   &   V236   $\!\!\!\!$&$\!\!\!\!$   85
$\!\!\!\!$&$\!\!\!\!$   282601   &   NV314  $\!\!\!\!$&$\!\!\!\!$   85
$\!\!\!\!$&$\!\!\!\!$   163946   &   NV374  $\!\!\!\!$&$\!\!\!\!$   99
$\!\!\!\!$&$\!\!\!\!$   312254   \\   V22   $\!\!\!\!$&$\!\!\!\!$   99
$\!\!\!\!$&$\!\!\!\!$    86072   &   V92    $\!\!\!\!$&$\!\!\!\!$   99
$\!\!\!\!$&$\!\!\!\!$   327526   &   V157   $\!\!\!\!$&$\!\!\!\!$   98
$\!\!\!\!$&$\!\!\!\!$   221081   &   V237   $\!\!\!\!$&$\!\!\!\!$   99
$\!\!\!\!$&$\!\!\!\!$   286453   &   NV315  $\!\!\!\!$&$\!\!\!\!$   60
$\!\!\!\!$&$\!\!\!\!$   267539   &   NV375   $\!\!\!\!$&$\!\!\!\!$   0
$\!\!\!\!$&$\!\!\!\!$   349760   \\   V23  $\!\!\!\!$&$\!\!\!\!$   100
$\!\!\!\!$&$\!\!\!\!$   273446    &   V94   $\!\!\!\!$&$\!\!\!\!$   98
$\!\!\!\!$&$\!\!\!\!$   305927   &   V158   $\!\!\!\!$&$\!\!\!\!$   31
$\!\!\!\!$&$\!\!\!\!$   153087   &   V238   $\!\!\!\!$&$\!\!\!\!$   99
$\!\!\!\!$&$\!\!\!\!$   179489   &   NV316  $\!\!\!\!$&$\!\!\!\!$   99
$\!\!\!\!$&$\!\!\!\!$   142842   &   NV376   $\!\!\!\!$&$\!\!\!\!$   0
$\!\!\!\!$&$\!\!\!\!$   189233   \\   V24   $\!\!\!\!$&$\!\!\!\!$   99
$\!\!\!\!$&$\!\!\!\!$    84169   &   V95    $\!\!\!\!$&$\!\!\!\!$   94
$\!\!\!\!$&$\!\!\!\!$   188847   &   V161   $\!\!\!\!$&$\!\!\!\!$   76
$\!\!\!\!$&$\!\!\!\!$   149737   &   V239   $\!\!\!\!$&$\!\!\!\!$   99
$\!\!\!\!$&$\!\!\!\!$   179912  &   NV317   $\!\!\!\!$&$\!\!\!\!$  100
$\!\!\!\!$&$\!\!\!\!$   361100   &   NV377   $\!\!\!\!$&$\!\!\!\!$   0
$\!\!\!\!$&$\!\!\!\!$   324933   \\   V25   $\!\!\!\!$&$\!\!\!\!$   99
$\!\!\!\!$&$\!\!\!\!$   199162    &   V96   $\!\!\!\!$&$\!\!\!\!$   63
$\!\!\!\!$&$\!\!\!\!$   225607   &   V162   $\!\!\!\!$&$\!\!\!\!$   97
$\!\!\!\!$&$\!\!\!\!$   108440   &   V240   $\!\!\!\!$&$\!\!\!\!$   99
$\!\!\!\!$&$\!\!\!\!$   239985   &   NV318  $\!\!\!\!$&$\!\!\!\!$   98
$\!\!\!\!$&$\!\!\!\!$   246801   &   NV378  $\!\!\!\!$&$\!\!\!\!$   31
$\!\!\!\!$&$\!\!\!\!$   341736   \\   V26   $\!\!\!\!$&$\!\!\!\!$   98
$\!\!\!\!$&$\!\!\!\!$   226616    &   V97   $\!\!\!\!$&$\!\!\!\!$   99
$\!\!\!\!$&$\!\!\!\!$   256623   &   V163  $\!\!\!\!$&$\!\!\!\!$   100
$\!\!\!\!$&$\!\!\!\!$   337663    &   V241   $\!\!\!\!$&$\!\!\!\!$   1
$\!\!\!\!$&$\!\!\!\!$   289203   &   NV319  $\!\!\!\!$&$\!\!\!\!$   94
$\!\!\!\!$&$\!\!\!\!$   145544   &   NV379  $\!\!\!\!$&$\!\!\!\!$   99
$\!\!\!\!$&$\!\!\!\!$   328605   \\   V27   $\!\!\!\!$&$\!\!\!\!$   96
$\!\!\!\!$&$\!\!\!\!$   201410    &   V98   $\!\!\!\!$&$\!\!\!\!$   99
$\!\!\!\!$&$\!\!\!\!$   227653   &   V164   $\!\!\!\!$&$\!\!\!\!$   99
$\!\!\!\!$&$\!\!\!\!$   334394   &   V242   $\!\!\!\!$&$\!\!\!\!$   98
$\!\!\!\!$&$\!\!\!\!$   171250   &   NV320  $\!\!\!\!$&$\!\!\!\!$   99
$\!\!\!\!$&$\!\!\!\!$   80363   &   NV380   $\!\!\!\!$&$\!\!\!\!$   96
$\!\!\!\!$&$\!\!\!\!$   148727   \\   V29   $\!\!\!\!$&$\!\!\!\!$   62
$\!\!\!\!$&$\!\!\!\!$   191401    &   V99   $\!\!\!\!$&$\!\!\!\!$   99
$\!\!\!\!$&$\!\!\!\!$   210796   &   V165   $\!\!\!\!$&$\!\!\!\!$   85
$\!\!\!\!$&$\!\!\!\!$   227981   &   V249   $\!\!\!\!$&$\!\!\!\!$   99
$\!\!\!\!$&$\!\!\!\!$   236550   &   NV321  $\!\!\!\!$&$\!\!\!\!$   87
$\!\!\!\!$&$\!\!\!\!$   209026   &   NV381   $\!\!\!\!$&$\!\!\!\!$   0
$\!\!\!\!$&$\!\!\!\!$   311380   \\   V30   $\!\!\!\!$&$\!\!\!\!$   99
$\!\!\!\!$&$\!\!\!\!$   168094   &   V100   $\!\!\!\!$&$\!\!\!\!$   77
$\!\!\!\!$&$\!\!\!\!$   215927   &   V166   $\!\!\!\!$&$\!\!\!\!$   72
$\!\!\!\!$&$\!\!\!\!$   241584   &   V250   $\!\!\!\!$&$\!\!\!\!$   92
$\!\!\!\!$&$\!\!\!\!$   62410   &   NV322   $\!\!\!\!$&$\!\!\!\!$   93
$\!\!\!\!$&$\!\!\!\!$   214126   &   NV382  $\!\!\!\!$&$\!\!\!\!$   96
$\!\!\!\!$&$\!\!\!\!$   343538   \\   V32  $\!\!\!\!$&$\!\!\!\!$   100
$\!\!\!\!$&$\!\!\!\!$   322448   &   V101   $\!\!\!\!$&$\!\!\!\!$   99
$\!\!\!\!$&$\!\!\!\!$   169863   &   V167   $\!\!\!\!$&$\!\!\!\!$   23
$\!\!\!\!$&$\!\!\!\!$   87332    &   V251   $\!\!\!\!$&$\!\!\!\!$   97
$\!\!\!\!$&$\!\!\!\!$   231812   &   NV323  $\!\!\!\!$&$\!\!\!\!$   65
$\!\!\!\!$&$\!\!\!\!$   265232   &   NV383  $\!\!\!\!$&$\!\!\!\!$   99
$\!\!\!\!$&$\!\!\!\!$   336713   \\   V33   $\!\!\!\!$&$\!\!\!\!$   98
$\!\!\!\!$&$\!\!\!\!$   184952   &   V102  $\!\!\!\!$&$\!\!\!\!$   100
$\!\!\!\!$&$\!\!\!\!$   162564    &   V168   $\!\!\!\!$&$\!\!\!\!$   0
$\!\!\!\!$&$\!\!\!\!$   124934   &   V252   $\!\!\!\!$&$\!\!\!\!$   98
$\!\!\!\!$&$\!\!\!\!$   187128   &   NV324  $\!\!\!\!$&$\!\!\!\!$   68
$\!\!\!\!$&$\!\!\!\!$   154689   &   NV384  $\!\!\!\!$&$\!\!\!\!$   93
$\!\!\!\!$&$\!\!\!\!$   46225    \\   V34   $\!\!\!\!$&$\!\!\!\!$   99
$\!\!\!\!$&$\!\!\!\!$   103189   &   V103   $\!\!\!\!$&$\!\!\!\!$   98
$\!\!\!\!$&$\!\!\!\!$   195140   &   V169   $\!\!\!\!$&$\!\!\!\!$   99
$\!\!\!\!$&$\!\!\!\!$   285765   &   V253   $\!\!\!\!$&$\!\!\!\!$   99
$\!\!\!\!$&$\!\!\!\!$   209800   &   NV325  $\!\!\!\!$&$\!\!\!\!$   80
$\!\!\!\!$&$\!\!\!\!$   163821   &   NV385  $\!\!\!\!$&$\!\!\!\!$   99
$\!\!\!\!$&$\!\!\!\!$   35533    \\   V35   $\!\!\!\!$&$\!\!\!\!$   99
$\!\!\!\!$&$\!\!\!\!$   309218   &   V104   $\!\!\!\!$&$\!\!\!\!$   99
$\!\!\!\!$&$\!\!\!\!$   92705   &   V184   $\!\!\!\!$&$\!\!\!\!$   100
$\!\!\!\!$&$\!\!\!\!$   134524   &   V254   $\!\!\!\!$&$\!\!\!\!$   99
$\!\!\!\!$&$\!\!\!\!$   214031   &   NV326  $\!\!\!\!$&$\!\!\!\!$   98
$\!\!\!\!$&$\!\!\!\!$   268304   &   NV386  $\!\!\!\!$&$\!\!\!\!$   98
$\!\!\!\!$&$\!\!\!\!$   218806   \\   V38   $\!\!\!\!$&$\!\!\!\!$   98
$\!\!\!\!$&$\!\!\!\!$   52122    &   V105   $\!\!\!\!$&$\!\!\!\!$   98
$\!\!\!\!$&$\!\!\!\!$   111797   &   V185   $\!\!\!\!$&$\!\!\!\!$   99
$\!\!\!\!$&$\!\!\!\!$   320506   &   V258   $\!\!\!\!$&$\!\!\!\!$   99
$\!\!\!\!$&$\!\!\!\!$   253920   &   NV327  $\!\!\!\!$&$\!\!\!\!$   14
$\!\!\!\!$&$\!\!\!\!$   235193   &   NV387  $\!\!\!\!$&$\!\!\!\!$   96
$\!\!\!\!$&$\!\!\!\!$   190420   \\   V39   $\!\!\!\!$&$\!\!\!\!$   99
$\!\!\!\!$&$\!\!\!\!$   76656    &   V106   $\!\!\!\!$&$\!\!\!\!$   99
$\!\!\!\!$&$\!\!\!\!$   202932   &   V186   $\!\!\!\!$&$\!\!\!\!$   94
$\!\!\!\!$&$\!\!\!\!$   200024   &   V259   $\!\!\!\!$&$\!\!\!\!$   99
$\!\!\!\!$&$\!\!\!\!$   177256   &   NV328  $\!\!\!\!$&$\!\!\!\!$   69
$\!\!\!\!$&$\!\!\!\!$   129813   &   NV388   $\!\!\!\!$&$\!\!\!\!$   0
$\!\!\!\!$&$\!\!\!\!$   205940   \\   V40   $\!\!\!\!$&$\!\!\!\!$   99
$\!\!\!\!$&$\!\!\!\!$   150793   &   V107  $\!\!\!\!$&$\!\!\!\!$   100
$\!\!\!\!$&$\!\!\!\!$   147114   &   V187   $\!\!\!\!$&$\!\!\!\!$   91
$\!\!\!\!$&$\!\!\!\!$   107554   &   V261   $\!\!\!\!$&$\!\!\!\!$   98
$\!\!\!\!$&$\!\!\!\!$   324469   &   NV329  $\!\!\!\!$&$\!\!\!\!$   51
$\!\!\!\!$&$\!\!\!\!$   235596   &   NV389  $\!\!\!\!$&$\!\!\!\!$   49
$\!\!\!\!$&$\!\!\!\!$   169916   \\   V41   $\!\!\!\!$&$\!\!\!\!$   72
$\!\!\!\!$&$\!\!\!\!$   145915   &   V108   $\!\!\!\!$&$\!\!\!\!$   99
$\!\!\!\!$&$\!\!\!\!$   178628   &   V188   $\!\!\!\!$&$\!\!\!\!$   95
$\!\!\!\!$&$\!\!\!\!$   160942   &   V263  $\!\!\!\!$&$\!\!\!\!$   100
$\!\!\!\!$&$\!\!\!\!$   243242   &   NV330  $\!\!\!\!$&$\!\!\!\!$   64
$\!\!\!\!$&$\!\!\!\!$   271511  &   NV390   $\!\!\!\!$&$\!\!\!\!$  100
$\!\!\!\!$&$\!\!\!\!$   262766   \\   V42   $\!\!\!\!$&$\!\!\!\!$   88
$\!\!\!\!$&$\!\!\!\!$   177086   &   V109   $\!\!\!\!$&$\!\!\!\!$   99
$\!\!\!\!$&$\!\!\!\!$   174888   &   V189   $\!\!\!\!$&$\!\!\!\!$   99
$\!\!\!\!$&$\!\!\!\!$   175684   &   V264   $\!\!\!\!$&$\!\!\!\!$   98
$\!\!\!\!$&$\!\!\!\!$   157202   &   NV331  $\!\!\!\!$&$\!\!\!\!$   69
$\!\!\!\!$&$\!\!\!\!$   244602   &   NV391  $\!\!\!\!$&$\!\!\!\!$   99
$\!\!\!\!$&$\!\!\!\!$   46483    \\   V43   $\!\!\!\!$&$\!\!\!\!$   99
$\!\!\!\!$&$\!\!\!\!$   227377   &   V110  $\!\!\!\!$&$\!\!\!\!$   100
$\!\!\!\!$&$\!\!\!\!$   164532   &   V190   $\!\!\!\!$&$\!\!\!\!$   99
$\!\!\!\!$&$\!\!\!\!$   243006   &   V265   $\!\!\!\!$&$\!\!\!\!$   98
$\!\!\!\!$&$\!\!\!\!$   192064   &   NV332  $\!\!\!\!$&$\!\!\!\!$   84
$\!\!\!\!$&$\!\!\!\!$   137605   &   NV393  $\!\!\!\!$&$\!\!\!\!$   99
$\!\!\!\!$&$\!\!\!\!$   36249   \\   V44   $\!\!\!\!$&$\!\!\!\!$   100
$\!\!\!\!$&$\!\!\!\!$   78476    &   V111   $\!\!\!\!$&$\!\!\!\!$   95
$\!\!\!\!$&$\!\!\!\!$   193829   &   V191   $\!\!\!\!$&$\!\!\!\!$   99
$\!\!\!\!$&$\!\!\!\!$   301765   &   V266   $\!\!\!\!$&$\!\!\!\!$   85
$\!\!\!\!$&$\!\!\!\!$   206486   &   NV333  $\!\!\!\!$&$\!\!\!\!$   62
$\!\!\!\!$&$\!\!\!\!$   192856   &   NV394  $\!\!\!\!$&$\!\!\!\!$   95
$\!\!\!\!$&$\!\!\!\!$   81813    \\   V45   $\!\!\!\!$&$\!\!\!\!$   98
$\!\!\!\!$&$\!\!\!\!$   219445   &   V112   $\!\!\!\!$&$\!\!\!\!$   78
$\!\!\!\!$&$\!\!\!\!$   158996   &   V192  $\!\!\!\!$&$\!\!\!\!$   100
$\!\!\!\!$&$\!\!\!\!$   319396   &   V267   $\!\!\!\!$&$\!\!\!\!$   99
$\!\!\!\!$&$\!\!\!\!$   234518   &   NV334  $\!\!\!\!$&$\!\!\!\!$   88
$\!\!\!\!$&$\!\!\!\!$   183248   &   NV395  $\!\!\!\!$&$\!\!\!\!$   88
$\!\!\!\!$&$\!\!\!\!$   177086   \\   V46   $\!\!\!\!$&$\!\!\!\!$   99
$\!\!\!\!$&$\!\!\!\!$   249013   &   V113   $\!\!\!\!$&$\!\!\!\!$   99
$\!\!\!\!$&$\!\!\!\!$   130486   &   V193   $\!\!\!\!$&$\!\!\!\!$   99
$\!\!\!\!$&$\!\!\!\!$   318939   &   V268   $\!\!\!\!$&$\!\!\!\!$   99
$\!\!\!\!$&$\!\!\!\!$   243048   &   NV335  $\!\!\!\!$&$\!\!\!\!$   99
$\!\!\!\!$&$\!\!\!\!$   362118   &   NV397  $\!\!\!\!$&$\!\!\!\!$   94
$\!\!\!\!$&$\!\!\!\!$   244021   \\   V47   $\!\!\!\!$&$\!\!\!\!$   99
$\!\!\!\!$&$\!\!\!\!$   281501   &   V114   $\!\!\!\!$&$\!\!\!\!$   97
$\!\!\!\!$&$\!\!\!\!$   159704   &   V194   $\!\!\!\!$&$\!\!\!\!$   99
$\!\!\!\!$&$\!\!\!\!$   128211   &   V270   $\!\!\!\!$&$\!\!\!\!$   98
$\!\!\!\!$&$\!\!\!\!$   164847   &   NV336  $\!\!\!\!$&$\!\!\!\!$   96
$\!\!\!\!$&$\!\!\!\!$   106751   &   NV398  $\!\!\!\!$&$\!\!\!\!$   96
$\!\!\!\!$&$\!\!\!\!$   267455   \\   V48   $\!\!\!\!$&$\!\!\!\!$   98
$\!\!\!\!$&$\!\!\!\!$   158259   &   V115   $\!\!\!\!$&$\!\!\!\!$   99
$\!\!\!\!$&$\!\!\!\!$   83123    &   V195   $\!\!\!\!$&$\!\!\!\!$   81
$\!\!\!\!$&$\!\!\!\!$   274973   &   V271   $\!\!\!\!$&$\!\!\!\!$   98
$\!\!\!\!$&$\!\!\!\!$   165504   &   NV337  $\!\!\!\!$&$\!\!\!\!$   97
$\!\!\!\!$&$\!\!\!\!$   118163   &   NV399  $\!\!\!\!$&$\!\!\!\!$   98
$\!\!\!\!$&$\!\!\!\!$   165850   \\   V49   $\!\!\!\!$&$\!\!\!\!$   99
$\!\!\!\!$&$\!\!\!\!$   36832    &   V116   $\!\!\!\!$&$\!\!\!\!$   56
$\!\!\!\!$&$\!\!\!\!$   204728   &   V197   $\!\!\!\!$&$\!\!\!\!$   79
$\!\!\!\!$&$\!\!\!\!$   126226   &   V272   $\!\!\!\!$&$\!\!\!\!$   99
$\!\!\!\!$&$\!\!\!\!$   248016   &   NV338  $\!\!\!\!$&$\!\!\!\!$   98
$\!\!\!\!$&$\!\!\!\!$   136414   &   NV400  $\!\!\!\!$&$\!\!\!\!$   99
$\!\!\!\!$&$\!\!\!\!$   129220   \\   V50   $\!\!\!\!$&$\!\!\!\!$   99
$\!\!\!\!$&$\!\!\!\!$   214772   &   V117   $\!\!\!\!$&$\!\!\!\!$   75
$\!\!\!\!$&$\!\!\!\!$   179957   &   V198   $\!\!\!\!$&$\!\!\!\!$   55
$\!\!\!\!$&$\!\!\!\!$   145299   &   V273   $\!\!\!\!$&$\!\!\!\!$   99
$\!\!\!\!$&$\!\!\!\!$   223751   &   NV339  $\!\!\!\!$&$\!\!\!\!$   99
$\!\!\!\!$&$\!\!\!\!$   169580   &   NV401  $\!\!\!\!$&$\!\!\!\!$   98
$\!\!\!\!$&$\!\!\!\!$   130105   \\   V51   $\!\!\!\!$&$\!\!\!\!$   99
$\!\!\!\!$&$\!\!\!\!$   278945   &   V118   $\!\!\!\!$&$\!\!\!\!$   96
$\!\!\!\!$&$\!\!\!\!$   160390   &   V199   $\!\!\!\!$&$\!\!\!\!$   87
$\!\!\!\!$&$\!\!\!\!$   194465   &   V274   $\!\!\!\!$&$\!\!\!\!$   77
$\!\!\!\!$&$\!\!\!\!$   305738   &   NV340  $\!\!\!\!$&$\!\!\!\!$   97
$\!\!\!\!$&$\!\!\!\!$   216161   &   NV402  $\!\!\!\!$&$\!\!\!\!$   87
$\!\!\!\!$&$\!\!\!\!$   52179    \\   V52   $\!\!\!\!$&$\!\!\!\!$   53
$\!\!\!\!$&$\!\!\!\!$   205670   &   V119   $\!\!\!\!$&$\!\!\!\!$   91
$\!\!\!\!$&$\!\!\!\!$   140617   &   V200   $\!\!\!\!$&$\!\!\!\!$   75
$\!\!\!\!$&$\!\!\!\!$   243370   &   V275   $\!\!\!\!$&$\!\!\!\!$   94
$\!\!\!\!$&$\!\!\!\!$   214740   &   NV341  $\!\!\!\!$&$\!\!\!\!$   94
$\!\!\!\!$&$\!\!\!\!$   191217   &   NV403   $\!\!\!\!$&$\!\!\!\!$   0
$\!\!\!\!$&$\!\!\!\!$   229293   \\   V53   $\!\!\!\!$&$\!\!\!\!$   99
$\!\!\!\!$&$\!\!\!\!$   56123    &   V120   $\!\!\!\!$&$\!\!\!\!$   99
$\!\!\!\!$&$\!\!\!\!$   110236   &   V201   $\!\!\!\!$&$\!\!\!\!$   99
$\!\!\!\!$&$\!\!\!\!$   369578   &   V276   $\!\!\!\!$&$\!\!\!\!$   99
$\!\!\!\!$&$\!\!\!\!$   101192   &   NV342  $\!\!\!\!$&$\!\!\!\!$   99
$\!\!\!\!$&$\!\!\!\!$   199357   &   NV404  $\!\!\!\!$&$\!\!\!\!$   99
$\!\!\!\!$&$\!\!\!\!$   244779   \\   V54   $\!\!\!\!$&$\!\!\!\!$   98
$\!\!\!\!$&$\!\!\!\!$   352756   &   V121   $\!\!\!\!$&$\!\!\!\!$   97
$\!\!\!\!$&$\!\!\!\!$   129294   &   V203  $\!\!\!\!$&$\!\!\!\!$   100
$\!\!\!\!$&$\!\!\!\!$   372023   &   V277   $\!\!\!\!$&$\!\!\!\!$   99
$\!\!\!\!$&$\!\!\!\!$   217316   &   NV343  $\!\!\!\!$&$\!\!\!\!$   99
$\!\!\!\!$&$\!\!\!\!$   174660   &   NV405  $\!\!\!\!$&$\!\!\!\!$   88
$\!\!\!\!$&$\!\!\!\!$   252122   \\   V56   $\!\!\!\!$&$\!\!\!\!$   98
$\!\!\!\!$&$\!\!\!\!$   38620    &   V122   $\!\!\!\!$&$\!\!\!\!$   99
$\!\!\!\!$&$\!\!\!\!$   105971   &   V204   $\!\!\!\!$&$\!\!\!\!$   99
$\!\!\!\!$&$\!\!\!\!$   45537    &   V280   $\!\!\!\!$&$\!\!\!\!$   99
$\!\!\!\!$&$\!\!\!\!$   300999   &   NV344  $\!\!\!\!$&$\!\!\!\!$   99
$\!\!\!\!$&$\!\!\!\!$   271739  &   NV406   $\!\!\!\!$&$\!\!\!\!$  100
$\!\!\!\!$&$\!\!\!\!$   292866   \\   V57  $\!\!\!\!$&$\!\!\!\!$   100
$\!\!\!\!$&$\!\!\!\!$   48194   &   V123   $\!\!\!\!$&$\!\!\!\!$   100
$\!\!\!\!$&$\!\!\!\!$   44090    &   V205   $\!\!\!\!$&$\!\!\!\!$   96
$\!\!\!\!$&$\!\!\!\!$   302520   &   V285   $\!\!\!\!$&$\!\!\!\!$   99
$\!\!\!\!$&$\!\!\!\!$   74726   &   NV345   $\!\!\!\!$&$\!\!\!\!$   87
$\!\!\!\!$&$\!\!\!\!$   174082   &   NV407  $\!\!\!\!$&$\!\!\!\!$   88
$\!\!\!\!$&$\!\!\!\!$   309849   \\   V58   $\!\!\!\!$&$\!\!\!\!$   96
$\!\!\!\!$&$\!\!\!\!$   284348   &   V124   $\!\!\!\!$&$\!\!\!\!$   99
$\!\!\!\!$&$\!\!\!\!$   26731    &   V206   $\!\!\!\!$&$\!\!\!\!$   97
$\!\!\!\!$&$\!\!\!\!$   57769    &   V288   $\!\!\!\!$&$\!\!\!\!$   99
$\!\!\!\!$&$\!\!\!\!$   289087   &   NV346  $\!\!\!\!$&$\!\!\!\!$   90
$\!\!\!\!$&$\!\!\!\!$   202477   &   NV408  $\!\!\!\!$&$\!\!\!\!$   50
$\!\!\!\!$&$\!\!\!\!$   234623   \\   V59   $\!\!\!\!$&$\!\!\!\!$   96
$\!\!\!\!$&$\!\!\!\!$   171211   &   V125   $\!\!\!\!$&$\!\!\!\!$   99
$\!\!\!\!$&$\!\!\!\!$   14416    &   V207   $\!\!\!\!$&$\!\!\!\!$   98
$\!\!\!\!$&$\!\!\!\!$   297708   &   V289   $\!\!\!\!$&$\!\!\!\!$   99
$\!\!\!\!$&$\!\!\!\!$   324714   &   NV347  $\!\!\!\!$&$\!\!\!\!$   79
$\!\!\!\!$&$\!\!\!\!$   211972   &   NV409  $\!\!\!\!$&$\!\!\!\!$   73
$\!\!\!\!$&$\!\!\!\!$   322122   \\   V60   $\!\!\!\!$&$\!\!\!\!$   99
$\!\!\!\!$&$\!\!\!\!$   110456   &   V126   $\!\!\!\!$&$\!\!\!\!$   99
$\!\!\!\!$&$\!\!\!\!$   15927    &   V208   $\!\!\!\!$&$\!\!\!\!$   97
$\!\!\!\!$&$\!\!\!\!$   244408   &   V291   $\!\!\!\!$&$\!\!\!\!$   98
$\!\!\!\!$&$\!\!\!\!$   97939   &   NV349   $\!\!\!\!$&$\!\!\!\!$   88
$\!\!\!\!$&$\!\!\!\!$   212595   &   NV410   $\!\!\!\!$&$\!\!\!\!$   0
$\!\!\!\!$&$\!\!\!\!$   180157   \\   V61  $\!\!\!\!$&$\!\!\!\!$   100
$\!\!\!\!$&$\!\!\!\!$   216835   &   V127   $\!\!\!\!$&$\!\!\!\!$   99
$\!\!\!\!$&$\!\!\!\!$   193936   &   V209   $\!\!\!\!$&$\!\!\!\!$   99
$\!\!\!\!$&$\!\!\!\!$   297731   &   V292   $\!\!\!\!$&$\!\!\!\!$   98
$\!\!\!\!$&$\!\!\!\!$   129991   &   NV350  $\!\!\!\!$&$\!\!\!\!$   84
$\!\!\!\!$&$\!\!\!\!$  149574 &  & &  \\ V62  $\!\!\!\!$&$\!\!\!\!$ 99
$\!\!\!\!$&$\!\!\!\!$   208493   &   V128  $\!\!\!\!$&$\!\!\!\!$   100
$\!\!\!\!$&$\!\!\!\!$   162142   &   V210   $\!\!\!\!$&$\!\!\!\!$   15
$\!\!\!\!$&$\!\!\!\!$   58394    &   V293   $\!\!\!\!$&$\!\!\!\!$   99
$\!\!\!\!$&$\!\!\!\!$   85152   &   NV351   $\!\!\!\!$&$\!\!\!\!$   84
$\!\!\!\!$&$\!\!\!\!$  215235 &  & &  \\ V64  $\!\!\!\!$&$\!\!\!\!$ 99
$\!\!\!\!$&$\!\!\!\!$   54131    &   V129   $\!\!\!\!$&$\!\!\!\!$   19
$\!\!\!\!$&$\!\!\!\!$   185687   &   V211   $\!\!\!\!$&$\!\!\!\!$   90
$\!\!\!\!$&$\!\!\!\!$   298164   &   NV294  $\!\!\!\!$&$\!\!\!\!$   97
$\!\!\!\!$&$\!\!\!\!$   265591   &   NV352  $\!\!\!\!$&$\!\!\!\!$   99
$\!\!\!\!$&$\!\!\!\!$  183506 &  &  & \\  V65 $\!\!\!\!$&$\!\!\!\!$  0
$\!\!\!\!$&$\!\!\!\!$   50869   &   V131   $\!\!\!\!$&$\!\!\!\!$   100
$\!\!\!\!$&$\!\!\!\!$   173358   &   V212   $\!\!\!\!$&$\!\!\!\!$   99
$\!\!\!\!$&$\!\!\!\!$   293439   &   NV295  $\!\!\!\!$&$\!\!\!\!$   99
$\!\!\!\!$&$\!\!\!\!$   340398   &   NV353  $\!\!\!\!$&$\!\!\!\!$   72
$\!\!\!\!$&$\!\!\!\!$          208251          &          &          &
\\ \multicolumn{18}{c}{\phantom{z}}\\ \hline
\end{tabular}}
\label{tab_kalu}
\end{table*}

\begin{table}[b!]
\caption{Membership probability of XMM-Newton X-Ray counterparts
  candidates. The IDs refer to Kaluzny et al. work.}
\centering
\begin{tabular}{lcclc}
\hline                   
\hline
ID$_{K}$&$P_{\mu}$&\phantom{hhh}&ID$_{K}$&$P_{\mu}$\\
\hline
&&&\\ NV367 & 0 & & NV376 & 0 \\  NV375 & 0 & & NV377 & 0 \\ V167 & 23
& &NV369 & 94 \\ V223 & 0 & & NV383 & 99 \\ NV378 & 31 & &&\\
\end{tabular}
\label{tab_xmm}
\end{table}

Using  the 1.0$m$  Telescope  of the  Australian National  University,
Weldrake et al.  (\cite{weldrake07})  detected a total of 187 variable
stars covering a wide area around $\omega$~Cen.
These  stars were matched with the  Kaluzny et al.  (\cite{kaluzny04})
catalog, and 81 variable stars were found to be new discoveries.
A  cross-check  of our  catalog   with that of  Weldrake et  al.
(\cite{weldrake07})  provided  102  variable stars  in  common.  
Their location in the $V$ versus $V-I$ CMD is shown in the right panel
of Fig.~\ref{fig_variable}.
As  completed  for  the  Kaluzny et  al.   (\cite{kaluzny04})  sample,
different  symbols mark different  membership probability  ranges, and
Table~\ref{tab_weldra}  reports  our  membership probability  for  the
Weldrake et al. (\cite{weldrake07}) variable catalog.
Of the 81 Weldrake et  al.  (\cite{weldrake07}) new variable stars, 16
have counterparts in our proper-motion catalog.
Four of these new variables are clearly not cluster members.
These  field objects include   a detached eclipsing binary  (V59), and
three long-period variables (V31, V125, and V126).

\begin{table*}[ht!]
\centering
\caption{Membership probability for the Weldrake et
  al.  (\cite{weldrake07}) variable star  catalog. ID$_W$ are Weldrake
  et  al. identification labels, ID$_{tw}$  come  from this~work.  The
  symbol (*) marks new Weldrake et al. identifications.}
\tiny{
\begin{tabular}{rclrclrclrclrclrcl}
\hline                                                           \hline
\multicolumn{18}{c}{\phantom{z}}\\
\multicolumn{1}{l}{ID$_W$}&\multicolumn{1}{c}{$P_{\mu}$}&\multicolumn{1}{l}{ID$_{tw}$\phantom{zzz}}&
\multicolumn{1}{l}{ID$_W$}&\multicolumn{1}{c}{$P_{\mu}$}&\multicolumn{1}{l}{ID$_{tw}$\phantom{zzz}}&
\multicolumn{1}{l}{ID$_W$}&\multicolumn{1}{c}{$P_{\mu}$}&\multicolumn{1}{l}{ID$_{tw}$\phantom{zzz}}&
\multicolumn{1}{l}{ID$_W$}&\multicolumn{1}{c}{$P_{\mu}$}&\multicolumn{1}{l}{ID$_{tw}$\phantom{zzz}}&
\multicolumn{1}{l}{ID$_W$}&\multicolumn{1}{c}{$P_{\mu}$}&\multicolumn{1}{l}{ID$_{tw}$\phantom{zzz}}&
\multicolumn{1}{l}{ID$_W$}&\multicolumn{1}{c}{$P_{\mu}$}&\multicolumn{1}{l}{ID$_{tw}$}\\
\multicolumn{18}{c}{\phantom{z}}\\                               \hline
\multicolumn{18}{c}{\phantom{z}}\\      V31*     $\!\!\!\!$&$\!\!\!\!$
0$\!\!\!\!$&$\!\!\!\!$234574     &     V48     $\!\!\!\!$&$\!\!\!\!$
100$\!\!\!\!$&$\!\!\!\!$360819    &    V102    $\!\!\!\!$&$\!\!\!\!$
15$\!\!\!\!$&$\!\!\!\!$58394     &     V119    $\!\!\!\!$&$\!\!\!\!$
51$\!\!\!\!$&$\!\!\!\!$126839     &    V142    $\!\!\!\!$&$\!\!\!\!$
42$\!\!\!\!$&$\!\!\!\!$208115    &    V160*    $\!\!\!\!$&$\!\!\!\!$
73$\!\!\!\!$&$\!\!\!\!$292364     \\    V32    $\!\!\!\!$&$\!\!\!\!$
99$\!\!\!\!$&$\!\!\!\!$236315     &    V59*    $\!\!\!\!$&$\!\!\!\!$
0$\!\!\!\!$&$\!\!\!\!$23863     &     V103     $\!\!\!\!$&$\!\!\!\!$
99$\!\!\!\!$&$\!\!\!\!$54131     &     V120    $\!\!\!\!$&$\!\!\!\!$
68$\!\!\!\!$&$\!\!\!\!$154689     &    V143    $\!\!\!\!$&$\!\!\!\!$
98$\!\!\!\!$&$\!\!\!\!$201164     &    V161    $\!\!\!\!$&$\!\!\!\!$
73$\!\!\!\!$&$\!\!\!\!$322122     \\    V33    $\!\!\!\!$&$\!\!\!\!$
100$\!\!\!\!$&$\!\!\!\!$205260     &    V60    $\!\!\!\!$&$\!\!\!\!$
48$\!\!\!\!$&$\!\!\!\!$40133     &     V104    $\!\!\!\!$&$\!\!\!\!$
93$\!\!\!\!$&$\!\!\!\!$46225     &     V121    $\!\!\!\!$&$\!\!\!\!$
53$\!\!\!\!$&$\!\!\!\!$159019    &    V144*    $\!\!\!\!$&$\!\!\!\!$
100$\!\!\!\!$&$\!\!\!\!$247143    &    V162    $\!\!\!\!$&$\!\!\!\!$
99$\!\!\!\!$&$\!\!\!\!$311494     \\    V34    $\!\!\!\!$&$\!\!\!\!$
99$\!\!\!\!$&$\!\!\!\!$239985     &     V61    $\!\!\!\!$&$\!\!\!\!$
100$\!\!\!\!$&$\!\!\!\!$48194     &    V105    $\!\!\!\!$&$\!\!\!\!$
99$\!\!\!\!$&$\!\!\!\!$83123     &     V122    $\!\!\!\!$&$\!\!\!\!$
100$\!\!\!\!$&$\!\!\!\!$162685    &    V145*   $\!\!\!\!$&$\!\!\!\!$
99$\!\!\!\!$&$\!\!\!\!$270697     &    V163    $\!\!\!\!$&$\!\!\!\!$
93$\!\!\!\!$&$\!\!\!\!$321474     \\    V35    $\!\!\!\!$&$\!\!\!\!$
56$\!\!\!\!$&$\!\!\!\!$214167     &     V62    $\!\!\!\!$&$\!\!\!\!$
97$\!\!\!\!$&$\!\!\!\!$42989     &     V106    $\!\!\!\!$&$\!\!\!\!$
100$\!\!\!\!$&$\!\!\!\!$78476     &    V123    $\!\!\!\!$&$\!\!\!\!$
99$\!\!\!\!$&$\!\!\!\!$155556     &    V146    $\!\!\!\!$&$\!\!\!\!$
100$\!\!\!\!$&$\!\!\!\!$243242    &    V164    $\!\!\!\!$&$\!\!\!\!$
99$\!\!\!\!$&$\!\!\!\!$327526     \\    V36    $\!\!\!\!$&$\!\!\!\!$
98$\!\!\!\!$&$\!\!\!\!$267824     &     V63    $\!\!\!\!$&$\!\!\!\!$
99$\!\!\!\!$&$\!\!\!\!$86072     &     V107    $\!\!\!\!$&$\!\!\!\!$
99$\!\!\!\!$&$\!\!\!\!$103189     &    V124    $\!\!\!\!$&$\!\!\!\!$
75$\!\!\!\!$&$\!\!\!\!$137352     &    V147    $\!\!\!\!$&$\!\!\!\!$
99$\!\!\!\!$&$\!\!\!\!$246994     &    V165    $\!\!\!\!$&$\!\!\!\!$
99$\!\!\!\!$&$\!\!\!\!$320506     \\    V37    $\!\!\!\!$&$\!\!\!\!$
0$\!\!\!\!$&$\!\!\!\!$243506     &     V64     $\!\!\!\!$&$\!\!\!\!$
99$\!\!\!\!$&$\!\!\!\!$84169     &     V108    $\!\!\!\!$&$\!\!\!\!$
48$\!\!\!\!$&$\!\!\!\!$92977     &    V125*    $\!\!\!\!$&$\!\!\!\!$
0$\!\!\!\!$&$\!\!\!\!$127167     &     V148    $\!\!\!\!$&$\!\!\!\!$
77$\!\!\!\!$&$\!\!\!\!$305738     &    V166    $\!\!\!\!$&$\!\!\!\!$
31$\!\!\!\!$&$\!\!\!\!$341736     \\    V38    $\!\!\!\!$&$\!\!\!\!$
99$\!\!\!\!$&$\!\!\!\!$270509     &     V65    $\!\!\!\!$&$\!\!\!\!$
99$\!\!\!\!$&$\!\!\!\!$128211     &    V109    $\!\!\!\!$&$\!\!\!\!$
99$\!\!\!\!$&$\!\!\!\!$110456    &    V126*    $\!\!\!\!$&$\!\!\!\!$
0$\!\!\!\!$&$\!\!\!\!$85789     &     V150     $\!\!\!\!$&$\!\!\!\!$
50$\!\!\!\!$&$\!\!\!\!$302398     &    V167    $\!\!\!\!$&$\!\!\!\!$
100$\!\!\!\!$&$\!\!\!\!$345639    \\    V39    $\!\!\!\!$&$\!\!\!\!$
100$\!\!\!\!$&$\!\!\!\!$292866     &    V66    $\!\!\!\!$&$\!\!\!\!$
98$\!\!\!\!$&$\!\!\!\!$111797     &    V110    $\!\!\!\!$&$\!\!\!\!$
99$\!\!\!\!$&$\!\!\!\!$105971     &    V133    $\!\!\!\!$&$\!\!\!\!$
94$\!\!\!\!$&$\!\!\!\!$201061     &    V151    $\!\!\!\!$&$\!\!\!\!$
2$\!\!\!\!$&$\!\!\!\!$283256     &     V168    $\!\!\!\!$&$\!\!\!\!$
64$\!\!\!\!$&$\!\!\!\!$347475     \\    V40    $\!\!\!\!$&$\!\!\!\!$
99$\!\!\!\!$&$\!\!\!\!$289620     &     V67    $\!\!\!\!$&$\!\!\!\!$
98$\!\!\!\!$&$\!\!\!\!$93459     &     V111    $\!\!\!\!$&$\!\!\!\!$
99$\!\!\!\!$&$\!\!\!\!$110236     &    V134    $\!\!\!\!$&$\!\!\!\!$
53$\!\!\!\!$&$\!\!\!\!$234700     &    V152    $\!\!\!\!$&$\!\!\!\!$
100$\!\!\!\!$&$\!\!\!\!$298983    &    V169    $\!\!\!\!$&$\!\!\!\!$
100$\!\!\!\!$&$\!\!\!\!$337663    \\    V41    $\!\!\!\!$&$\!\!\!\!$
99$\!\!\!\!$&$\!\!\!\!$293439     &    V68*    $\!\!\!\!$&$\!\!\!\!$
99$\!\!\!\!$&$\!\!\!\!$120134     &    V112    $\!\!\!\!$&$\!\!\!\!$
86$\!\!\!\!$&$\!\!\!\!$153744     &    V135    $\!\!\!\!$&$\!\!\!\!$
85$\!\!\!\!$&$\!\!\!\!$205826     &    V153    $\!\!\!\!$&$\!\!\!\!$
65$\!\!\!\!$&$\!\!\!\!$297730     &    V170    $\!\!\!\!$&$\!\!\!\!$
99$\!\!\!\!$&$\!\!\!\!$352882     \\    V42    $\!\!\!\!$&$\!\!\!\!$
99$\!\!\!\!$&$\!\!\!\!$285765     &    V69*    $\!\!\!\!$&$\!\!\!\!$
98$\!\!\!\!$&$\!\!\!\!$108406     &    V113    $\!\!\!\!$&$\!\!\!\!$
98$\!\!\!\!$&$\!\!\!\!$136502     &    V136    $\!\!\!\!$&$\!\!\!\!$
50$\!\!\!\!$&$\!\!\!\!$228136     &    V154    $\!\!\!\!$&$\!\!\!\!$
98$\!\!\!\!$&$\!\!\!\!$287148     &    V171    $\!\!\!\!$&$\!\!\!\!$
98$\!\!\!\!$&$\!\!\!\!$352756     \\    V43    $\!\!\!\!$&$\!\!\!\!$
99$\!\!\!\!$&$\!\!\!\!$324714     &     V70    $\!\!\!\!$&$\!\!\!\!$
100$\!\!\!\!$&$\!\!\!\!$162564    &    V114    $\!\!\!\!$&$\!\!\!\!$
55$\!\!\!\!$&$\!\!\!\!$160877     &    V137    $\!\!\!\!$&$\!\!\!\!$
42$\!\!\!\!$&$\!\!\!\!$225221     &    V155    $\!\!\!\!$&$\!\!\!\!$
64$\!\!\!\!$&$\!\!\!\!$283990     &    V172    $\!\!\!\!$&$\!\!\!\!$
96$\!\!\!\!$&$\!\!\!\!$358729    \\    V44*    $\!\!\!\!$&$\!\!\!\!$
93$\!\!\!\!$&$\!\!\!\!$328996     &    V71*    $\!\!\!\!$&$\!\!\!\!$
88$\!\!\!\!$&$\!\!\!\!$132518     &    V115    $\!\!\!\!$&$\!\!\!\!$
57$\!\!\!\!$&$\!\!\!\!$158023     &    V138    $\!\!\!\!$&$\!\!\!\!$
99$\!\!\!\!$&$\!\!\!\!$215061     &    V156    $\!\!\!\!$&$\!\!\!\!$
97$\!\!\!\!$&$\!\!\!\!$279540     &    V173    $\!\!\!\!$&$\!\!\!\!$
98$\!\!\!\!$&$\!\!\!\!$219445     \\    V45    $\!\!\!\!$&$\!\!\!\!$
99$\!\!\!\!$&$\!\!\!\!$316247     &    V72*    $\!\!\!\!$&$\!\!\!\!$
73$\!\!\!\!$&$\!\!\!\!$133191     &    V116    $\!\!\!\!$&$\!\!\!\!$
60$\!\!\!\!$&$\!\!\!\!$141176     &    V139    $\!\!\!\!$&$\!\!\!\!$
56$\!\!\!\!$&$\!\!\!\!$204728    &    V157*    $\!\!\!\!$&$\!\!\!\!$
96$\!\!\!\!$&$\!\!\!\!$273996     &    V174    $\!\!\!\!$&$\!\!\!\!$
99$\!\!\!\!$&$\!\!\!\!$249013    \\    V46*    $\!\!\!\!$&$\!\!\!\!$
96$\!\!\!\!$&$\!\!\!\!$320258     &    V81*    $\!\!\!\!$&$\!\!\!\!$
96$\!\!\!\!$&$\!\!\!\!$    6069    &   V117    $\!\!\!\!$&$\!\!\!\!$
97$\!\!\!\!$&$\!\!\!\!$129294     &    V140    $\!\!\!\!$&$\!\!\!\!$
57$\!\!\!\!$&$\!\!\!\!$224813     &    V158    $\!\!\!\!$&$\!\!\!\!$
98$\!\!\!\!$&$\!\!\!\!$273997     &    V176    $\!\!\!\!$&$\!\!\!\!$
98$\!\!\!\!$&$\!\!\!\!$292312     \\    V47    $\!\!\!\!$&$\!\!\!\!$
99$\!\!\!\!$&$\!\!\!\!$352935     &    V101    $\!\!\!\!$&$\!\!\!\!$
99$\!\!\!\!$&$\!\!\!\!$36832     &     V118    $\!\!\!\!$&$\!\!\!\!$
35$\!\!\!\!$&$\!\!\!\!$151268     &    V141    $\!\!\!\!$&$\!\!\!\!$
51$\!\!\!\!$&$\!\!\!\!$218172    &    V159*    $\!\!\!\!$&$\!\!\!\!$
93$\!\!\!\!$&$\!\!\!\!$287079     &    V178    $\!\!\!\!$&$\!\!\!\!$
36$\!\!\!\!$&$\!\!\!\!$309751 \\
 \multicolumn{18}{c}{\phantom{z}}\\ 
\hline
\end{tabular}}
\label{tab_weldra}
\end{table*}

\section{Summary}
\label{sec_final}

We have applied the photometric and astrometric technique developed by
Anderson et al.  (2006, Paper~I) to the most puzzling globular cluster
of the Milky Way: $\omega$~Centauri.

\begin{figure*}[ht!]
\centering
\includegraphics[width=18.0cm]{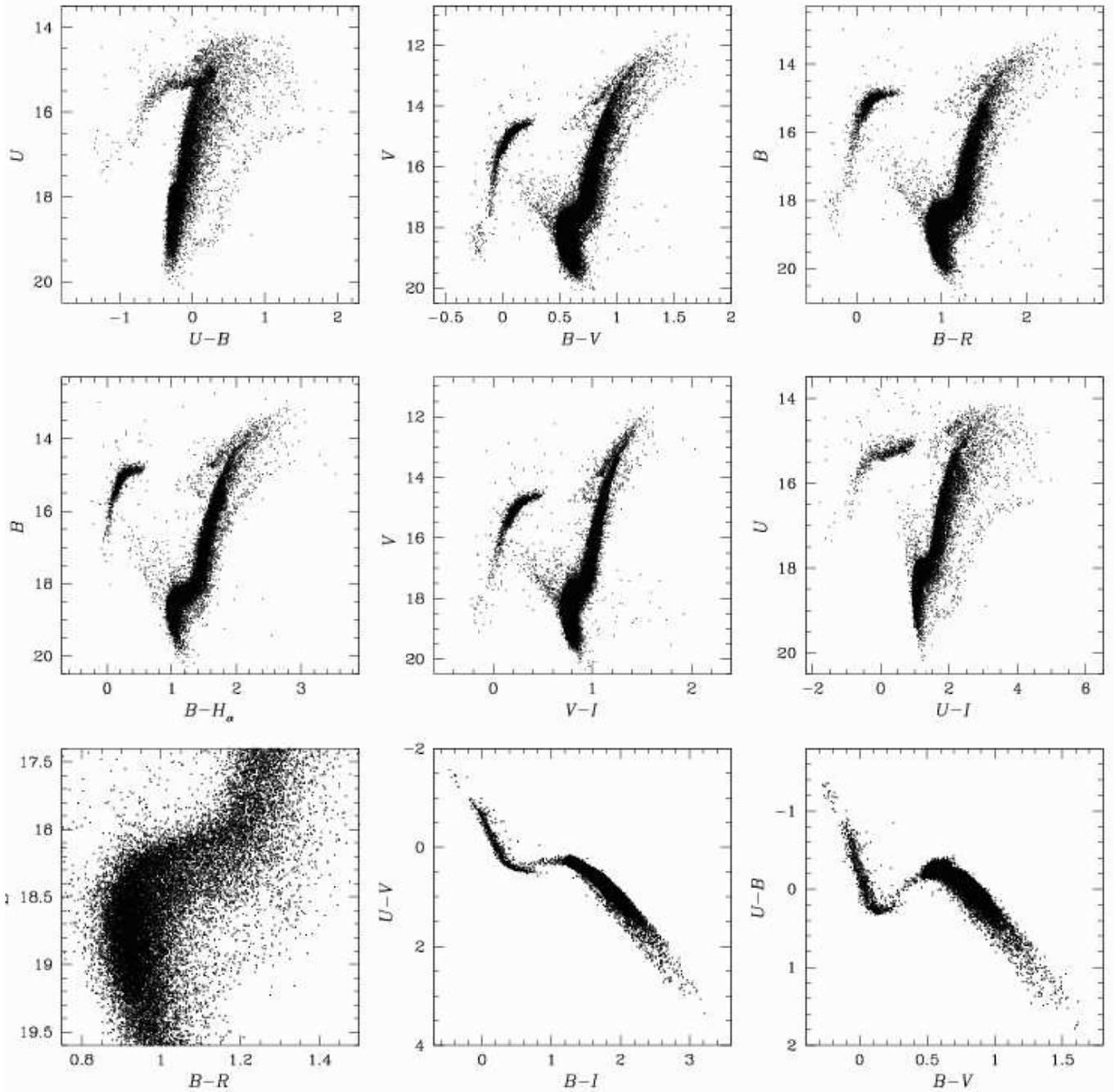}
\caption{(\textit{Top)}: membership probability selected
 ($P_{\mu}>90\%$) \om  CMDs. From left to right  $U$ vs.  ($U-B$), $V$
 vs. ($B-V$),  and  $B$ vs.  ($B-R$).  Plotted stars  have photometric
 errors going from 0.02 mag for the  brighter to 0.05 for the faintest
 ones. (\textit{Middle}):  same star selection criteria. From left to
 right: $B$ vs. ($B-H_{\alpha}$), $V$ vs. ($V-I$), and $U$ vs. ($U-I$).
(\textit{Bottom}): the $B$  vs. ($B-R$) CMD, zoomed in  the SGB region,
 is shown  on   the left.  The    two  plots on   the  right  show two
 color-color diagrams obtained from different filters.}
\label{fig:cmd_all}
\end{figure*}

%
Based on CCD observations taken with only four years of temporal
base-line, our measurements provide accurate proper motions to
$B\sim20$, four magnitudes deeper than the photographic catalog of
vL00.
We have minimized the sky-concentration effects in our photometry.  We
provide a membership probability  for all stars.  Our catalog contains
almost $360\,000$ stars with measured proper motion, and covers a wide
area ($\sim33\times33$ arcmin$^2$) around the cluster center.
In Fig.~\ref{fig:cmd_all}, we   show   a summary of    our  photometric
catalog:  we  plot several \om  CMDs,  derived with all  the available
filters  and   different   color-baselines (top and  middle  rows).
Plotted          stars  have         a    membership       probability
of $P_{\mu}>90\%$. 
Photometric errors range from 0.02  mag for brighter stars to 0.05 mag
for the fainter ones.  In the bottom panels of Fig.~\ref{fig:cmd_all},
we show the SGB region of \om in the $B$ versus $B-R$ CMD on the left,
and two color-color diagrams with different colors on the right.

The high precision of our astrometry and multi-band photometry once
again emphasizes the importance of accurate representation of the PSF
across the entire field-of view, especially for wide-field imagers,
exemplified by the concept of empirical PSF (Paper~I).

The primary   aim  of this  work  is,  of  course, to  provide wide-field
membership  probability measurements   for  spectroscopic    follow-up
studies, down to the turn-off region of the cluster. However, the high
quality of our photometric and astrometric measurements also provide a
crucial observational constraint of the multiple \om sub-populations.
Due  to our  proper-motion-selected  RGB  sub-populations, we  can
confirm  that the metal-poor,  metal-intermediate,  and metal-rich
components  have the same  proper motions which is that of
$\omega$~Cen,  within our measurements uncertainties.

We finally  provide   membership  probability determinations   for  the
Kaluzny  et     al.    (\cite{kaluzny04})    and Weldrake     et   al.
(\cite{weldrake07}) \om variable star catalogs.

\subsection{Electronic catalog}
\label{subsec_catalog}

The     catalog    is    available     at    the     SIMBAD    on-line
database\footnote{\texttt{http://simbad.u-strasbg.fr/simbad/}}.
Description  of the catalog: column  (1)  contains the ID; columns (2)
and (3)  give the J2003.29  equatorial  coordinates in  decimal  degrees;
columns (4) and (5)  provide the pixel coordinates $x$  and $y$ of the
distortion-corrected reference meta-chip.  
Columns   (6)  and   (7)   contain  the   proper-motion values along
$\mu_{\alpha} \cos \delta$ in units of  \masyr, with the corresponding
r.m.s.; columns (8)  and  (9) provide the  proper-motion values along
$\mu_{\delta}$, with the corresponding r.m.s., in  the same units.  
Columns from (10)  to (21) give the  photometric data, i.e.  $U, B, V,
R_C, I_C, H_{\alpha}$ magnitudes and their errors.
The  last two columns  give the  proper-motion  membership probability
$P_\mu(1)$ for  all the  stars (22), and  for a sub-sample  of 120,259
stars  with   the  alternative  membership   determination  $P_\mu(2)$
(23).  In this  case, if  the second  membership determination  is not
provided, we flag it with the value -1.

%
If photometry in a  specific band is  not  available, a flag  equal to
99.999 is set for magnitude and error.

Table~6 contains the first lines of the on-line catalog.


\begin{acknowledgements}
A.Bellini  acknowledges the support  by the  CA.RI.PA.RO.  foundation,
and  the  STScI  under the  2008  graduate research  assistantship
program.
I. Platais  gratefully acknowledges support from  the National Science
Foundation through grant AST  04-06689 to Johns Hopkins University and
and  by NASA  grant \textit{HST}-AR-09958.01-A,  awarded by  the Space
Telescope  Institute,   which  is  operated  by   the  Association  of
Universities  for Research  in  Astronomy, Inc.,  under NASA  contract
NAS5-26555.
We  thank Floor van Leeuwen for  his help in   providing us a complete
version of his catalog as well as stimulating discussions.
We  thank  the anonymous  referee  for  the  careful reading  of  the
manuscript, and for the useful comments.
\end{acknowledgements}


{}


\clearpage

\begin{sidewaystable}

\label{tab_cat}
\tiny{
\begin{tabular}{ccccccccccccccccccccccc}
&&&&&&&&&&&&&&&&&&&&&&\\
\hline
\hline
$\!\!\!$&$\!\!\!$$\!\!\!$&$\!\!\!$$\!\!\!$&$\!\!\!$$\!\!\!$&$\!\!\!$$\!\!\!$&
$\!\!\!$$\!\!\!$&$\!\!\!$$\!\!\!$&$\!\!\!$$\!\!\!$&$\!\!\!$$\!\!\!$&$\!\!\!$$
\!\!\!$&$\!\!\!$$\!\!\!$&$\!\!\!$&\\
ID$\!\!\!$&$\!\!\!$$\alpha$$\!\!\!$&$\!\!\!$$\delta$$\!\!\!$&$\!\!\!$$x$$\!\!\!$&
$\!\!\!$$y$$\!\!\!$&$\!\!\!$$\mu_{\alpha}\cos
\delta$$\!\!\!$&$\!\!\!$$\sigma_{\mu_{\alpha}\cos
\delta}$$\!\!\!$&$\!\!\!$
$\mu_\delta$$\!\!\!$&$\!\!\!$$\sigma_{\mu_\delta}$$\!\!\!$&$\!\!\!$$U$$\!\!\!$&$\!\!\!$
$\sigma_U$$\!\!\!$&$\!\!\!$$B$$\!\!\!$&$\!\!\!$$\sigma_B$$\!\!\!$&$\!\!\!$$V$$\!\!\!$&
$\!\!\!$$\sigma_V$$\!\!\!$&$\!\!\!$$R$$\!\!\!$&$\!\!\!$$\sigma_R$$\!\!\!$&$\!\!\!$$I$
$\!\!\!$&$\!\!\!$$\sigma_I$$\!\!\!$&$\!\!\!$
$H_{\alpha}$$\!\!\!$&$\!\!\!$$\sigma_{H_{\alpha}}$$\!\!\!$&$\!\!\!\!$$P_\mu(1)$&$\!\!\!\!$$P_\mu(2)$\\
$\!\!\!$&$\!\!\!$$\!\!\!$&$\!\!\!$$\!\!\!$&$\!\!\!$$\!\!\!$&$\!\!\!$$\!\!\!$&$\!\!\!$
$\!\!\!$&$\!\!\!$$\!\!\!$&$\!\!\!$$\!\!\!$&$\!\!\!$$\!\!\!$&$\!\!\!$$\!\!\!$&$\!\!\!$$\!\!\!$&$\!\!\!$\\
(1)$\!\!\!$&$\!\!\!$(2)$\!\!\!$&$\!\!\!$(3)$\!\!\!$&$\!\!\!$(4)$\!\!\!$&$\!\!\!$(5)$\!\!\!$&$\!\!\!$(6)
$\!\!\!$&$\!\!\!$(7)$\!\!\!$&$\!\!\!$(8)$\!\!\!$&$\!\!\!$(9)$\!\!\!$&$\!\!\!$(10)$\!\!\!$&$\!\!\!$(11)
$\!\!\!$&$\!\!\!$(12)$\!\!\!$&$\!\!\!$(13)$\!\!\!$&$\!\!\!$(14)$\!\!\!$&$\!\!\!$(15)$\!\!\!$&$\!\!\!$(16)
$\!\!\!$&$\!\!\!$(17)$\!\!\!$&$\!\!\!$(18)$\!\!\!$&$\!\!\!$(19)$\!\!\!$&$\!\!\!$(20)$\!\!\!$&$\!\!\!$
(21)$\!\!\!$&$\!\!\!$(22)&$\!\!\!$(23)\\
$\!\!\!$&$\!\!\!$$\!\!\!$&$\!\!\!$$\!\!\!$&$\!\!\!$$\!\!\!$&$\!\!\!$$\!\!\!$&$\!\!\!$
$\!\!\!$&$\!\!\!$$\!\!\!$&$\!\!\!$$\!\!\!$&$\!\!\!$$\!\!\!$&$\!\!\!$$\!\!\!$&$\!\!\!$$\!\!\!$&$\!\!\!$&\\
$[$                                                                 \#
$]$$\!\!\!$&$\!\!\!$$[$$^\circ$$]$$\!\!\!$&$\!\!\!$$[$$^\circ$$]$$\!\!\!$&$\!\!\!$$[$pixel$]$
$\!\!\!$&$\!\!\!$$[$pixel$]$$\!\!\!$&$\!\!\!$
$[$mas/yr$]$$\!\!\!$&$\!\!\!$$[$mas/yr$]$$\!\!\!$&$\!\!\!$$[$mas/yr$]$$\!\!\!$&$\!\!\!$
$[$mas/yr$]$$\!\!\!$&$\!\!\!$
$[$mag$]$$\!\!\!$&$\!\!\!$$[$mag$]$$\!\!\!$&$\!\!\!$$[$mag$]$$\!\!\!$&$\!\!\!$$[$mag$]$
$\!\!\!$&$\!\!\!$$[$mag$]$$\!\!\!$&$\!\!\!$$[$mag$]$$\!\!\!$&$\!\!\!$$[$mag$]$$\!\!\!$&$\!\!\!$
$[$mag$]$$\!\!\!$&$\!\!\!$
$[$mag$]$$\!\!\!$&$\!\!\!$$[$mag$]$$\!\!\!$&$\!\!\!$$[$mag$]$$\!\!\!$&$\!\!\!$$[$mag$]$
$\!\!\!$&$\!\!\!$$[$\%$]$&$\!\!\!$$[$\%$]$\\
$\!\!\!$&$\!\!\!$$\!\!\!$&$\!\!\!$$\!\!\!$&$\!\!\!$$\!\!\!$&$\!\!\!$$\!\!\!$&$\!\!\!$$\!\!\!$&
$\!\!\!$$\!\!\!$&$\!\!\!$$\!\!\!$&$\!\!\!$$\!\!\!$&$\!\!\!$$\!\!\!$&$\!\!\!$$\!\!\!$&$\!\!\!$&\\
\hline
$\!\!\!$&$\!\!\!$$\!\!\!$&$\!\!\!$$\!\!\!$&$\!\!\!$$\!\!\!$&$\!\!\!$$\!\!\!$&$\!\!\!$$\!\!\!$&
$\!\!\!$$\!\!\!$&$\!\!\!$$\!\!\!$&$\!\!\!$$\!\!\!$&$\!\!\!$$\!\!\!$&$\!\!\!$$\!\!\!$&$\!\!\!$&\\
   1 $\!\!\!$&$\!\!\!$ 201.802078 $\!\!\!$&$\!\!\!$ -47.750278 $\!\!\!$&$\!\!\!$  3016.264 $\!\!\!$&$\!\!\!$   -93.270 $\!\!\!$&$\!\!\!$  -3.25 $\!\!\!$&$\!\!\!$   0.52 $\!\!\!$&$\!\!\!$  11.78 $\!\!\!$&$\!\!\!$   6.12	$\!\!\!$&$\!\!\!$  99.999 $\!\!\!$&$\!\!\!$ 99.999 $\!\!\!$&$\!\!\!$ 20.535 $\!\!\!$&$\!\!\!$  0.056 $\!\!\!$&$\!\!\!$ 19.842 $\!\!\!$&$\!\!\!$  0.155 $\!\!\!$&$\!\!\!$ 19.332  $\!\!\!$&$\!\!\!$ 0.219 $\!\!\!$&$\!\!\!$ 18.729 $\!\!\!$&$\!\!\!$  0.071 $\!\!\!$&$\!\!\!$ 99.999 $\!\!\!$&$\!\!\!$ 99.999  $\!\!\!$&$\!\!\!$    0	$\!\!\!$&$\!\!\!$   28\\
    2 $\!\!\!$&$\!\!\!$ 201.763730 $\!\!\!$&$\!\!\!$ -47.750272 $\!\!\!$&$\!\!\!$  3407.004 $\!\!\!$&$\!\!\!$   -93.189 $\!\!\!$&$\!\!\!$   7.26 $\!\!\!$&$\!\!\!$   8.23 $\!\!\!$&$\!\!\!$   1.98 $\!\!\!$&$\!\!\!$   3.16	$\!\!\!$&$\!\!\!$  99.999 $\!\!\!$&$\!\!\!$ 99.999 $\!\!\!$&$\!\!\!$ 19.254 $\!\!\!$&$\!\!\!$  0.064 $\!\!\!$&$\!\!\!$ 18.681 $\!\!\!$&$\!\!\!$  0.044 $\!\!\!$&$\!\!\!$ 18.214  $\!\!\!$&$\!\!\!$ 0.063 $\!\!\!$&$\!\!\!$ 17.909 $\!\!\!$&$\!\!\!$  0.033 $\!\!\!$&$\!\!\!$ 99.999 $\!\!\!$&$\!\!\!$ 99.999  $\!\!\!$&$\!\!\!$   69	$\!\!\!$&$\!\!\!$   -1\\
    3 $\!\!\!$&$\!\!\!$ 201.799369 $\!\!\!$&$\!\!\!$ -47.750242 $\!\!\!$&$\!\!\!$  3043.872 $\!\!\!$&$\!\!\!$   -92.734 $\!\!\!$&$\!\!\!$  -0.41 $\!\!\!$&$\!\!\!$  14.48 $\!\!\!$&$\!\!\!$   7.24 $\!\!\!$&$\!\!\!$   6.54	$\!\!\!$&$\!\!\!$  99.999 $\!\!\!$&$\!\!\!$ 99.999 $\!\!\!$&$\!\!\!$ 21.173 $\!\!\!$&$\!\!\!$  0.080 $\!\!\!$&$\!\!\!$ 20.441 $\!\!\!$&$\!\!\!$  0.046 $\!\!\!$&$\!\!\!$ 19.854  $\!\!\!$&$\!\!\!$ 0.044 $\!\!\!$&$\!\!\!$ 19.437 $\!\!\!$&$\!\!\!$  0.058 $\!\!\!$&$\!\!\!$ 99.999 $\!\!\!$&$\!\!\!$ 99.999  $\!\!\!$&$\!\!\!$   39	$\!\!\!$&$\!\!\!$   -1\\
    4 $\!\!\!$&$\!\!\!$ 201.641896 $\!\!\!$&$\!\!\!$ -47.750155 $\!\!\!$&$\!\!\!$  4648.364 $\!\!\!$&$\!\!\!$   -92.553 $\!\!\!$&$\!\!\!$  11.20 $\!\!\!$&$\!\!\!$   8.53 $\!\!\!$&$\!\!\!$   9.50 $\!\!\!$&$\!\!\!$   6.65	$\!\!\!$&$\!\!\!$  99.999 $\!\!\!$&$\!\!\!$ 99.999 $\!\!\!$&$\!\!\!$ 20.938 $\!\!\!$&$\!\!\!$  0.060 $\!\!\!$&$\!\!\!$ 20.242 $\!\!\!$&$\!\!\!$  0.083 $\!\!\!$&$\!\!\!$ 19.708  $\!\!\!$&$\!\!\!$ 0.051 $\!\!\!$&$\!\!\!$ 19.308 $\!\!\!$&$\!\!\!$  0.081 $\!\!\!$&$\!\!\!$ 99.999 $\!\!\!$&$\!\!\!$ 99.999  $\!\!\!$&$\!\!\!$   43	$\!\!\!$&$\!\!\!$   -1\\
    5 $\!\!\!$&$\!\!\!$ 201.611638 $\!\!\!$&$\!\!\!$ -47.750119 $\!\!\!$&$\!\!\!$  4956.656 $\!\!\!$&$\!\!\!$   -92.562 $\!\!\!$&$\!\!\!$   0.43 $\!\!\!$&$\!\!\!$   0.63 $\!\!\!$&$\!\!\!$  11.93 $\!\!\!$&$\!\!\!$   3.89	$\!\!\!$&$\!\!\!$  99.999 $\!\!\!$&$\!\!\!$ 99.999 $\!\!\!$&$\!\!\!$ 17.145 $\!\!\!$&$\!\!\!$  0.015 $\!\!\!$&$\!\!\!$ 16.632 $\!\!\!$&$\!\!\!$  0.016 $\!\!\!$&$\!\!\!$ 16.238  $\!\!\!$&$\!\!\!$ 0.007 $\!\!\!$&$\!\!\!$ 15.953 $\!\!\!$&$\!\!\!$  0.035 $\!\!\!$&$\!\!\!$ 99.999 $\!\!\!$&$\!\!\!$ 99.999  $\!\!\!$&$\!\!\!$   19	$\!\!\!$&$\!\!\!$    0\\
    6 $\!\!\!$&$\!\!\!$ 201.609219 $\!\!\!$&$\!\!\!$ -47.750121 $\!\!\!$&$\!\!\!$  4981.298 $\!\!\!$&$\!\!\!$   -92.651 $\!\!\!$&$\!\!\!$  -1.55 $\!\!\!$&$\!\!\!$   6.82 $\!\!\!$&$\!\!\!$   1.60 $\!\!\!$&$\!\!\!$   7.53	$\!\!\!$&$\!\!\!$  99.999 $\!\!\!$&$\!\!\!$ 99.999 $\!\!\!$&$\!\!\!$ 20.856 $\!\!\!$&$\!\!\!$  0.036 $\!\!\!$&$\!\!\!$ 20.202 $\!\!\!$&$\!\!\!$  0.064 $\!\!\!$&$\!\!\!$ 19.660  $\!\!\!$&$\!\!\!$ 0.040 $\!\!\!$&$\!\!\!$ 19.263 $\!\!\!$&$\!\!\!$  0.047 $\!\!\!$&$\!\!\!$ 99.999 $\!\!\!$&$\!\!\!$ 99.999  $\!\!\!$&$\!\!\!$   57	$\!\!\!$&$\!\!\!$   -1\\
    7 $\!\!\!$&$\!\!\!$ 201.506294 $\!\!\!$&$\!\!\!$ -47.749911 $\!\!\!$&$\!\!\!$  6029.910 $\!\!\!$&$\!\!\!$   -92.178 $\!\!\!$&$\!\!\!$   6.71 $\!\!\!$&$\!\!\!$   5.49 $\!\!\!$&$\!\!\!$  21.84 $\!\!\!$&$\!\!\!$   4.47	$\!\!\!$&$\!\!\!$  99.999 $\!\!\!$&$\!\!\!$ 99.999 $\!\!\!$&$\!\!\!$ 19.905 $\!\!\!$&$\!\!\!$  0.033 $\!\!\!$&$\!\!\!$ 19.095 $\!\!\!$&$\!\!\!$  0.031 $\!\!\!$&$\!\!\!$ 18.552  $\!\!\!$&$\!\!\!$ 0.039 $\!\!\!$&$\!\!\!$ 18.136 $\!\!\!$&$\!\!\!$  0.060 $\!\!\!$&$\!\!\!$ 99.999 $\!\!\!$&$\!\!\!$ 99.999  $\!\!\!$&$\!\!\!$    0	$\!\!\!$&$\!\!\!$   -1\\
    8 $\!\!\!$&$\!\!\!$ 201.721527 $\!\!\!$&$\!\!\!$ -47.750125 $\!\!\!$&$\!\!\!$  3837.015 $\!\!\!$&$\!\!\!$   -91.161 $\!\!\!$&$\!\!\!$   1.23 $\!\!\!$&$\!\!\!$   3.45 $\!\!\!$&$\!\!\!$   7.76 $\!\!\!$&$\!\!\!$   9.23	$\!\!\!$&$\!\!\!$  99.999 $\!\!\!$&$\!\!\!$ 99.999 $\!\!\!$&$\!\!\!$ 20.844 $\!\!\!$&$\!\!\!$  0.057 $\!\!\!$&$\!\!\!$ 20.116 $\!\!\!$&$\!\!\!$  0.017 $\!\!\!$&$\!\!\!$ 19.692  $\!\!\!$&$\!\!\!$ 0.068 $\!\!\!$&$\!\!\!$ 19.347 $\!\!\!$&$\!\!\!$  0.069 $\!\!\!$&$\!\!\!$ 99.999 $\!\!\!$&$\!\!\!$ 99.999  $\!\!\!$&$\!\!\!$   62	$\!\!\!$&$\!\!\!$   -1\\
    9 $\!\!\!$&$\!\!\!$ 201.714951 $\!\!\!$&$\!\!\!$ -47.750153 $\!\!\!$&$\!\!\!$  3904.018 $\!\!\!$&$\!\!\!$   -91.634 $\!\!\!$&$\!\!\!$   5.27 $\!\!\!$&$\!\!\!$   3.52 $\!\!\!$&$\!\!\!$   2.33 $\!\!\!$&$\!\!\!$   6.89	$\!\!\!$&$\!\!\!$  99.999 $\!\!\!$&$\!\!\!$ 99.999 $\!\!\!$&$\!\!\!$ 20.354 $\!\!\!$&$\!\!\!$  0.050 $\!\!\!$&$\!\!\!$ 19.696 $\!\!\!$&$\!\!\!$  0.028 $\!\!\!$&$\!\!\!$ 19.261  $\!\!\!$&$\!\!\!$ 0.051 $\!\!\!$&$\!\!\!$ 18.909 $\!\!\!$&$\!\!\!$  0.032 $\!\!\!$&$\!\!\!$ 99.999 $\!\!\!$&$\!\!\!$ 99.999  $\!\!\!$&$\!\!\!$   70	$\!\!\!$&$\!\!\!$   -1\\
   10 $\!\!\!$&$\!\!\!$ 201.629034 $\!\!\!$&$\!\!\!$ -47.750053 $\!\!\!$&$\!\!\!$  4779.417 $\!\!\!$&$\!\!\!$   -91.229 $\!\!\!$&$\!\!\!$   4.24 $\!\!\!$&$\!\!\!$   3.80 $\!\!\!$&$\!\!\!$   7.46 $\!\!\!$&$\!\!\!$   1.69	$\!\!\!$&$\!\!\!$  99.999 $\!\!\!$&$\!\!\!$ 99.999 $\!\!\!$&$\!\!\!$ 19.275 $\!\!\!$&$\!\!\!$  0.024 $\!\!\!$&$\!\!\!$ 18.694 $\!\!\!$&$\!\!\!$  0.024 $\!\!\!$&$\!\!\!$ 18.310  $\!\!\!$&$\!\!\!$ 0.030 $\!\!\!$&$\!\!\!$ 17.953 $\!\!\!$&$\!\!\!$  0.057 $\!\!\!$&$\!\!\!$ 99.999 $\!\!\!$&$\!\!\!$ 99.999  $\!\!\!$&$\!\!\!$    0	$\!\!\!$&$\!\!\!$   55\\
$[\dots]$$\!\!\!$&$\!\!\!$$[\dots]$$\!\!\!$&$\!\!\!$$[\dots]$$\!\!\!$&$\!\!\!$$[\dots]$$\!\!\!$&$\!\!\!$$[\dots]$$\!\!\!$&$\!\!\!$$[\dots]$$\!\!\!$&$\!\!\!$$[\dots]$
$\!\!\!$&$\!\!\!$$[\dots]$$\!\!\!$&$\!\!\!$$[\dots]$$\!\!\!$&$\!\!\!$$[\dots]$$\!\!\!$&$\!\!\!$$[\dots]$$\!\!\!$&$\!\!\!$$[\dots]$$\!\!\!$&$\!\!\!$$[\dots]$
$\!\!\!$&$\!\!\!$$[\dots]$$\!\!\!$&$\!\!\!$$[\dots]$$\!\!\!$&$\!\!\!$$[\dots]$$\!\!\!$&$\!\!\!$$[\dots]$$\!\!\!$&$\!\!\!$$[\dots]$$\!\!\!$&$\!\!\!$$[\dots]$
$\!\!\!$&$\!\!\!$$[\dots]$$\!\!\!$&$\!\!\!$$[\dots]$$\!\!\!$&$\!\!\!$$[\dots]$$\!\!\!$&$\!\!\!$$[\dots]$\\
\hline
\end{tabular}}
\caption{First lines of the electronically available catalog.}
\end{sidewaystable}

\end{document}